\documentclass[ fleqn,usenatbib]{mnras}

\usepackage{newtxtext,newtxmath}

\usepackage[T1]{fontenc}
\usepackage{ae,aecompl}
\usepackage[utf8]{inputenc}
\usepackage{hyperref}
\hypersetup{
    colorlinks=true,
    linkcolor=blue,
    citecolor=blue,
}
\usepackage{mn2e-breakabs}
\usepackage{graphicx}	
\usepackage{amsmath}	
\usepackage{amssymb}	
\usepackage{booktabs, blkarray}
\usepackage[dvipsnames]{xcolor}
\usepackage[normalem]{ulem}
\usepackage{multirow}

\newcommand{\aperp}{\alpha_\perp}	
\newcommand{\apara}{\alpha_\parallel}
\newcommand{\fsig}{f{\sigma_8}}
\newcommand{\hmpc }{$h^{-1}$Mpc}

\title[BAO and RSD analysis from eBOSS LRG]{
The Completed SDSS-IV extended Baryon Oscillation Spectroscopic Survey: measurement of the BAO and growth rate of structure of the luminous red galaxy sample from the anisotropic correlation function between redshifts 0.6 and 1}

\author[Bautista et al.]{\parbox{\textwidth}{
Julian E. Bautista$^1$\thanks{julian.bautista@port.ac.uk}, 
Romain Paviot$^{2,3}$\thanks{romain.paviot@lam.fr},
Mariana Vargas Maga\~na$^{4}$\thanks{mmaganav@fisica.unam.mx}, 
Sylvain de la Torre$^{2}$,
Sebastien Fromenteau$^{5}$,
Hector Gil-Mar\'{i}n$^{6, 7}$,
Ashley J. Ross$^{8}$, 
Etienne Burtin$^{9}$,
Kyle S. Dawson$^{10}$,
Jiamin Hou$^{11}$,
Jean-Paul Kneib$^{12}$, 
Arnaud de Mattia$^{9}$,
Will J. Percival$^{13, 14, 15}$,
Graziano Rossi$^{16}$,
Rita Tojeiro$^{17}$, 
Cheng Zhao$^{12}$,
Gong-Bo Zhao$^{18, 19, 1}$, 
Shadab Alam$^{20}$,
Joel Brownstein$^{10}$,
Michael J. Chapman$^{13, 14}$,
Peter D. Choi$^{16}$,
Chia-Hsun Chuang$^{21}$,
St\'ephanie Escoffier$^{3}$,
Axel de la Macorra$^{4}$,
H\'elion du Mas des Bourboux$^{10}$,
Faizan G. Mohammad$^{13, 14}$,
Jeongin Moon$^{16}$,
Eva-Maria M\"uller$^{22}$,
Seshadri Nadathur$^{1}$,
Jeffrey A. Newman$^{23}$,
Donald Schneider$^{24, 25}$,
Hee-Jong Seo$^{26}$,
Yuting Wang$^{18}$
} \vspace*{10pt} \\ 
$^{1}${
Institute of Cosmology \& Gravitation, 
Dennis Sciama Building, University of Portsmouth, 
Portsmouth, PO1 3FX, UK
}\\
$^{2}${
Aix Marseille Univ, CNRS, CNES, 
LAM, Marseille, France
}\\
$^{3}${
Aix Marseille Univ, CNRS/IN2P3,
CPPM, Marseille, France
}\\
$^{4}${
Instituto de F\'isica, 
Universidad Nacional Aut\'onoma de M\'exico, 
Apdo. Postal 20-364, Ciudad de M\'exico, M\'exico
}\\
$^{5}${ 
Instituto de Ciencias F\'isicas, 
Universidad Nacional Aut\'onoma de M\'exico, 
Av. Universidad s/n, 62210 Cuernavaca, Mor., Mexico
}\\
$^{6}${
Institut de Ci\`encies del Cosmos,  
Universitat  de  Barcelona,  ICCUB,  
Mart\'i  i  Franqu\`es  1,  E08028  Barcelona,  Spain
}\\
$^{7}${
Institut  d’Estudis  Espacials  de  Catalunya  (IEEC),  
E08034  Barcelona,  Spain
}\\
$^{8}${
Center for Cosmology and Astro-Particle Physics,
Ohio State University, Columbus, OH 43210
}\\
$^{9}${
CEA, Centre de Saclay, Irfu/SPP,  F-91191 Gif-sur-Yvette, France.
}\\
$^{10}${
Department of Physics and Astronomy, 
University of Utah, Salt Lake City, UT 84112, USA.
}\\
$^{11}${
Max-Planck-Institut f\"ur Extraterrestrische Physik, 
Postfach 1312, Giessenbachstr., 85748 Garching bei M\"unchen, Germany
}\\
$^{12}${
Laboratoire d\'astrophysique, Ecole Polytechnique F\'ed\'erale de Lausanne
Observatoire de Sauverny, 1290 Versoix, Switzerland
}\\
$^{13}${
Waterloo Centre for Astrophysics, 
University of Waterloo, Waterloo, ON N2L 3G1, Canada
}\\
$^{14}${ 
Department of Physics and Astronomy, University of Waterloo, 
Waterloo, ON N2L 3G1, Canada
}\\
$^{15}${
Perimeter Institute for Theoretical Physics, 
31 Caroline St. North, Waterloo, ON N2L 2Y5, Canada
}\\
$^{16}${
Department of Physics and Astronomy, 
Sejong University, Seoul, 143-747, Korea 
}\\
$^{17}${
School of Physics and Astronomy, 
University of St Andrews, 
St Andrews, KY16 9SS, UK
}\\
$^{18}${
National Astronomy Observatories, 
Chinese Academy of Science, 
Beijing, 100012, P. R. China
}\\
$^{19}${
College of Astronomy and Space Sciences, 
University of Chinese Academy of Sciences, Beijing 100049,
China
}\\
$^{20}${
Institute for Astronomy,
University of Edinburgh, Royal Observatory,
Edinburgh EH9 3HJ, UK
}\\
$^{21}${
Kavli Institute for Particle Astrophysics and Cosmology, 
Stanford University, 452 Lomita Mall, Stanford, CA 94305, USA
}\\
$^{22}${
Sub-department of Astrophysics, Department of Physics, 
University of Oxford, Denys Wilkinson Building, Keble Road, Oxford OX1 3RH, UK
}\\
$^{23}${
Department of Physics and Astronomy and PITT PACC,
University of Pittsburgh, Pittsburgh, PA 15260, USA
}\\
$^{24}${
Department of Astronomy and Astrophysics, The Pennsylvania State University,
University Park, PA 16802, USA
}\\
$^{25}${
Institute for Gravitation and the Cosmos, The Pennsylvania State University, 
University Park, PA 16802, USA
}\\
$^{26}${
Department of Physics and Astronomy,
Ohio University,
251B Clippinger Labs, Athens, OH 45701
}
}

\date{Accepted XXX. Received YYY; in original form ZZZ}

\pubyear{2020}

\hypersetup{draft}

\begin{document}
\label{firstpage}
\pagerange{\pageref{firstpage}--\pageref{lastpage}}
\maketitle

\begin{abstract}

We present the cosmological analysis of the configuration-space anisotropic clustering in the completed Sloan Digital Sky Survey IV (SDSS-IV) extended Baryon Oscillation Spectroscopic Survey (eBOSS) DR16 galaxy sample. This sample consists of luminous red galaxies (LRGs) spanning the redshift range $0.6 < z < 1$, at an effective redshift of $z_{\rm eff}=0.698$. It combines 174\,816 eBOSS LRGs and 202\,642 BOSS CMASS galaxies. We extract and model the baryon acoustic oscillations (BAO) and redshift-space distortions (RSD) features from the galaxy two-point correlation function to infer geometrical and dynamical cosmological constraints. The adopted methodology is extensively tested on a set of realistic simulations. The correlations between the inferred parameters from the BAO and full-shape correlation function analyses are estimated. This allows us to derive joint constraints on the three cosmological parameter combinations: $D_M(z)/r_d$, $D_H(z)/r_d$ and $f\sigma_8(z)$, where $D_M$ is the comoving angular diameter distance, $D_H$ is Hubble distance, $r_d$ is the comoving BAO scale, $f$ is the linear growth rate of structure, and $\sigma_8$ is the amplitude of linear matter perturbations. After combining the results with those from the parallel power spectrum analysis of Gil-Marin et al. 2020, we obtain the constraints: 
  $D_M/r_d =  17.65 \pm 0.30$,
     $D_H/r_d = 19.77 \pm 0.47$, 
     $\fsig = 0.473 \pm 0.044$.
These measurements are consistent with a flat $\Lambda$CDM model with standard gravity.

\end{abstract}




\section{Introduction}

The large-scale structure (LSS) in the late Universe is a fundamental probe of the cosmological model, sensitive to both universal expansion and structure growth, and complementary to early Universe observations from the cosmic microwave background. The LSS can be mapped by large redshift surveys through systematic measurements of the three-dimensional positions of matter tracers such as galaxies or quasars. Because the observed LSS is the result of the growth of initial matter perturbations through gravity in an expanding universe, it gives the possibility of both testing the expansion and structure growth histories, which in turn put us in a unique position to solve the question of the origin of the late acceleration of the expansion and dark energy \citep{clifton_modified_2012, weinberg_observational_2013, 
zhai_evaluation_2017, ferreira_cosmological_2019}.

Over the last two decades, redshift surveys have explored increasingly larger volumes of the Universe at different cosmic times. The methodology to extract the cosmological information from those redshift surveys has evolved and has now reached maturity. Particularly, the baryon acoustic oscillations (BAO) and the redshift-space distortions (RSD) in the two-point and three-point statistics of the galaxy spatial distribution are now key observables to constrain cosmological models. The BAO horizon scale imprinted in the matter distribution was frozen in the LSS at the drag epoch, slightly after matter-radiation decoupling. This characteristic scale can still be seen in the large-scale distribution of galaxies at late times and be used as a standard ruler to measure the expansion history. At the same time, the galaxy peculiar velocities distorting the line-of-sight cosmological distances based on observed redshifts, are sensitive on large scales to the coherent motions induced by the growth rate of structure, which in turn depends on the strength of gravity. BAO and RSD are highly complementary, as they allow both geometrical and dynamical cosmological constraints from the same observations.

The signature of baryons in the clustering of galaxies was first detected in the Sloan  Digital  Sky  Survey (SDSS; \citealt{eisenstein_detection_2005}) 
and 2dF Galaxy Redshift Survey (2dFGRS; \citealt{percival_2df_2001,cole_2df_2005}). Since then, further measurements using the 2dFGRS, SDSS and additional surveys have improved the accuracy of BAO measurements and extended the range of redshifts covered from $z=0$ to $z=1$. Examples of analyses include those of the SDSS-II \citep{percival_baryon_2010}, 6dFGS \citep{beutler_6df_2011}, WiggleZ, \citep{kazin_wigglez_2014} and SDSS-MGS \citep{ross_clustering_2015} galaxy surveys. An important milestone was achieved with the Baryon Oscillation Spectroscopic Survey (BOSS; \citealt{dawson_baryon_2013}), part of the third generation of the Sloan Digital Sky Survey \citep{eisenstein_sdss-iii:_2011}. This
allowed the most precise measurements of BAO using galaxies achieved to date using galaxies as direct tracers \citep{alam_clustering_2017} and Lyman-$\alpha$ forest measurements \citep{bautista_measurement_2017, 
du_mas_des_bourboux_baryon_2017}, reaching a relative precision of 
1 per cent on the distance relative to the sound horizon at the drag epoch.

Although RSD have been understood and measured since the late 1980s \citep{kaiser_clustering_1987}, it is only in the last decade when there has been significant interest in deviations from standard gravity that would explain the apparent late-time acceleration of the expansion of the Universe, that the ability of RSD measurements to provide such tests has been explored \citep{guzzo_test_2008,song_reconstructing_2009}. This has resulted in renewed interest in RSD with examples of RSD measurement from the WiggleZ \citep{blake_wigglez_2011}, 6dFGRS \citep{beutler_6df_2012}, SDSS-II \citep{samushia_interpreting_2012}, SDSS-MGS \citep{Howlett_clustering_2015}, FastSound \citep{okumura_subaru_2016}, and VIPERS \citep{pezzotta_vimos_2017} galaxy surveys, with BOSS achieving the best precision of $\sim6$\% on the parameter combination $f\sigma_8$ \citep{beutler_clustering_2017, grieb_clustering_2017, sanchez_clustering_2017, satpathy_clustering_2017}, which is commonly used to quantify the amplitude of the velocity power spectrum.

The extended Baryon Oscillation Spectroscopic Survey 
(eBOSS; \citealt{dawson_sdss-iv_2016}) program is the successor of 
BOSS in the fourth generation of the SDSS \citep{blanton_sloan_2017}. 
It maps the LSS using four main tracers: Luminous Red Galaxies (LRGs),
Emission Line Galaxies (ELGs), quasars used as direct tracers of the
density field, and quasars from whose spectra we can measure the
Ly$\alpha$ forest. With respect to BOSS, it explores galaxies at higher
redshifts, covering the range $0.6<z<2.2$. Using the first two years of
data from Data Release 14 (DR14), BAO and RSD measurements have been
performed using different tracers and methods: 
LRG BAO \citep{bautista_sdss-iv_2018}, 
LRG RSD \citep{icaza-lizaola_clustering_2020}, 
quasar BAO \citep{ata_clustering_2018}, 
quasar BAO with redshift weights \citep{zhu_clustering_2018}, 
quasar BAO Fourier-space \citep{wang_clustering_2018}, 
quasar RSD Fourier-space \citep{gil-marin_clustering_2018}, 
quasar RSD Fourier-space with redshift weights \citep{ruggeri_optimal_2017, ruggeri_clustering_2019},
quasar RSD in configuration space 
\citep{hou_clustering_2018, zarrouk_clustering_2018}, 
and quasar tomographic RSD in Fourier space with redshift weights \citep{zhao_clustering_2019}.

In this paper we perform the BAO and RSD analyses in configuration space of the completed eBOSS LRG sample, part of Data Release 16. This work is part of a series of papers using different tracers and methods\footnote{A summary of all SDSS BAO and RSD measurements with accompanying legacy figures can be found at \\
\url{sdss.org/science/final-bao-and-rsd-measurements/} \\
and the cosmological interpretation of these measurements can be found at \\
\url{sdss.org/science/cosmology-results-from-eboss/}}. The official SDSS-IV DR16 quasar catalog is described in \citet{lyke_2020}. The production of the catalogs specific for large-scale clustering measurements of the quasar and LRG sample (input for this work) is described in \citet{ross_2020}, while the analogous work for the ELG sample is described in \cite{raichoor_2020}. From the same LRG catalog, \citet{gil-marin_2020} report the BAO and RSD analyses in Fourier space. The BAO and RSD constraints from the quasar sample are presented by \citet{hou_2020} in configuration space and by \citet{neveux_2020} in Fourier space. The clustering from the ELG sample is described by \citet{de_mattia_2020} in Fourier space and by \citet{amelie_2020} in configuration space. Finally, a series of articles describes the simulations used to test the different methodologies for each tracer. The approximate mocks used to estimate covariance matrices and assess observational systematics for the LRG, ELG, and quasar samples are described in \citet{zhao_2020} (see also \citet{lin_2020} for an alternative method for ELGs), while realistic N-body simulations were produced by \citet{rossi_2020} for the LRG sample, by \citet{smith_2020} for the quasar sample, and by \citet{alam_2020} for the ELG sample. In \citet{avila_2020}, halo occupation models for ELGs are studied. A machine-learning method to remove systematics caused by photometry was applied to the ELG sample \citep{kong_2020} and a new method to account for fiber collisions in the eBOSS sample is described in \citet{mohammad_2020}. The BAO analysis of the Lyman-$\alpha$ forest sample is presented by \citet{du_mas_des_bourboux_2020}. The final cosmological implications from all these clustering analyses are presented in \citet{mueller_2020}. 

The paper is organized as follows. Section \ref{sec:dataset} describes the LRG dataset and simulations used in this analysis. Section \ref{sec:method} presents the adopted methodology and particularly BAO and RSD theoretical models. We estimate biases and systematic errors from different sources in section \ref{sec:robustness}. We present BAO and RSD results in Section \ref{sec:results} and finally conclude in
Section \ref{sec:conclusion}.

\section{Dataset} 
\label{sec:dataset}

In this section, we summarize the observations, catalogs, and mock datasets that are used to test our methodology, as well as  the clustering statistics used in this work.

\subsection{Spectroscopic observations and reductions}

The fourth generation of the Sloan Digital Sky Survey \citep[SDSS-IV][]{blanton_sloan_2017} employed the two multi-object BOSS spectrographs 
\citep{smee_multi-object_2013} installed on the 2.5-meter telescope
\citep{gunn_2.5_2006} at the Apache Point Observatory in New Mexico,
USA, to carry out spectroscopic observations for eBOSS. 
The target sample of LRGs, the analysis of which is our focus, was selected from the optical SDSS photometry from DR13 \citep{albareti_13th_2017}, with
additional infrared information from the WISE satellite 
\citep{lang_wise_2014}. The final targeting algorithm is described in detail in 
\citet{prakash_sdss-iv_2016} and produced about 60 deg$^{-2}$ LRG targets over the 7500 deg$^{2}$ of the eBOSS footprint, of which 50 deg$^{-2}$ were observed spectroscopically. The selection was tested over 466 deg$^2$ covered during the Sloan Extended Quasar, ELG, and LRG Survey (SEQUELS), confirming 
that more than 41 deg$^{-2}$ LRGs have $0.6 < z < 1.0$ 
\citep{dawson_sdss-iv_2016}. 

The raw CCD images were converted to one-dimensional, wavelength and flux
calibrated spectra using version \textsc{v5\_13\_0} of the SDSS spectroscopic
pipeline \textsc{idlspec2d}\footnote{Publicly available at
\url{sdss.org/dr16/software/products}}. Two main improvements of this pipeline
since its previous release \citep[DR14;][]{abolfathi_fourteenth_2018}  include
a new library of stellar templates for flux calibration and a more stable 
extraction procedure. \citet{ahumada_sixteenth_2019} provide a summary of 
all improvements of the spectroscopic pipeline since SDSS-III. 

The redshift of each LRG was estimated with the \textsc{redrock}
algorithm\footnote{Publicly available at \url{github.com/desihub/redrock}}. 
This algorithm improves classification rates with respect to its predecessor
\textsc{redmonster} \citep{hutchinson_redshift_2016}. \textsc{redrock} 
uses templates derived from principal component analysis of SDSS data to 
classify spectra, which is followed by a redshift refinement procedure 
that uses stellar population models for galaxies. On average, 96.5 per cent of
spectra yield a confident redshift estimate with \textsc{redrock} compared to
90 per cent with \textsc{redmonster}, with less than 1 per cent of catastrophic redshift 
errors (details can be found in Ross et al., 2020). 

\subsection{Survey geometry and observational features}

The full procedure to model the survey geometry and correct for
observational features is described in detail in the companion paper
\citet{ross_2020}. We summarize it in the following. 

The random catalog allows estimating the survey 
geometry and number density of galaxies in the 
observed sample. It contains a random population of objects
with the same radial and angular selection functions as the data. 
A random uniform sample of points is drawn over the angular footprint 
of eBOSS targets to model its geometry. We use random samples with 
50 times more objects than in the data to minimize the shot noise 
contribution in the estimated correlation function, and redshifts are 
randomly taken from galaxy redshifts in the data. 
A series of masks are then applied to both data and random samples 
in order to eliminate regions with bad photometric properties, 
targets that collide with quasar spectra (which had priority in 
fiber assignement), and the centerpost region of the plates where it 
is physically impossible to put a fiber. All masks combined cover 17 
per cent of the initial footprint, with the quasar collision mask
accounting for 11 per cent. The spectroscopic information is finally 
matched to the remaining targets. 

About 4 per cent of the LRG targets were not observed due to 
\textit{fiber collisions}, i.e., when a group of two or more 
galaxies are closer than 
62$^{\prime\prime}$ they cannot all receive a fiber. On regions of 
the sky observed more than once, some collisions could be resolved. 
These collisions can bias the clustering measurements so we applied 
the following correction: $N_{\rm targ}$ objects in a given collision 
group for which $N_{\rm spec}$ have a spectrum, all objects are 
up-weighted by $w_{\rm cp} = N_{\rm  targ}/N_{\rm spec}$. 
This is different compared to \citet{bautista_sdss-iv_2018}, 
where the weight of the collided object without spectrum was 
transferred to its nearest neighbor with valid spectrum. Both 
corrections are only approximations valid on scales larger than
62$^{\prime\prime}$. An unbiased correction method is described in
\cite{bianchi_unbiased_2017} and applied to eBOSS samples in 
\citet{mohammad_2020}. We show in Appendix \ref{app:pip_weights} 
that our results are insensitive to the correction method
since it affects mostly the smallest scales. 

A similar procedure as in \citet{bautista_sdss-iv_2018} was used to
account for the $3.5$ per cent of LRG targets without reliable redshift 
estimate. The \textit{redshift-failure} weight $w_{\rm noz}$ 
acts as an inverse probability weight, boosting galaxies with good 
redshifts such that this weighted sample is an unbiased sampling of the 
full population. This assumes that the probability of a given galaxy 
being selected is a function of both its trace position on the CCD and 
the overall signal-to-noise ratio of the spectrograph in which this 
target was observed, and that the galaxies not observed are 
statistically equivalent to the observed galaxies.
Spurious fluctuations in the target selection caused by the photometry
are corrected by weighting each galaxy by $w_{\rm sys}$. These weights 
are computed with a multi-linear regression on the observed relations 
between the angular over-densities of galaxies versus stellar density, 
seeing and galactic extinction. Fitting all quantities simultaneously
automatically accounts for their correlations. 
The weights $w_{\rm noz}$ and $w_{\rm sys}$ are computed independently. 

The observational completeness creates artificial angular variations 
of the density that are accounted for using the random catalog. 
The completeness is defined as the ratio of the number of 
weighted spectra (including those classified as stars or quasars) to
the number of targets (Eq.~11 in Ross et al., 2020). This quantity is
computed per sky \textit{sector}, i.e., a connected region observed by 
a unique set of plates. We downweight each point in the random catalog 
by the completeness of its corresponding sky sector.  

Optimal weights for large-scale correlations, known as FKP weights
\citep{feldman_power-spectrum_1994}, are computed with the estimated
comoving density of tracers $\bar{n}(z)$ as a function of redshift 
using our fiducial cosmology in Table~\ref{tab:cosmologies}. The final 
weight for each galaxy is defined\footnote{Note that this definition 
differs from the one used in BOSS, 
where $w = (w_{\rm noz}+ w_{\rm cp}-1)w_{\rm sys} w_{\rm FKP}$.}
as $w = w_{\rm noz}w_{\rm cp}w_{\rm syst} w_{\rm FKP}$. 
The weight for each galaxy from the random catalogue is the same, with the completeness information already included in $w_{\rm sys}$.

\begin{figure} 
    \centering
    \includegraphics[width=\columnwidth]{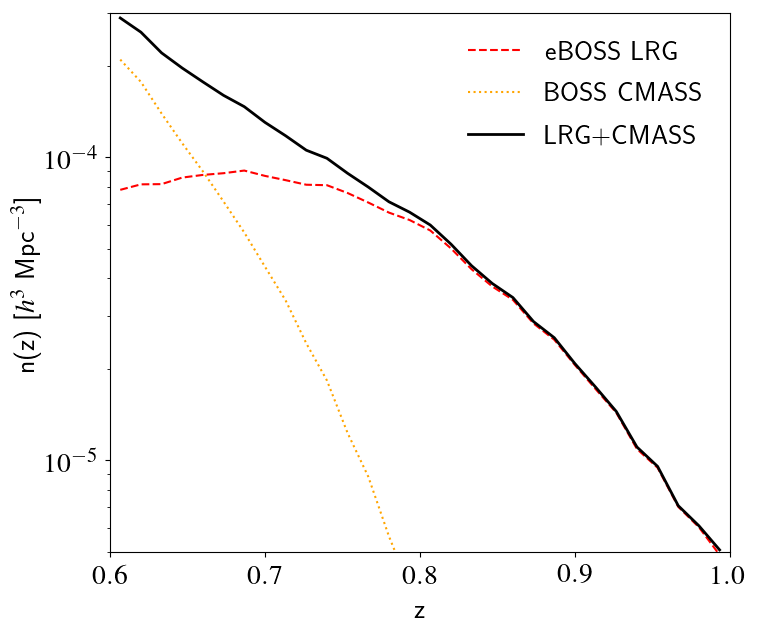}
    \caption{The observed number density of eBOSS LRGs (dashed curve), BOSS CMASS galaxies (dotted curve), and combined CMASS+LRG sample galaxies (solid curve) at $0.6<z<1$. This combines NGC and SGC fields.}
  \label{fig:nz}
\end{figure}

The eBOSS sample of LRGs overlaps in area and redshift range with the 
highest-redshift bin of the CMASS sample ($0.5<z<0.75$). We combine the 
eBOSS LRG sample with all the $z > 0.6$ BOSS CMASS galaxies and their 
corresponding random catalog (including the non-overlapping with eBOSS), 
making sure that the data-to-random number ratio is the same for both 
samples. This combination is beneficial for two reasons. First, the 
combined sample supersedes the last redshift bin of BOSS measurements 
while being completely independent of the first two lower redshift bins. 
Second, the reconstruction technique applied to this sample (see next 
section) benefits from a higher density of tracers, reducing potential 
noise introduced by the procedure. 
The new eBOSS LRG sample covers 4,242 deg$^{2}$ of the total 
BOSS CMASS footprint of 9,494 deg$^2$ (NGC and SGC combined). 
Considering their spectroscopic weights, the new eBOSS sample has 
185,295 new redshifts over $0.6 < z < 1.0$ while CMASS contributes with 
104,865 redshifts in the overlapping area and 111,892 in the 
non-overlapping area. A total of 402,052 LRGs over $0.6 < z < 1.0$ 
contribute to this measurement, with a total effective comoving volume of 2.72 Gpc$^3$ (1.43 Gpc$^3$ from the CMASS sample and 1.28 Gpc$^3$ from the new eBOSS sample).
A detailed description of these numbers is given in \citet{ross_2020}. 
In the following, 
we simply refer to the combined CMASS+LRG sample as the eBOSS LRG 
sample. The number density of CMASS galaxies, LRGs, and combined 
CMASS+LRG sample are presented in Fig. \ref{fig:nz}.

\subsection{Reconstruction}
\label{sec:reconstruction}

While constraints on the growth rate of structure are obtained using the information from the full shape of the correlation function, 
BAO analyses extract the cosmological information only from the position of the BAO peak. In our BAO analysis, 
we applied the reconstruction technique of
\citet{burden_efficient_2014, burden_reconstruction_2015} to the observed galaxy density field in order to remove a fraction of the redshift-space distortions, as well as non-linear motions of galaxies that smeared out the BAO peak. This technique sharpens the BAO feature in the two-point statistics in Fourier and configuration space, increasing the precision of the measurement of the acoustic scale. Reconstruction is applied on actual data and on mock catalogs using a publicly available\footnote{\url{https://github.com/julianbautista/eboss_clustering}} code \citep{bautista_sdss-iv_2018}. Our final BAO results are solely based on reconstructed catalogs, while full-shape results use the pre-reconstruction sample. 

We apply reconstruction to the full eBOSS+CMASS final LRG catalog. We use our fiducial cosmology from Table~\ref{tab:cosmologies} to convert redshifts to comoving distances. For the reconstruction, we fix the bias value to $b = 2.3$ and assume the standard gravity relation between the growth rate of structure and $\Omega_m$, i.e. $f = \Omega_m^{6/11}(z=0.7) = 0.815$. We use a smoothing scale of 
$15$ \hmpc. The BAO results are not sensitive to small variations of those parameter choices as studied in \cite{carter_impact_2019}.

\subsection{Mocks}
\label{sec:mocks_description}

In order to test the overall methodology and study the impact of systematic effects, we have constructed several sets of mock samples. Approximate methods are considered to be sufficient for covariance matrix estimates and to derive systematic biases in BAO measurements. However, the full-shape analysis of the 
correlation function requires more realistic N-body simulations, particularly in order to test the modeling. In this study, our synthetic datasets are the following:

\begin{itemize}

\item 1000 realisations of the LRG eBOSS+CMASS survey geometry using the \textsc{EZmock} method 
\citep{chuang_ezmocks:_2015}, which employs the Zel'dovich approximation to compute the density field at a given redshift and populate it with galaxies. This method is fast and has been calibrated to reproduce the two- and three-point statistics of the given galaxy sample, to a good approximation and up to mildly non-linear scales. The angular and redshift distributions of the eBOSS LRG sample in combination with the $z>0.6$ CMASS sample were reproduced in these mock catalogs. The full description of the \textsc{EZmock} LRG samples can be found in the companion paper Zhao et al. 2020. We use these mocks in several steps of our analysis: to infer the error covariance matrix of our clustering measurements in the data, to study the impact of observational systematic effects on cosmology, and to estimate the correlations between different methods for the calculation of the consensus results.  

\item 84 realisations of the \textsc{Nseries} mocks, which are N-body simulation snapshots populated with a single Halo Occupation distribution (HOD) model. These mock catalogs reproduce the angular and redshift distributions of the North Galactic Cap of the BOSS CMASS sample within the redshift range $0.43<z<0.70$  \citep{alam_clustering_2017}. While this dataset is not fully representative of the eBOSS LRG sample, we use these N-body mocks to test the RSD models down to the non-linear regime.  The number of available realisations and their large volume are ideal to test model accuracy in the high-precision regime. The covariance matrix for these mocks were computed from 2048 realisations of the same volume with the \textsc{MD-Patchy} approximated method \citep{kitaura_modelling_2014}. The redshift of those mocks is $z = 0.55$.

\item 27 realisations extracted from the {\sc OuterRim} N-body simulation \citep{heitmann_outer_2019}, and corresponding to cubical mocks of $1~h^{-3}~{\rm Gpc}^3$ each. The dark matter haloes have been populated with galaxies using four different HOD \citep{zheng_galaxy_2007, leauthaud_theoretical_2011, tinker_evolution_2013, hearin_beyond_2015} at 3 different luminosity thresholds to cover a large range of galaxy populations. These mocks are part of our internal \textsc{MockChallenge} and aimed at quantifying potential systematic errors originating from the HODs. A detailed description of these simulations and the \textsc{MockChallenge} can be found in the companion paper \citet{rossi_2020}. The redshift of those mocks is $z=0.695$.
\end{itemize}

\subsection{Fiducial cosmologies}
\label{sec:fiducial_cosmology}

The redshift of each galaxy is converted into radial comoving 
distances for clustering measurements by means of a fiducial cosmology. 
The fiducial cosmologies employed in this work are shown in 
Table~\ref{tab:cosmologies}. Our baseline choice, named ``Base'',
is a flat $\Lambda$CDM model matching the cosmology used in 
previous BOSS analyses \citep{alam_clustering_2017} with parameters within 1$\sigma$ of Planck best-fit parameters \citep{planck_collaboration_planck_2018-1}. 
Some of these cosmologies were used to produce the mock datasets
described in Section \ref{sec:mocks_description}. A choice of fiducial 
cosmology is also needed when computing the linear power spectrum 
$P_{\rm lin}(k)$, input for all our correlation function models in this
work (see Sections~\ref{sec:bao_modelling} and \ref{sec:rsd_modelling}). 
In Section~\ref{sec:systematics_bao} and 
Section~\ref{sec:systematics_rsd} 
we study the dependence of our results to the choice of fiducial
cosmology.

\begin{table}
    \centering
    \caption{Sets of cosmological models used in this work. All models are parameterised by their fraction of the total energy density in form of total matter $\Omega_m$, cold dark matter $\Omega_c$, baryons $\Omega_b$, and neutrinos $\Omega_\nu$, the Hubble constant $h = H_0/(100 {\rm km/s/Mpc})$, the primordial spectral index $n_s$ and primordial amplitude of power spectrum $A_s$. With these parameters we compute the normalisation of the linear
    power spectrum $\sigma_8$ at $z=0$ and the comoving sound horizon scale at drag epoch $r_{\rm drag}$. The different labels refer to our baseline choice
    (Base), the {\sc EZmocks} (EZ), the {\sc Nseries}  (NS), the {\sc OuterRim} (OR) cosmologies, and an additional model (X) with larger value for $\Omega_m$.
    }
    \label{tab:cosmologies}
    \begin{tabular}{cccccc}
    \hline
    \hline
     & Base & \sc EZ  & \sc NS & \sc OR &  $X$ \\
    \hline
$\Omega_m$& 0.310 & 0.307 & 0.286 & 0.265 & 0.350  \\
$\Omega_c$& 0.260 & 0.259 & 0.239 & 0.220 & 0.300  \\
$\Omega_b$& 0.048 & 0.048 & 0.047 & 0.045 & 0.048  \\
$\Omega_\nu$& 0.0014 & 0 & 0 & 0 & 0.0014 \\
$h$& 0.676 & 0.678 & 0.700 & 0.710 & 0.676  \\
$n_s$& 0.970 & 0.961 & 0.960 & 0.963 & 0.970  \\
$A_s$ [$10^{-9}$]& 2.041& 2.116& 2.147& 2.160& 2.041 \\
$\sigma_8(z=0)$ & 0.800 & 0.823 & 0.820 & 0.800 & 0.874  \\
$r_{\rm drag}$ [Mpc] & 147.78 & 147.66 & 147.15 & 149.35 & 143.17  \\
    \hline
    \hline
    \end{tabular}
\end{table}

\begin{table}
    \centering
    \caption{Values for the comoving angular diameter distance $D_M$
    and the Hubble distance $D_H = c/H(z)$ in units of the sound horizon scale 
    at drag epoch $r_d$, and the normalised growth rate of structures $\fsig$.
    These values are predictions from the cosmological models in 
    Table~\ref{tab:cosmologies} computed at typical redshifts used 
    in this work. }
    \label{tab:cosmo_derived_parameters}
    \begin{tabular}{lcccc}
    \hline
    \hline
    Model & $z_{\rm eff}$ &  $\frac{D_M}{r_{\rm drag}}$ & $\frac{D_H}{r_{\rm drag}}$ & $\fsig$ \\ 
    \hline
    Base & 0.698 & 17.436 & 20.194 & 0.456 \\
    Base & 0.560 & 14.529 & 21.960 & 0.465 \\
    {\sc EZ} & 0.698 & 17.429 & 20.211 & 0.467 \\
    {\sc NS} & 0.560 & 14.221 & 21.692 & 0.469 \\
    {\sc OR} & 0.695 & 16.717 & 19.866 & 0.447 \\
    X & 0.698 & 17.685 & 20.146 & 0.504 \\
    X & 0.560 & 14.778 & 22.019 & 0.518 \\
\hline
\hline
    \end{tabular}
\end{table}

We define the effective redshift of our data and mock catalogs as the weighted mean redshift of galaxy pairs,
\begin{equation}
z_{\rm eff} =\frac{\sum_{i>j}w_i w_j(z_i+z_j)/2}{\sum_{i>j}w_i w_j},
\label{eq:zeff}
\end{equation}
where $w_i$ is the total weight of the galaxy $i$ and the indices $i$,$j$ run over the galaxies in the considered catalog. We only include the pairs
of galaxies with separations comprised between $25$ and $130$ $h^{-1}~{\rm Mpc}$, which correspond to those effectively used in our full-shape analysis (see Section~\ref{sec:rsd_modelling}). 
By doing so, we obtain $z_{\rm eff}=0.698$ for the combined sample. The {\sc EZmocks} were constructed to mimic our data sample and thus have the same $z_{\rm eff}$. The {\sc Nseries} mocks were constructed to match the BOSS CMASS NGC sample and we obtain $z_{\rm eff} = 0.56$. The {\sc MockChallenge} mocks were produced with a snapshot at $z=0.695$ and we use this value as their effective redshift.

\subsection{Galaxy clustering estimation}

We estimate the redshift-space galaxy clustering in configuration space by measuring the galaxy anisotropic two-point correlation function $\xi(r,\mu)$. 
This measurement is performed with the standard \citet{landy_bias_1993} 
estimator:
\begin{equation}
\xi(r,\mu)=\frac{GG(r,\mu)-2GR(r,\mu)+RR(r,\mu)}{RR(r,\mu)}, \label{eq:ls}
\end{equation}
where $GG(r,\mu)$, $GR(r,\mu)$, and $RR(r,\mu)$ are respectively the
normalized galaxy-galaxy, galaxy-random, and random-random number of
pairs with separation $(r,\mu)$. For the post-reconstruction, 
we employ the same estimator except that in the numerator, displaced galaxy and random catalogs are used instead.
Since we are interested in quantifying RSD effects, we decompose the three-dimensional
galaxy separation vector $\vec{r}$ into polar coordinates $(r,\mu)$ 
aligned with the line-of-sight direction, where $r$ is the norm of the separation vector and $\mu$ is the cosine
of the angle between the line-of-sight and separation vector
directions. The pair counts are binned in  
5\hmpc\ bins in $r$ and 0.01 bins in $\mu$. 

\begin{figure*}
    \centering    
    \includegraphics[width=2\columnwidth]{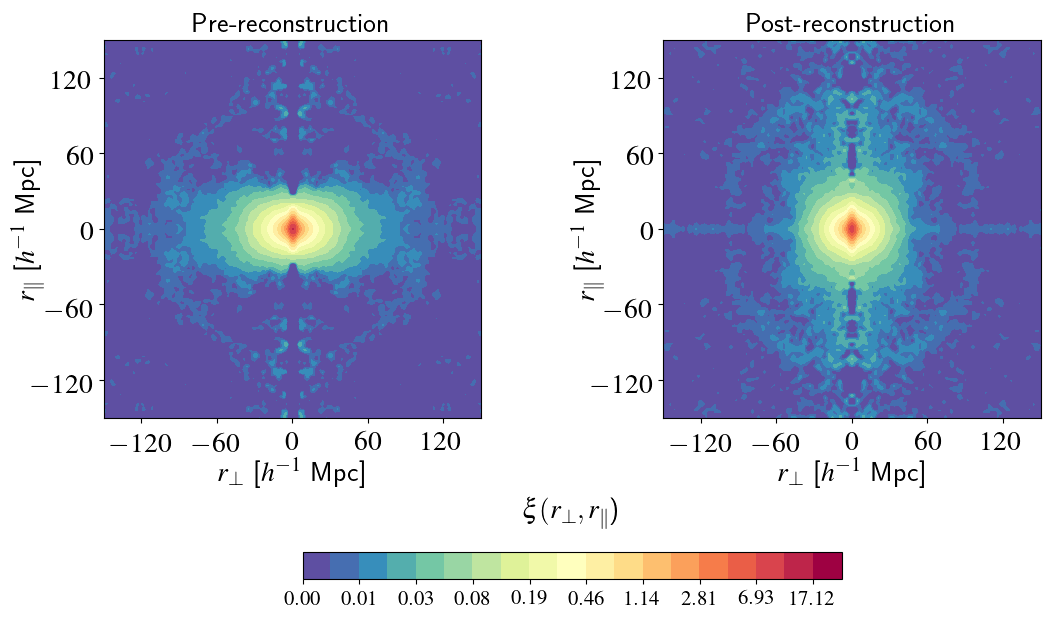}
    \caption{
      Anisotropic two-point correlation function of eBOSS LRG+CMASS galaxies at $0.6<z<1$. The left (right) panel shows the pre-reconstruction (post-reconstruction) two-point correlation function in bins of $r_\perp$ and $r_\parallel$. Bins of size $1.25~h^{-1}~$Mpc and a bi-cubic spline interpolation have been used to produce the contours.
      }
    \label{fig:cfs}  
\end{figure*}

The measured anisotropic correlation function, where the galaxy separation vector $\vec{r}$ has been decomposed into line-of-sight and transverse separations $(r_\perp,r_\parallel)$, is presented in the left panel of
Fig.~\ref{fig:cfs}. A clear BAO feature is seen at $r \approx 100$\hmpc\ as well as the impact of RSD, which squash the contours along the line of sight on large scales. In the right panel of Fig.~\ref{fig:cfs} we show the post-reconstruction correlation function where some of the isotropy is recovered and the BAO feature is sharpened. 

For the cosmological analysis, we compress the information contained in the full anisotropic correlation function. We define the multipole moments of the correlation function by decomposing $\xi(r, \mu)$ on the basis of Legendre polynomials. Since we are working with binned data, the discrete decomposition is written as:
\begin{equation}
    \hat{\xi}_\ell(r) = (2\ell+1)\sum_{i} \xi(r, \mu_i) L_\ell(\mu_i) {\rm d}\mu,
\end{equation}
where only even multipoles do not vanish given the symmetry of galaxy pairs and our choice of line of sight. We note that in the previous equation there is a factor of 2 cancellation due to the imposed symmetry between negative and positive $\mu$. Throughout this work, we only consider $\ell = 0$, 2 and 4 multipoles, referred to as monopole, quadrupole, and hexadecapole, respectively in the following. 

\begin{figure*}
    \centering
    \textbf{Pre-reconstruction}\par\medskip
    \includegraphics[width=.9\textwidth]{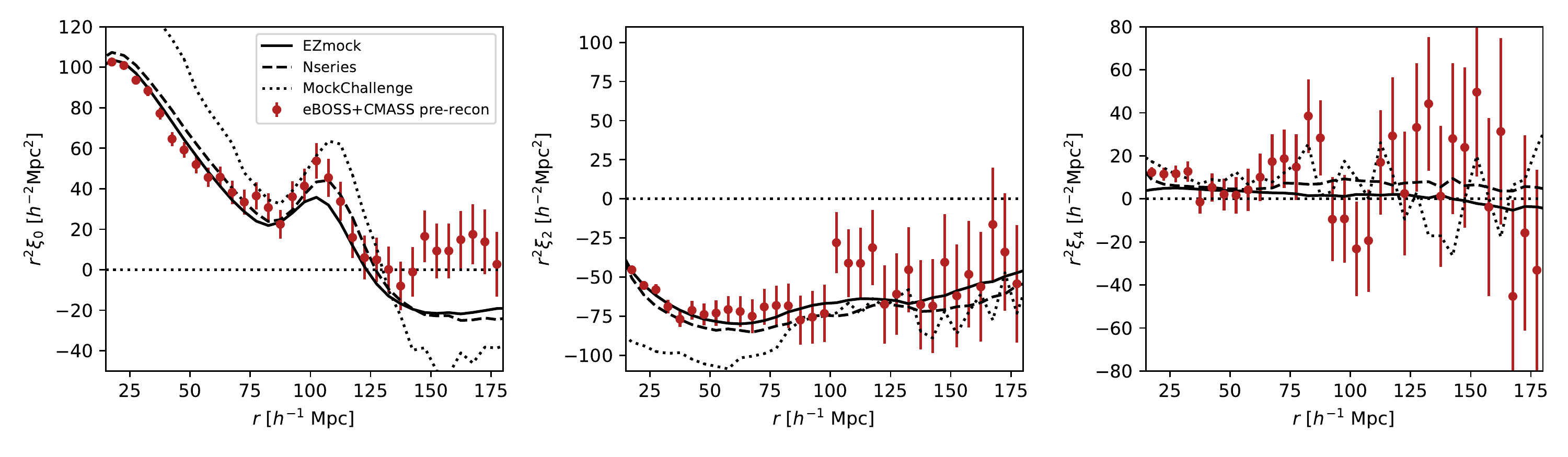}
    \textbf{Post-reconstruction}\par\medskip

    \includegraphics[width=.9\textwidth]{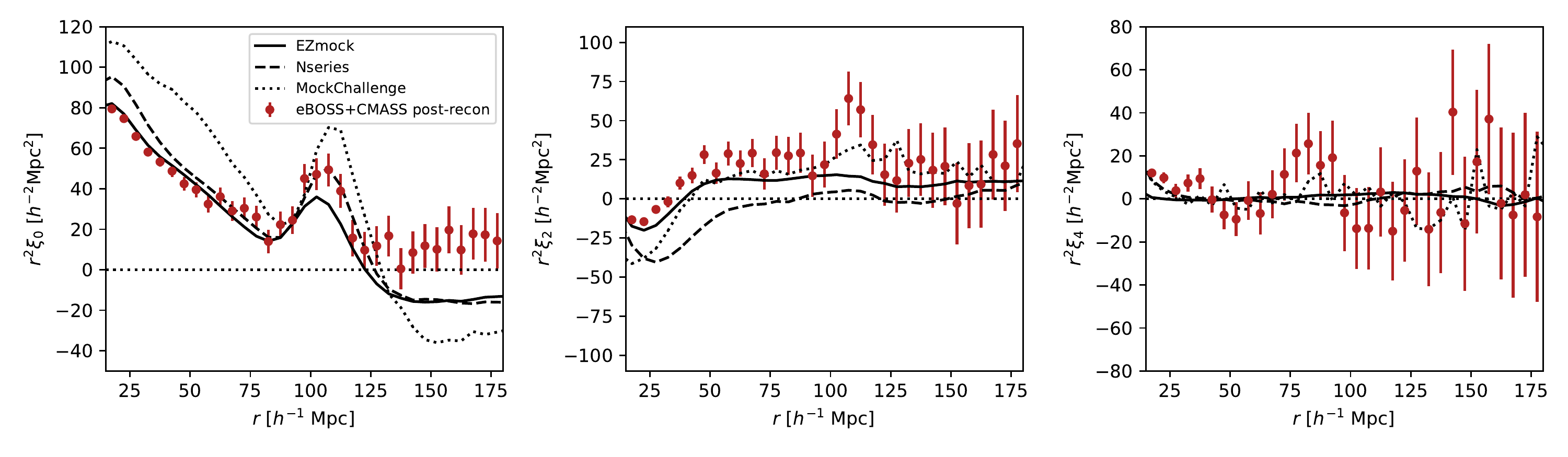}
    \caption{Multipoles of the correlation function of data compared to the mock catalogs. The data is the combined eBOSS LRG + CMASS (NGC+SGC) samples and the mocks are the average multipoles of 1000 {\sc EZmocks} realisations (solid line), 84 {\sc Nseries} realisations (dashed line) and 27 {\sc MockChallenge} mocks populated with L11 HOD model (dotted lines). Top panels show the monopole, quadrupole and hexadecapole of the pre-reconstruction samples while bottom panels show the same for the post-reconstruction case.}
    \label{fig:multipoles_data_versus_mocks}
\end{figure*}

The red points with error bars in Fig.~\ref{fig:multipoles_data_versus_mocks} show the even multipoles of the correlation function from the eBOSS LRG 
sample. The solid, dashed, and dotted black curves display the average multipoles in the different mock datasets used in this 
study: {\sc EZmocks}, {\sc Nseries}, and {\sc MockChallenge}. The error bars are obtained from the dispersion of the 1000 {\sc EZmocks} multipoles around their mean. By construction, the amplitude of the EZmock multipoles matches the data at separations $s<70$\hmpc. A slight mismatch in the BAO peak amplitudes between data and {\sc EZmocks} is visible. 
This mismatch does not impact cosmological results from the data since the covariance matrix dependency on the peak amplitude is small. However, the comparison of the precision of BAO peak measurements between mocks and data needs to account for this mismatch: the expected errors of our BAO measurement are smaller for data than for the ensemble of {\sc EZmocks}. For comparison, the average multipoles of the {\sc Nseries} mocks, also shown in Fig.~\ref{fig:multipoles_data_versus_mocks}, are a better match to the peak amplitude seen in the data.

\section{Methodology} \label{sec:method}

In this section we describe the BAO and RSD modelling, 
fitting procedure, and how errors on cosmological parameters are estimated.

\subsection{BAO modelling}
\label{sec:bao_modelling}

We employ the standard approach used in previous SDSS publications for measuring the baryon acoustic oscillations scale in configuration
space (e.g.,
\citealt{anderson_clustering_2014, ross_clustering_2017, alam_clustering_2017, bautista_sdss-iv_2018}).
The code that produces the model and perform the fitting to the data is publicly available\footnote{\url{https://github.com/julianbautista/eboss_clustering}}.

The aim is to model the correlation function multipoles $\xi_\ell(r)$ as a function of separations $r$ relevant for BAO ($30<r<180~h^{-1}$Mpc). The starting point is the model for the redshift-space anisotropic galaxy power-spectrum $P(k, \mu)$, 
\begin{multline}
    P(k, \mu)  = \frac{b^2 \left[1+\beta(1-S(k))\mu^2\right]^2}
{(1+ k^2\mu^2\Sigma_s^2/2)} \times \\  
\times \left[ P_{\rm no \ peak}(k) + P_{\rm peak}(k)
e^{-k^2\Sigma_{\rm nl}^2(\mu)/2}   \right]
\label{eq:pk2d}
\end{multline}
where $b$ is the linear bias, $\beta = f/b$ is the redshift-space
distortions parameter, $k$ is the modulus of the wave-vector and $\mu$
is the cosine of the angle between the wave-vector and the
line of sight. The non-linear broadening of the BAO peak is modelled by multiplying the ``peak-only'' power spectrum $P_{\rm
  peak}$ (see below) by a Gaussian distribution with $\Sigma_{\rm nl}^2(\mu) =
\Sigma_\parallel^2 \mu^2 + \Sigma^2_\perp(1-\mu^2)$.  The non-linear random motions on small scales are modeled by a Lorentzian distribution parametrized by $\Sigma_s$.  When performing fits to the multipoles of a single realisation of the survey, the values of 
$(\Sigma_\parallel, \Sigma_\perp, \Sigma_s)$ are held fixed
to improve convergence.  The values chosen for these damping terms were obtained from fits to the average correlation function of the {\sc Nseries} mocks, which are full N-body simulations. We show in Section~\ref{sec:systematics_bao} that our results are insensitive to small changes to those values. 
Following \citet{seo_modeling_2016} theoretical considerations,  we apply a term $S(k) = e^{-k^2\Sigma_r^2/2}$ to the post-reconstruction modeling of the correlation function ($S(k)=0$ for the pre-reconstruction BAO model).
This term models the smoothing used in our reconstruction technique, where $\Sigma_r = 15$\hmpc\ (see Section~\ref{sec:reconstruction}). 

We follow the procedure from \cite{kirkby_fitting_2013} to decompose the BAO peak component $P_{\rm peak}$ from the linear power-spectrum $P_{\rm lin}$. We start by computing the correlation function by Fourier transforming 
$P_{\rm lin}$, then we replace the correlations over the peak region by a polynomial function fitted using information outside the peak region ($50 < r < 80$ and $160 < r < 190$\hmpc). The resulting correlation function is then Fourier transformed back to get $P_{\rm no \ peak}$.
The linear power spectrum $P_{\rm lin}$ is computed using the code CAMB\footnote{\url{camb.info}} \citep{lewis_efficient_2000} with cosmological parameters of our fiducial cosmology (Table~\ref{tab:cosmologies}). The analysis in Fourier space uses the same 
procedure \citep[see][]{gil-marin_2020}. Previous BOSS \& eBOSS analyses making BAO measurements from direct tracer galaxies, used the approximate formulae from \citet{eisenstein_cosmic_1998} for decomposing the peak. We have checked that both methods yield only negligibly different results.

The correlation function multipoles $\xi_\ell(s)$ are obtained from the multipoles of the power-spectrum $P_\ell(k)$, defined as:
\begin{equation} \label{eq:pkmultipoles}
P_\ell(k) = \frac{2\ell+1}{2} \int_{-1}^{1} P(k, \mu) L_\ell(\mu)
~{\rm d}\mu
\end{equation}
where $L_\ell$ are Legendre polynomials. The $P_\ell$ are then Hankel transformed to $\xi_\ell$ using:
\begin{equation}
\xi_\ell(r) = \frac{i^\ell}{2\pi^2}\int_0^\infty k^2 j_\ell(kr)
P_\ell(k) ~ {\rm d}k
\end{equation}
where $j_\ell$ are the spherical Bessel functions. These transforms are computed using a Python implementation\footnote{\url{https://github.com/julianbautista/eboss_clustering}} of the FFTLog algorithm
described in \citet{hamilton_uncorrelated_2000}.

We parameterise the BAO peak position in our model via two dilation parameters that scale separations into transverse, $\aperp$, and radial, $\apara$, directions.  These quantities are related, respectively, to the comoving angular diameter distance, $D_M =
(1+z)D_A(z)$, and to the Hubble distance, $D_H = c/H(z)$, by
\begin{equation}
\aperp = \frac{D_M(z_{\rm eff})/r_d}{D_M^{\rm fid}(z_{\rm eff})/ r_d^{\rm fid}}
\label{eq:aperp}
\end{equation}

\begin{equation} 
\apara =  \frac{D_H(z_{\rm eff})/r_d}{D_H^{\rm fid}(z_{\rm eff})/ r_d^{\rm fid}}
\label{eq:apara}
\end{equation}
In our implementation, we apply the scaling factors exclusively to the peak 
component of the power spectrum. As shown by \citet{kirkby_fitting_2013}, 
the decoupling between the peak and full-shape of the correlation function makes the constraints on the dilation parameters to be only dependent on the BAO peak position, with no information coming from the full-shape as it is the case for RSD analysis. 

The final BAO model is a combination of the cosmological multipoles $\xi_\ell$ and a smooth function of separation. The smooth function is meant to account for unknown systematic effects in the survey that potentially create
large-scale correlations that could contaminate our measurements. Furthermore, there are currently no accurate analytical models for the post-reconstruction multipoles to date (the $S(k)$ term in Eq.~\ref{eq:pk2d} is generally not sufficient). Our final template is written as:
\begin{equation}
\xi^t_\ell(r) = \xi_\ell(\aperp, \apara, r) + 
    \sum_{i=i_{\rm min}}^{i_{\rm max}} a_{\ell, i}{r^i}.
\label{eq:template}
\end{equation}
Our baseline analysis uses $i_{\rm min} = -2$ and $i_{\rm max}=0$, corresponding to three nuisance parameters per multipole. We find that increasing the numbers of nuisance terms does not impact significantly the results. Note
that this smooth function cannot be used in the full-shape RSD analysis since these terms would be completely degenerate with the growth rate of structure parameter. 

Our baseline BAO analysis uses the monopole $\xi_0$ 
and the quadrupole $\xi_2$ of the correlation function.
We performed fits on mock multipoles including the hexadecapole
$\xi_4$, finding that it does not add information (see 
Table~\ref{tab:bao_ezmock_stats_errors}). We fix $\beta = 0.35$ and fitting $b$ with a flat prior between $b=1.0$ and 4.  For all fits, the broadband parameters are free, while both dilation parameters are allowed to vary between 0.5 and 1.5. A total of 9 parameters are fitted simultaneously.

\subsection{RSD modelling}
\label{sec:rsd_modelling}

We describe the apparent distortions introduced by galaxy peculiar velocities in the redshift-space galaxy clustering pattern using two different analytical models: the combined Gaussian streaming and 
Convolutional Lagrangian Perturbation Theory (CLPT) formalism developed by \citet{reid_towards_2011, carlson_convolution_2013, wang_analytic_2014}, and the \citet{taruya_baryon_2010} model (TNS) supplemented with a 
 non-linear galaxy bias prescription. These two models, frequently used in the literature, partially account for RSD non-linearities and describe the anisotropic clustering down to the quasi-linear regime.  We use both models to fit the multipoles of the correlation function and later combine their results to provide more robust estimates of the growth rate of structure and geometrical parameters. This procedure should reduce the residual theoretical systematic errors. In this section, we briefly describe the two models and assess in Section~\ref{sec:systematics_rsd} their performance in the recovery of unbiased cosmological parameters using mock datasets.

\subsubsection{Convolutional Lagrangian Perturbation Theory with Gaussian Streaming}

CLPT provides a non-perturbative resummation of Lagrangian perturbation to the two-point statistic in configuration space for biased tracers. The Lagrangian coordinates $\vec q$ of
a given tracer are related to their Eulerian coordinates $\vec x$ through the following equation:

\begin{equation}
\vec x (\vec q,t)=\vec q+\vec \Psi(\vec q,t),
\end{equation}
where $ \Psi(\vec q,t)$ refers to the displacement field evaluated at the Lagrangian position at each time $t$.
 The two-point correlation function  is expanded in its Lagrangian coordinates considering the tracer $X$, in our case the LRG, to be locally biased with respect to the matter overdensity $\delta (\vec q)$.  The expansion is performed over different orders of the Lagrangian bias function $F[\delta (\vec q) ]$, defined as:
\begin{equation}
1+\delta_X(\vec q,t)=F[\delta(\vec q)].
\end{equation}
The Eulerian density contrast field is computed by convolving with the displacement:
\begin{equation}
\label{delta_model}
1+\delta_X (\vec x)=\int d^3q\, F\left[ \delta(\vec q) \right]\int \frac{d^3 k}{(2 \pi)^3} e^{i \vec k (\vec x - \vec q - \vec \psi(\vec q))}.
\end{equation}
The local Lagrangian bias function $F$ is approximated by a non-local expansion using its first and second derivative, where the $n^{th}$ derivative is given by:
\begin{equation}
\label{F_der}
\langle F^n \rangle=\int \frac{d\delta}{\sqrt{2\pi} \sigma}e^{-\delta^2/2\sigma^2}\frac{d^n F}{d \delta^n}.
\end{equation}
The two-point correlation function is obtained by evaluating the expression $\xi_X(\vec r) = \left< \delta_X (\vec x) \delta_X (\vec x + \vec r)\right>$, corresponding to Eq 19 of  \cite{carlson_convolution_2013}, and that can be simplified as in their Eq. 46:
\begin{equation}
\label{xi_model}
1+\xi_X(\vec r)=\int d^3 q M(\vec r, \vec q),
\end{equation}
where $ M(\vec r, \vec q)$ is the kernel of convolution taking into account the
displacement and bias expansion up to its second derivative term. The bias derivative terms are computed using the linear power spectrum derived from the 
code CAMB \citep{lewis_efficient_2000} using the fiducial cosmology described in Table~\ref{tab:cosmologies}.

As we are interested in studying RSD, we need to model the impact of peculiar velocity. The CLPT provides the pairwise mean velocity
 $v_{12}(r)$ and the pairwise velocity dispersion $\sigma_{12}(r)$ as a function of the
 real-space separation. They are computed following the formalism developed in \cite{wang_analytic_2014},
which is similar to the one describe above but modifying the kernel to take into account the velocity rather than the density:
\begin{equation}
v_{12}(r)=(1+\xi(\vec{r}))^{-1}\int {M_1}(\vec{r}, \vec{q}) d^3 q,
\end{equation}
and
\begin{equation}
\sigma_{12}(r)=(1+\xi(\vec r))^{-1}\int M_2(\vec r, \vec q)d^3q.
\end{equation}

The kernels $M_{1,2}(\vec{r}, \vec{q})$ also depend on the first two non-local derivatives of the Lagrangian bias  $\langle F' \rangle$ and 
$\langle F'' \rangle$, which are free parameters in addition to the linear growth rate $f$ in our model. Hereafter, we eliminate the angle brackets around the Lagrangian bias terms to simplify the notation in the following sections. 

Although CLPT is more accurate than Lagrangian Resummation Theory from \cite{matsubara_resumming_2008} in real space,
we still have to improve the small-scale modelling in order to study redshift-space distortions. This is particularly 
important considering that part of peculiar velocities are 
generated by interactions that occur at the typical scales of clusters of galaxies ($\sim$1 Mpc). This is achieved by mapping the real-space CLPT model of the two-point statistics into redshift space with the Gaussian Streaming (GS) model proposed by \cite{reid_towards_2011}. The pairwise velocity distribution of tracers is assumed to have a Gaussian distribution that depends on both the separation $r$ and the angle between the separation vector and the line of sight $\mu$.

We use the \cite{wang_analytic_2014} implementation that uses CLPT results as input for the GS model. The redshift-space correlation 
function is finally computed as:
\begin{equation}
\label{gsrd_integral}
\begin{split}
1+\xi_{\rm X}(r_\perp,r_\parallel)= & \int \frac{1}{\sqrt{2\pi \left[\sigma_{12}^2(r)+\sigma^2_{\rm FoG}\right]}}[1+\xi_{\rm X}(r)]\\
& \times \exp{-\frac{[r_\parallel-y-\mu v_{12}(r)]^2}{2\left[\sigma_{12}^2(r)+\sigma^2_{\rm FoG}\right]}} dy,
\end{split}
\end{equation}
where $\xi(r)$, $v_{12}(r)$, and $\sigma_{12}(r)$ are obtained 
from CLPT. The last function in the integral takes into account the scale-dependent halo-halo pairwise velocity and we have to introduce an extra parameter $\sigma_{\rm FoG}$
describing the galaxy random motions with respect to their parent halo, also known as Fingers-of-God (FoG) effect. 
\cite{reid_towards_2011} demonstrated that the GS model can predict clustering with an accuracy of $\approx 2$ per cent when dark-matter halos are used as tracers. Using galaxies, the accuracy decreases as $\sigma_{\rm FoG}$ increases. Considering that about 85 per cent of the galaxies from the LRG sample are central galaxies \citep{zhai_clustering_2017}, the accuracy remains close to the one obtained using halos. In summary, given a fiducial cosmology, this RSD model has four free parameters $[f, F',F'', \sigma_{\rm FoG}]$.

\subsubsection{TNS model}

The other RSD model that we consider is the \cite{taruya_baryon_2010} model extended to non-linearly biased tracers. We refer to it as TNS in this work. 
Its implementation closely follows the one presented in 
\citet{de_la_torre_vimos_2017}. This model is based on the conservation of the 
number density in real- and redshift-space \citep{kaiser_clustering_1987}. In 
this framework, the anisotropic power spectrum for unbiased matter tracers 
follows the general form \citep{scoccimarro_power_1999}
\begin{eqnarray}
    P^s(k,\mu)&=&\int \frac{d^3 \vec{r}}{(2\pi)^3} e^{-i\vec{k} \cdot \vec{r}}\left<e^{-ikf\mu \Delta u_\parallel} \times \right. \nonumber \\
    && \left. [\delta(\vec{x})+f \partial_{_\parallel} u_{_\parallel}(\vec{x})][\delta(\vec{x}^\prime)+f \partial_{_\parallel} u_{_\parallel}(\vec{x}^\prime)]
    \right> \label{eq:rspk}
\end{eqnarray}
where $\mu=k_\parallel/k$, $u_\parallel(\vec{r})=-v_\parallel(\vec{r})/(f aH(a))$,
$v_\parallel(\vec{r})$ is the line-of-sight component of the peculiar
velocity, $\delta$ is the matter density field, $\Delta
u_\parallel=u_\parallel(\vec{x})-u_\parallel(\vec{x}^\prime)$ and
$\vec{r}=\vec{x}-\vec{x}^\prime$. The model by \cite{taruya_baryon_2010} for Eq. \ref{eq:rspk} can be written
\begin{multline}
    P^s(k,\mu)= D(k\mu\sigma_v)\big[ P_{\delta\delta}(k) +2\mu^2f P_{\delta\theta}(k) + \mu^4f^2P_{\theta\theta}(k)+ \\ C_A(k,\mu,f)+C_B(k,\mu,f) \big]\,,
\end{multline}
where $\theta$ is the divergence of the velocity field defined as 
$\theta = -\nabla {\bf \cdot v}/(aHf)$. $P_{\delta\delta}$,  $P_{\theta\theta}$ 
and $P_{\delta\theta}$ are respectively the non-linear matter density, 
velocity divergence, and density-velocity divergence power-spectra. 
$C_A(k,\mu,f)$ and $C_B(k,\mu,f)$ are two correction terms that reduce to 
integrals of the matter power spectrum given in 
\citet{taruya_baryon_2010}. The phenomenological damping function 
$D(k\mu\sigma_v)$, not only describes the FoG effect induced by random motions 
in virialized systems, but has also a damping effect on the power spectra. 
Several functional forms can be used, in particular Gaussian or Lorentzian 
forms have been extensively used in previous analyses. We opt for a Lorentzian 
damping function that provides a better agreement to the LRG data and mocks,
\begin{equation}
    D(k,\mu,\sigma_v) = (1+k^2\mu^2\sigma_v^2)^{-1},
\end{equation}
where $\sigma_v$ represents an effective pairwise velocity dispersion that is 
later treated as a nuisance parameter in the cosmological inference.
This model can be generalized to the case of biased tracers by including a 
galaxy biasing model. In that case, the anisotropic galaxy power spectrum can 
be rewritten as 
\begin{multline}
P^s_{\rm g}(k,\mu) = D(k\mu\sigma_v) \big[ P_{\rm gg}(k) + 2\mu^2fP_{\rm{g} \theta} + \mu^4f^2 P_{\theta\theta}(k) + \\
C_A(k,\mu,f,b_1) + C_B(k,\mu,f,b_1) \big]
\label{eq:psg}
\end{multline}
where $b_1$ is the galaxy linear bias. The explicit expressions for 
$C_A(k,\mu,f,b_1)$ and $C_B(k,\mu,f,b_1)$ are given in, e.g.,  
\citet{de_la_torre_modelling_2012}.
We adopt here a non-linear, non-local, prescription for galaxy biasing that follows the work of \citet{mcdonald_clustering_2009, chan_gravity_2012}. 
Specifically we use renormalized pertubative bias scheme presented in 
\citet{assassi_renormalized_2014} at 1-loop. In that case, the relation between 
the galaxy overdensity $\delta_{\mathrm{g}}$ and matter overdensity $\delta$ is 
written as
\begin{equation}
\delta_{\mathrm{g}} = b_{1}\delta + \frac{b_{2}}{2}\delta^{2} + 
b_{ \mathcal{G}_{2}} \mathcal{G}_{2} +  b_{\Gamma_{3}}\Gamma_{3}
\end{equation}
where the two operators $\mathcal{G}_{2}$ and $\Gamma_{3}$ are
defined as 
\begin{align}
    \mathcal{G}_{2}(\phi)   &\equiv(\partial_i \partial_j \phi)^{2} - (\partial^{2}\phi)^{2}, \\
    \Gamma_{3}(\phi,\phi_v  )&\equiv\mathcal{G}_{2}(\phi) - \mathcal{G}_{2}(\phi_v),
\end{align}
and $\phi$ and $\phi_v$ correspond to the gravitational and velocity potentials 
respectively. In the local Lagrangian picture, the non-local bias parameters 
$b_{\mathcal{G}_{2}}$ and $b_{\Gamma_{3}}$ are related to the linear bias 
parameter $b_1$ as 
\begin{align}
 b_{\mathcal{G}_{2}} &= -\frac{2}{7}(b_1 - 1) \\
 b_{\Gamma_{3}} &= \frac{11}{42}(b_1 - 1). \label{eq:nllbg3}
\end{align}
Bispectrum analyses in halo simulations show that those relations are 
reasonable approximations \citep{chan_gravity_2012,saito_understanding_2014}. 
However, as pointed out in \cite{sanchez_clustering_2017}, fixing 
$b_{\Gamma_{3}}$ to the local Lagrangian prediction is not necessary 
optimal because $b_{\Gamma_{3}}$ partially absorbs the scale dependence in $b_1$, which should in principle be present in the bias expansion. Moreover, local Lagrangian relation remains an approximation in the nonlinear regime \citep[e.g.][]{matsubara_nonlinear_2011}. We investigate in Section~\ref{sec:robustness} whether fixing $b_{\Gamma_{3}}$ or not is optimal for the specific case of LRG 
using {\sc Nseries} mocks. With this biasing model, the galaxy-galaxy and galaxy-velocity divergence power spectra read \citep{assassi_efficient_2017, simonovic_cosmological_2018}
\begin{eqnarray}
  P_{gg}(k)&=&b_1^2 P_{\delta\delta}(k)+ b_2b_1I_{\delta^{2}}(k)  +  2b_1b_{ \mathcal{G}_{2}}I_{{ \mathcal{G}_{2}}}(k) \nonumber \\
  &&+2\left(b_1b_{ \mathcal{G}_{2}} + \frac{2}{5}b_1 b_{\Gamma_{3}}\right)F_{\mathcal{G}_{2}}(k) +  \frac{1}{4}b_2^{2}I_{\delta^{2}\delta^{2}}(k)  \nonumber \\
  &&+b_{ \mathcal{G}_{2}}^{2} I_{\mathcal{G}_{2}\mathcal{G}_{2}}(k) 
  \frac{1}{2}b_2b_{ \mathcal{G}_{2}}I_{\delta_2 \mathcal{G}_{2}}(k) \\
  P_{g\theta}(k)&=&b_1P_{\delta\theta}(k)+\frac{b_2}{4} I_{\delta^{2}\theta}(k)  + b_{ \mathcal{G}_{2}}I_{{ \mathcal{G}_{2}}\theta}(k) \nonumber \\
  && + \left(b_{ \mathcal{G}_{2}} + \frac{2}{5} b_{\Gamma_{3}}\right)F_{\mathcal{G}_{2}\theta}(k). 
\end{eqnarray}
In the above equations, $I_{\delta^{2}}(k)$, $I_{{ \mathcal{G}_{2}}}(k) $, ${F_{\mathcal{G}_{2}}(k)}$, $I_{\delta^{2}\delta^{2}}(k)$, $I_{\mathcal{G}_{2}\mathcal{G}_{2}}(k)$, $I_{\delta_2 \mathcal{G}_{2}}(k)$,
are 1-loop integrals which expressions can be found in \citet{simonovic_cosmological_2018}. 
The expressions for $ I_{\delta^{2}\theta}(k)$, $I_{{ \mathcal{G}_{2}}\theta}(k)$, and $F_{\mathcal{G}_{2}\theta}(k)$ integrals are nearly identical as for $I_{\delta^{2}}(k)$, $I_{{ \mathcal{G}_{2}}}(k)$, and  ${F_{\mathcal{G}_{2}}(k)}$, except that the $G_2$ kernel replaces the $F_2$ kernel in $I_{\delta^{2}}(k)$, $I_{{ \mathcal{G}_{2}}}(k) $ and ${F_{\mathcal{G}_{2}}(k)}$.  Those 1-loop integrals are computed using the method described in \citet{simonovic_cosmological_2018}, which uses a power-law decomposition of the input linear power spectrum to perform the integrals. This allows a fast and robust computation of those integrals.

The input linear power spectrum $P_{\rm lin}$ is obtained with \textsc{CAMB}, while the non-linear power spectrum $P_{\delta\delta}$ is calculed from the {\sc RESPRESSO} code \citep{nishimichi_moving_2017}. This non-linear power spectrum prediction does agree very well with successful perturbation theory-based predictions such as RegPT, but extend their validity to $k\simeq0.4$ \citep{nishimichi_moving_2017}.
This is very relevant for configuration space analysis, where one needs to have both a correct BAO amplitude and a non-vanishing signal at high $k$ to avoid aliasing in the transformation from Fourier to configuration space.

To obtain $P_{\theta\theta}$ and $P_{\delta\theta}$ power spectra, we use the universal fitting functions obtained by \citet{bel_accurate_2019} and that depend on $\sigma_8(z)$, $P_{\delta\delta}$, and $P_{\rm lin}$ as
\begin{equation}
\begin{aligned}
P_{\theta \theta}(k)&=P_L(k) e^{-k\left(a_{1}+a_{2} k+a_{3} k^{2}\right)}, \\
P_{\delta \theta}(k)&=\left(P_{\delta \delta}(k) P_{\rm lin}(k)\right)^{\frac{1}{2}} e^{-\frac{k}{k_{\delta}}-b k^{6}}.
\label{eqn:pdv_fitting_function}
\end{aligned}
\end{equation}
The overall degree of nonlinear evolution is encoded by the amplitude of the matter fluctuation at the considered effective redshift. The explicit dependence of the fitting function coefficients on $\sigma_8$ is given by
\begin{equation}
\begin{aligned} a_{1} &=-0.817+3.198 \sigma_{8} \\ 
a_{2} &=0.877-4.191 \sigma_{8} \\ 
a_{3} &=-1.199+4.629 \sigma_{8} \\ 
1 / k_{\delta} &=-0.017+1.496 \sigma_{8}^{2} \\ 
b &=0.091+0.702 \sigma_{8}^{2}. \end{aligned}
\end{equation}

In total, this model has either four or five free parameters, $[f,b_1,b_2,\sigma_v]$ or $[f,b_1,b_2,b_{\Gamma_{3}}\sigma_v]$, depending on the number of bias parameters that are let free. Finally, the multipole moments of the anisotropic correlation function are obtained by performing the Hankel transform of the model $P_\ell^s(k)$.

\subsubsection{Alcock-Paczynski effect}

For both RSD models, the \citet{alcock_evolution_1979} effect implementation 
follows that of \citet{xu_measuring_2013}. The Alcock-Paczynski distortions are simplified if we define the
$\alpha$ and $\epsilon$ parameters, which characterize 
respectively the isotropic and anisotropic distortion components. These are 
related to $\alpha_{\perp}$  and $\alpha_\parallel$ (Eqs.~\ref{eq:aperp} and \ref{eq:apara}) as
\begin{eqnarray}
  \alpha &=& \alpha_\parallel^{1/3} \alpha^{2/3}_{\perp} \\ \label{eq:alpha}
  \epsilon &=& \left( \alpha_\parallel/\alpha_{\perp} \right)^{1/3} - 1, \label{eq:epsilon}
\end{eqnarray}
For model $\xi_0$, $\xi_2$, and $\xi_4$, the same quantities in the fiducial cosmology are given by \citep{xu_measuring_2013}:
\begin{eqnarray}
    \xi^{\rm fid}_0(r^{\rm fid}) &=& \xi_0(\alpha r) + \frac{2}{5}\epsilon
    \left[ 3\xi_2(\alpha r) + \frac{d\xi_2(\alpha r)}{d\ln(r)} \right] 
    \label{eqn:mono} \\
    \xi^{\rm fid}_2(r^{\rm fid}) &=& \bigg( 1 + \frac{6}{7}\epsilon \bigg)\xi_2(\alpha r) +2\epsilon \frac{d\xi_0(\alpha r)}{d\ln(r)}
    +\frac{4}{7}\epsilon \frac{d\xi_2(\alpha r)}{d\ln(r)} \nonumber \\
    && + \frac{4}{7}\epsilon \bigg[ 5\xi_4(\alpha r) + 
    \frac{d\xi_4(\alpha r)}{d\ln(r)} \bigg]. \\
    \xi^{\rm fid}_4(r^{\rm fid}) &=& \xi_4(\alpha r) + 
    \frac{36}{35}\epsilon \bigg[-2\xi_2(\alpha r) + \frac{d\xi_2(\alpha r)}{d\ln(r)} \bigg] \nonumber \\
    && + \frac{20}{77} \epsilon \bigg[3\xi_4(\alpha r) + 2\frac{d\xi_4(\alpha r)}{d\ln(r)} \bigg] \nonumber \\
    && + \frac{90}{143} \bigg[7\xi_6(\alpha r) + \frac{d\xi_6(\alpha r)}{d\ln(r)} \bigg].
    \label{eqn:quad}
\end{eqnarray}
We note that this is an approximation for small variations around $\alpha=1$ and $\epsilon=0$ \citep{xu_measuring_2013}. Nonetheless, for the observed values on those parameters and when comparing to the model prediction based on the exact transformation, the results are virtually the same.

\subsubsection{The fiducial scale at which $\sigma_8$ is measured}
\label{sec:fs8_scaling}

We perform an additional step in order to reduce the dependency of our $\fsig$ constraints on the choice of fiducial cosmology. When fitting the correlation function multipoles, $\sigma_8$ is kept fixed to its fiducial value defined as
\begin{equation}
    \sigma_R^2 = \int_0^\infty {\rm d}k \ k^2 P_{\rm lin}(k) W_{\rm TH}^2(Rk),
\end{equation}
where $P_{\rm lin}$ is the linear matter power-spectrum predicted by the fiducial cosmology, $W_{\rm TH}$ is the Fourier transform of a top-hat function with characteristic radius of $R=8$\hmpc. 
The resulting $f$ is scaled by $\sigma_8$. However, in Section~\ref{sec:systematics_rsd} we show that the recovered $f\sigma_8$ has a strong dependence on the fiducial cosmology when we have best-fit $\alpha$ not close to unity. We can reduce this dependency by recomputing $\sigma_8$ using $R=8\alpha$\hmpc, where $\alpha$ is the isotropic dilation factor (Eq.~\ref{eq:alpha}) obtained in the fit. In effect, this keeps the scale at which $\sigma_8$ is fitted fixed relative to the data in units of $h^{-1}$Mpc, which only depends on $\Omega_m^{\rm fid}$. This is an alternative approach to the recently proposed $\sigma_{12}$ parametrisation \citep{sanchez_let_2020}, where the radius of the top-hat function is set to $R=12$~Mpc instead of $R=8$\hmpc. Unless otherwise stated, all the reported values of $\fsig$ in this work provide $\fsig$ where the scale is fixed in this way.

\subsection{Parameter inference}

The cosmological parameter inference is performed by means of the likelihood analysis of the data. The likelihood $\mathcal{L}$ is defined such that
\begin{equation}
-2\ln\mathcal{L}(\theta) = \sum_{i,j}^{N_p}\Delta_i(\theta) \hat{\Psi}_{ij} \Delta_j(\theta),
\end{equation}
where $\theta$ is the vector of parameters, $\vec{\Delta}$ is the data-model difference vector, $N_p$ is the total number of data points. An estimate of the precision matrix $\hat{\Psi} = (1-D) \hat{C}^{-1}$ is obtained from the covariance $\hat{C}$ from 1000 realisation  of EZmocks, where $D = (N_p+1)/(N_{\rm mocks} -1)$ is a factor that accounts for the skewed nature of the Wishart distribution \citep{hartlap_why_2007}.
The data vector that enters in $\vec{\Delta}$ includes, in the baseline configuration, the monopole and quadrupole correlation functions for the BAO analysis, and the  monopole, quadrupole, and hexadecapole correlation functions for the RSD analysis.

In the BAO analysis, the best-fit parameters ($\aperp, \apara$) are found by minimizing $-2\ln\mathcal{L} = \chi^2$ using a quasi-Newton minimum finder algorithm 
{\sc iMinuit}\footnote{\url{https://iminuit.readthedocs.io/}}. 
The errors in $\apara$ and $\aperp$ are found by computing the intervals where $\chi^2$ increases by unity. Gaussianity is not assumed in the error calculation, but we find that on average, errors are symmetric and correctly described by a Gaussian.
The 2D errors in $(\aperp, \apara)$, such as those presented in
Figure~\ref{fig:consensus_bao}, are found by scanning $\chi^2$ values in a regular grid in $\aperp$ and $\apara$. In the case of the full-shape analysis, we explore the likelihood with the Markov chain Monte Carlo ensemble sampler 
{\sc emcee}\footnote{\url{https://emcee.readthedocs.io/}}. 
The input power spectrum shape parameters are fixed at the 
fiducial cosmology and any deviations are accounted for through
the Alcock-Paczynski parameters $\aperp$ and $\apara$. We assume the uniform priors on model parameters given in 
Table~\ref{tab:prior_ref}.

\begin{table}
\label{tab:prior_ref}
\caption{List of fitter parameters and their priors used in full-shape analysis for the two models.}
\centering
\begin{tabular}{cccc}
\hline
\hline
Par. TNS  &  Prior TNS &Par. CLPT-GS  &  Prior CLPT-GS\\ 
\hline
$\alpha_{\perp}$  & $[0.5, 1.5]$&
$\alpha_{\perp}$  & $[0.5, 1.5]$\\
 $\alpha_{\parallel}$  &   $[0.5, 1.5]$&
  $\alpha_{\parallel}$  &   $[0.5, 1.5]$\\
\hline
$f$ & $[0,2]$&$f$ & $[0,2]$ \\
$b_1$ & $[0.2,4]$& $\langle F' \rangle$& [0,3]\\
$b_2$  & $[-10, 10]$&$\langle F'' \rangle$& [-10,10]\\
$b_{\Gamma_3}$ & $[-2, 4]$&$\sigma_{\rm FoG}$ &[0,40]\\
$\sigma_v$ & $[0.1,8]$ \\
\hline
\hline
\end{tabular}
\end{table}

\begin{table}
\label{tab:percival_factors}
\caption{Characteristics of the baseline fits for all models in this work, where $N_{\rm mock}$ is the number of mocks used in the estimation of the covariance matrix, $N_{\rm par}$ is the total number of parameters fitted, $N_{\rm bins}$ is the total size of the data vector, $(1-D)$ is the correction factor to the precision matrix \citep{hartlap_why_2007}, $m_1$ is the factor to be applied to the 
estimated error matrix and $m_2$ is the factor that scales the scatter of best-fit parameters of a set of mocks (if these were used in the calculation of the covariance matrix). The derivation of $m_1$ and $m_2$ can be found in \citet{percival_clustering_2014}.
}
\centering
\begin{tabular}{cccc}
\hline
\hline
  & BAO & RSD TNS & RSD CLPT-GS \\
  \hline
$N_{\rm mock}$ & 1000 & 1000 & 1000 \\
$N_{\rm par}$ & 9 & 7 & 6 \\
$N_{\rm bins}$ & 40 & 65 & 63 \\
$(1-D)$ & 0.96 & 0.93 & 0.94 \\
$m_1$ & 1.022 & 1.053 & 1.053 \\
$m_2$ & 1.065 & 1.128 & 1.125 \\
\hline
\hline
\end{tabular}
\end{table}

The final parameter constraints are obtained by marginalizing 
the full posterior likelihood over the nuisance parameters. 
The marginal posterior is approximated by a multivariate Gaussian 
distribution with central values given by best-fitting parameter 
values $\theta^* = (\aperp, \apara, \fsig)$ and parameter covariance 
matrix $C_{\theta}$. Since the covariance matrix is computed from a 
finite number of mock realisations, we need to apply correction factors 
to the obtained $C_{\theta}$. These factors are Eq.~18 
and 22 from \citet{percival_clustering_2014} to be applied to 
uncertainties and to the scatter over best-fit values, respectively. 
These factors, which depend on the number of mocks, parameters and bins 
in the data vectors, are presented in Table~\ref{tab:percival_factors}. 
The final parameter constraints from this work are available 
to the public in this format\footnote{\url{sdss.org/}}.

\subsection{Combining BAO and RSD constraints}
\label{sec:combining_analysis}

From the same input LRG catalog, we produced BAO-only and 
full-shape RSD constraints, both in configuration and Fourier space \citep{gil-marin_2020}. Each measurement yields a marginal posterior 
on $(\aperp, \apara)$ for BAO-only or $(\aperp, \apara, \fsig)$ for the full-shape RSD analyses. 
In the following we describe the procedure to combine all these
posteriors into a single consensus constraint, while correctly accounting for their covariances. This consensus result is the one used for the final cosmological constraints described in \citet{mueller_2020}.

We follow closely the method presented in \citet{sanchez_clustering_2017} to derive the consensus result. The idea is to compress $M$ data vectors $x_m$ containing $p$ parameters and their $p\times p$ covariance matrices $C_{mm}$ from different methods into a single vector $x_c$ and covariance $C_c$, assuming that the $\chi^2$ between individual measurements is the same as the one from the compressed result.
The expression for the combined covariance matrix is
\begin{equation}
    C_c \equiv \left(\sum_{m=1}^M \sum_{n=1}^M C_{mn}^{-1} \right)^{-1}
\end{equation}
and the combined data vector is
\begin{equation}
    x_c = C_c \sum_{m=1}^M \left( \sum_{n=1}^M  C^{-1}_{nm} \right) x_m
\end{equation}
where $C_{mn}$ is a $p \times p$ block from 
the full covariance matrix between all parameters 
and methods $C$, defined as
\begin{equation}
    C = 
\begin{pmatrix}
C_{11} & C_{12} & \cdots & C_{1M} \\
C_{21} & C_{22} & \cdots & C_{2M} \\
\vdots  & \vdots  & \ddots & \vdots  \\
C_{M1} & C_{M2} & \cdots & C_{MM}
\end{pmatrix}
\label{eq:full_covariance}
\end{equation}
The diagonal blocks $C_{mm}$ are obtained from the Gaussian approximation 
of the marginal posterior from each method. 
The off-diagonal blocks $C_{mn}$ with $m\neq n$ cannot be estimated 
from our fits. We derive these
off-diagonal blocks from results from each method applied to 
the 1000 {\sc EZmocks} realisations. More precisely, we compute 
the correlation coefficients $\rho^{\rm mocks}_{p_1, p_2, m, n}$ 
between parameters $p_1$, $p_2$ and methods $m, n$ using the mocks 
and scale these coefficients by the diagonal errors from the data. 
It is worth emphasizing that the correlation coefficients between 
parameters depend on the given realisation of the data, while the 
ones derived from mock measurements are ensemble averaged coefficients. 
Therefore, we scale the correlations coefficients 
from the mocks in order to match the maximum correlation coefficient
that would be possible with the data \citep{ross_information_2015}.
For the same parameter $p_1$ measured by two different methods $m$ and $n$,
we assume that the maximum correlation between them is given by 
$\rho_{\rm max} = \sigma_{p1, m}/\sigma_{p1, n}$, 
where $\sigma_p$ is the error of parameter $p$. This number is computed
for the data realisation $\rho_{\rm max}^{\rm data}$ and for the ensemble of mocks
$\rho_{\rm max}^{\rm mocks}$. 
We can write the adjusted correlation coefficients as
\begin{equation}
    \rho^{\rm data}_{p_1, p_1, m, n} = \rho^{\rm mocks}_{p_1, p_1, m, n} 
                \frac{ \rho^{\rm data}_{\rm max} }{ \rho^{\rm mocks}_{\rm max} }
\end{equation}
The equation above accounts for the diagonal terms of the 
off-diagonal block $C_{mn}$. For the off-diagonal terms, we use
\begin{equation}
    \rho^{\rm data}_{p_1, p_2, m, n} = 
    \frac{1}{4}\left(\rho^{\rm data}_{p_1, p_1, m, n} + \rho^{\rm data}_{p_2, p_2, m, n}\right)
    \left(\rho^{\rm data}_{p_1, p_2, m, m} + \rho^{\rm data}_{p_1, p_2, n, n}\right)
\end{equation}

We use the method described above to perform all the constraint combinations, except for the combination of results from CLPT-GS and TNS RSD models, which use the same input data vector (pre-reconstruction multipoles in configuration space). For this particular combination, we simply assume that
$C_c^{-1} = 0.5(C_{mm}^{-1}+C_{nn}^{-1})$ and 
$x_c = 2 C_c^{-1}\left(C_{mm}^{-1} x_{m} +  C_{nn}^{-1} x_{n}\right)$.
For all combinations, we chose to use the results from at most two methods at once ($M=2$) in order to reduce the potential noise introduced by the procedure. 

Denoting $\xi_\ell$ the results from the configuration space analysis and $P_\ell$ that from the Fourier space analysis, our recipe to obtain the consensus result for the LRG sample is as follows:
\begin{itemize}
    \item Combine RSD $\xi_\ell$ TNS and RSD $\xi_\ell$ CLPT-GS results into RSD $\xi_\ell$,
    \item Combine BAO $\xi_\ell$ with BAO $P_\ell$ into BAO $(\xi_\ell+P_\ell)$,
    \item Combine RSD $\xi_\ell$ with RSD $P_\ell$ into RSD $(\xi_\ell+P_\ell)$,
    \item Combine BAO $(\xi_\ell+P_\ell)$ with RSD $(\xi_\ell+P_\ell)$ into BAO$+$RSD $(\xi_\ell+P_\ell)$
\end{itemize}
Alternatively, we can proceed as
\begin{itemize}
    \item Combine BAO $\xi_\ell$ with RSD $\xi_\ell$ into (BAO$+$RSD) $\xi_\ell$,
    \item Combine BAO $P_\ell$   with RSD $P_\ell$   into (BAO$+$RSD) $P_\ell$
    \item Combine BAO$+$RSD $\xi_\ell$ with BAO$+$RSD $P_\ell$ into (BAO$+$RSD) $\xi_\ell+P_\ell$
\end{itemize}
In Section~\ref{sec:statistical_properties} we test this procedure on the mock catalogues. 

\section{Robustness of the analysis and systematic errors}  
\label{sec:robustness}

In this section we perform a comprehensive set of tests of the adopted methodology using all the simulated datasets available. We estimate the biases in the measurement of the cosmological parameters ($\aperp, \apara, \fsig$) and derive the systematic errors for both BAO-only and full-shape RSD analyses. For a given parameter, we define the systematic error $\sigma_{p, \rm syst}$ as follows. We compare the estimated value of the parameter $x_p$ to a reference value $x_p^{\rm ref}$ and set the systematic error value to 
\begin{eqnarray}
    \sigma_{p, \rm syst} = 2\sigma_p, \ &{\rm if} \  | x_p - x_p^{\rm ref}| < 2\sigma_p, \label{eq:syst_error1} \\
    \sigma_{p, \rm syst} = |x_p - x_p^{\rm ref}|, \ & {\rm if} \ | x_p - x_p^{\rm ref}| > 2\sigma_p,
        \label{eq:syst_error2}
\end{eqnarray}
where $\sigma_p$ is the estimated statistical error on $x_p$. 
As a conservative approach, we use the maximum value of the bias amongst
the several cases studied. 

\subsection{Systematics in the BAO analysis}
\label{sec:systematics_bao}

The methodology described in Section~\ref{sec:bao_modelling} was tested using the 1000 {\sc EZmocks} mock survey realisations and 84 {\sc Nseries} realisations. For each realisation, we compute the correlation function and its multipoles,
and fit for the BAO peak position to determine the dilation parameters $\apara$, $\aperp$ and associated errors. We compare the best-fit $\aperp, \apara$ to their expected values, which are obtained from the cosmological models described in 
Table~\ref{tab:cosmologies}. The effective redshift of the \textsc{EZmocks} is $z_{\rm eff} = 0.698$ and $z_{\rm eff} = 0.56$ for \textsc{Nseries}. 

\begin{figure}
    \centering
    \textbf{Pre-reconstruction}\par\medskip
    \includegraphics[width=\columnwidth]{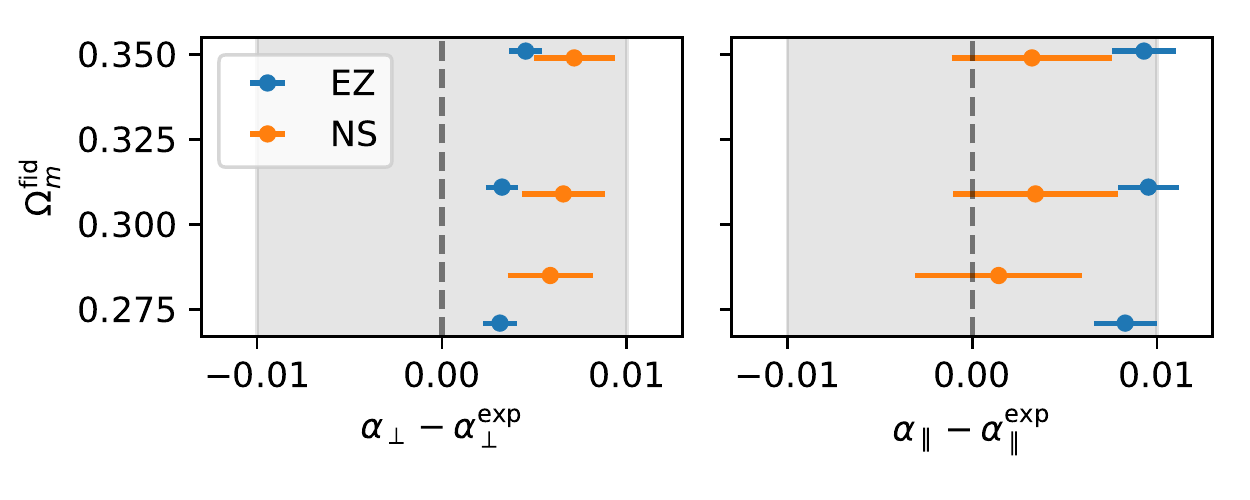}
    \textbf{Post-reconstruction}\par\medskip
     \includegraphics[width=\columnwidth]{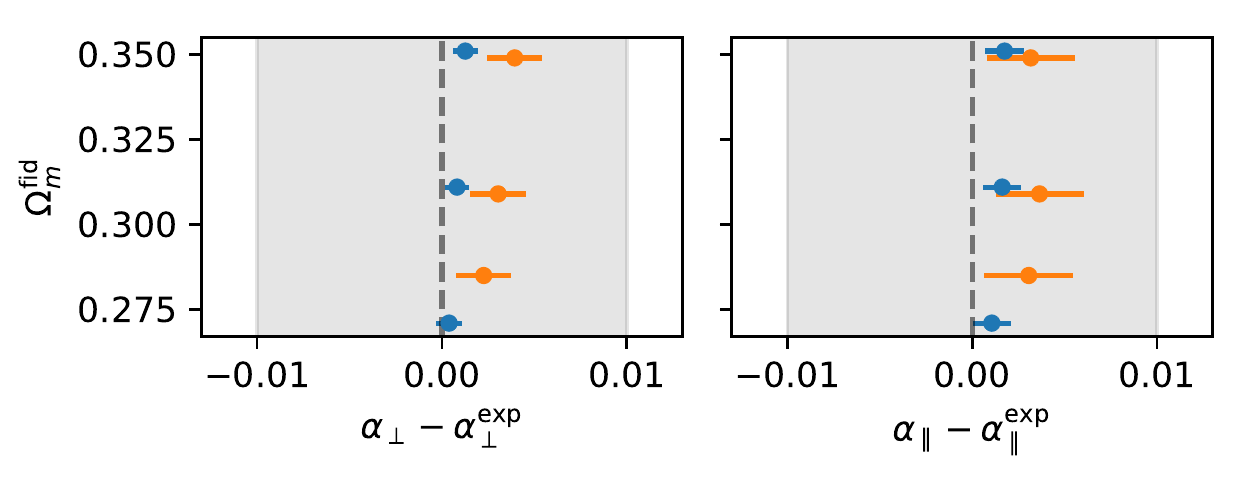}
    \caption{Impact of choice of fiducial cosmology in the recovered values of 
    $\apara$ and $\aperp$ from the stacks of 1000 multipoles from the {\sc EZmocks}
    (blue) and 84 {\sc Nseries} mocks (orange), for pre- (top panels) and post- 
    (bottom panels) reconstruction. Associated error bars correspond to the error on the mean of the mocks.   
    The gray shaded areas correspond to 
    one per cent errors. 
    For comparison, the error on real data is near 1.9 per cent 
    for $\aperp$ and 2.6 per cent for $\apara$ in the post-reconstruction case. }
    \label{fig:ezmock_bao_fiducial}
\end{figure}

In Figure~\ref{fig:ezmock_bao_fiducial} we summarize the systematic
biases from pre- and post-reconstruction mocks for a few choices
of fiducial cosmology, parameterised by $\Omega_m^{\rm fid}$.
In pre-reconstruction mocks, biases in the recovered $\alpha$ 
values reach up to 0.5 per cent in $\aperp$ and 1.0 per cent in
$\apara$. 
These biases are expected due to the impact of non-linear effects 
on the position of the peak that cannot be correctly accounted for
with the Gaussian damping terms in Eq.~\ref{eq:pk2d} at this level
of precision \citep{seo_modeling_2016}. We recall that we are 
fitting the average of all realisations. The reconstruction
procedure removes in part the non-linear effects and this is seen 
as a reduction of the biases to less than 0.2 per cent. The bias
reduction is also seen in the {\sc Nseries} mocks, particularly on
$\aperp$, confirming that the bias reduction is not related to 
a feature of the mocks induced by the approximate method used 
to build them.

Table~\ref{tab:bao_ezmock_stats_bias} shows results from
Figure~\ref{fig:ezmock_bao_fiducial} for the post-reconstruction 
case only, including the fits with the hexadecapole $\xi_{\ell=4}$. 
The impact of the hexadecapole is negligible even in this 
very low-noise regime, for both types of mocks. 
The reported dilation parameters for  almost all cases are 
consistent with 
expected value within 2$\sigma$. We see a 2.6$\sigma$ deviation on $\aperp$ for the {\sc Nseries} case analysed with $\Omega_m^{\rm fid} = 0.35$. However this choice of $\Omega_m^{\rm fid}$ is the most distant from the true value of the simulation and  its observed bias is still less than half a per cent, which is small compared to the statistical power of our sample. 
For the {\sc EZmocks}, which have smaller errors, the biases are 
up to 0.13 per cent for $\aperp$ and 0.18 per cent for $\apara$. 
These biases are much smaller than the expected statistical errors in our data, i.e.
$\sim$1.9 per cent for $\aperp$ and $\sim$2.6 per cent for $\apara$, 
showing that our methodology is robust at this statistical level. 
In these fits, all parameters except 
$\Sigma_{\rm rec} = 15$\hmpc\ were left free. The best-fit values of $\Sigma_\perp,\Sigma_\parallel$ and $\Sigma_s$ were used and held fixed in the fits of individual realisations. 

\begin{table}
  \centering
  \caption{Average biases from BAO fits on the stacked multipoles of 1000 {\sc EZmocks} and 84 {\sc Nseries} realisations. All results are based on post-reconstruction correlation functions. }
  \begin{tabular}{lcccc}
    \hline
    \hline
Sample  & $\Omega_m^{\rm fid}$ & $\ell_{\rm max}$ &  $\aperp - \aperp^{\rm exp} \ [10^{-3}]$ & $\apara - \apara^{\rm exp} \ [10^{-3}]$ \\ 
\hline 
{\sc EZ} & 0.27  & 2  & $0.4 \pm 0.7$ & $1.1 \pm 1.0$  \\
{\sc EZ} & 0.27  & 4  & $0.5 \pm 0.7$ & $1.4 \pm 1.0$  \\
{\sc EZ}  & 0.31  & 2  & $0.9 \pm 0.7$ & $0.3 \pm 1.1$  \\
{\sc EZ}  & 0.31  & 4  & $1.0 \pm 0.7$ & $0.4 \pm 1.1$  \\
{\sc EZ} & 0.35  & 2  & $1.3 \pm 0.7$ & $1.8 \pm 1.0$  \\
{\sc EZ} & 0.35  & 4  & $1.2 \pm 0.7$ & $1.5 \pm 1.0$  \\
{\sc NS} & 0.286 & 2  & $2.3 \pm 1.5$ & $3.1 \pm 2.4$  \\
{\sc NS} & 0.286 & 4  & $2.2 \pm 1.5$ & $3.0 \pm 2.4$  \\
{\sc NS} & 0.31 & 2  & $3.0 \pm 1.5$ & $3.6 \pm 2.4$  \\
{\sc NS} & 0.31 & 4  & $3.0 \pm 1.5$ & $3.7 \pm 2.4$  \\
{\sc NS} & 0.35 & 2  & $3.9 \pm 1.5$ & $3.2 \pm 2.4$  \\
{\sc NS} & 0.35 & 4  & $3.9 \pm 1.5$ & $3.5 \pm 2.4$  \\
\hline
  \end{tabular}
  \label{tab:bao_ezmock_stats_bias}
\end{table}

Results from Table~\ref{tab:bao_ezmock_stats_bias} and
Figure~\ref{fig:ezmock_bao_fiducial} show no statistically 
significant dependence of results with the choice of fiducial 
cosmology.
We derived the systematic errors for the BAO analysis using the values from Table~\ref{tab:bao_ezmock_stats_bias} and Eqs.~\ref{eq:syst_error1} and \ref{eq:syst_error2}. We used only the fits to the {\sc EZmocks} which have the better precision. The systematic errors are for $\aperp$ and $\apara$, respectively:
\begin{equation}
   {\rm BAO}: \  \sigma_{\rm syst, model} =  (0.0014,  0.0021)
\end{equation}
which are negligible compared to statistical errors of one realisation of our data. Note that the fiducial cosmologies considered are all flat and assume general relativity. \citet{carter_impact_2019} and \citet{bernal_robustness_2020} find that BAO measurements are robust to a larger variety of fiducial cosmologies (but all close to the assumed one). 
Additional systematic errors should be anticipated when extrapolating to cosmologies that are significantly different than the truth, for instance yielding dilation parameters significantly different than unity.

\begin{figure*}
    \centering
    \includegraphics[width=0.35\textwidth]{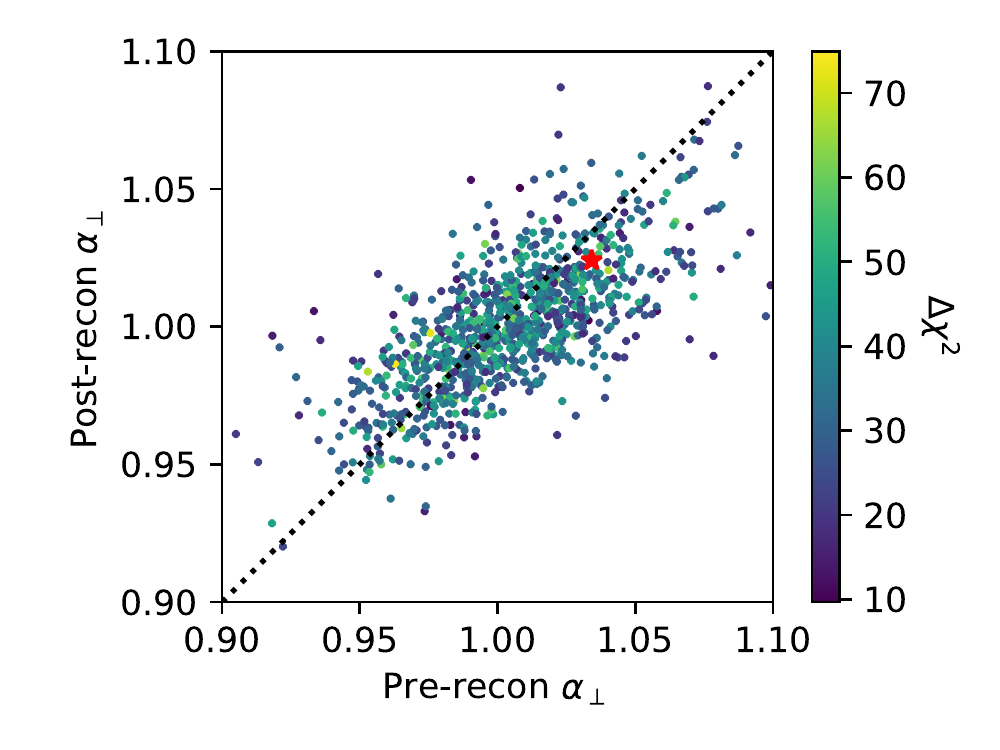}
    \includegraphics[width=0.35\textwidth]{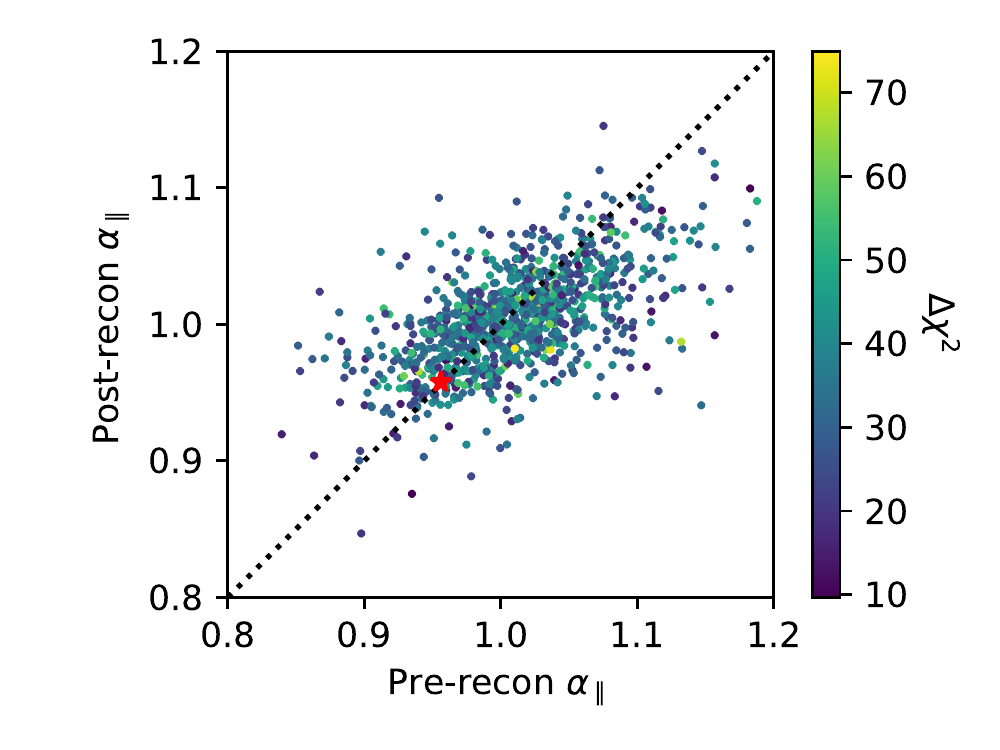}
    \includegraphics[width=0.35\textwidth]{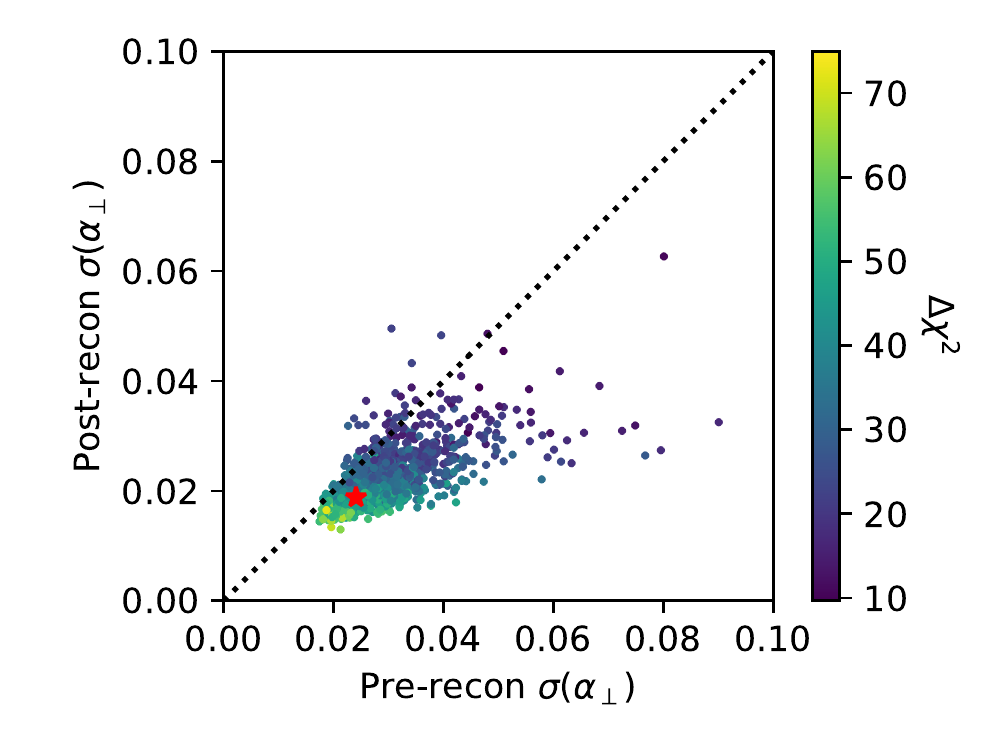}
    \includegraphics[width=0.35\textwidth]{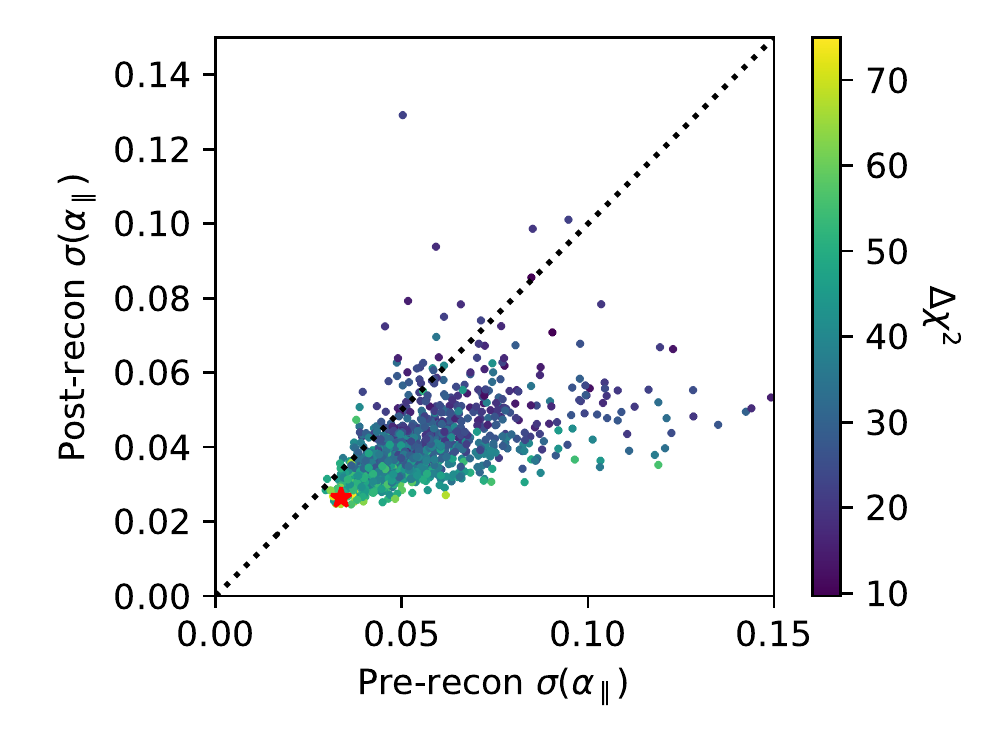}
    \caption{Distribution of dilation parameters $\aperp$ and $\apara$ and 
    its estimated errors for pre and post reconstruction \textsc{EZmock} 
    catalogs with systematic effects. The color scale indicates the 
    difference in $\chi^2$ values between a model with and without BAO peak.
    The red stars shows results with real data. There is a known mismatch in 
    the BAO peak amplitude between data and EZmocks causing the accuracy of 
    the data point to be slightly smaller than the error distribution in the 
    EZmocks (see Section~\ref{sec:mocks_description}).
    }
    \label{fig:ezmock_bao_alphas}
\end{figure*}

Figure~\ref{fig:ezmock_bao_alphas} displays the distribution of
recovered $\aperp$, $\apara$ and their respective errors measured 
from each of the individual {\sc EZmocks}. The error distribution 
shows that reconstruction improves the constraints on $\aperp$ or
$\apara$ in 94 per cent of the realisations (89 per cent have 
both errors improved). As expected, realisations with smaller 
errors generally exhibit larger values of $\Delta \chi^2 =
\chi^2_{\rm no \ peak} - \chi^2_{\rm peak}$, meaning a more 
pronounced BAO peak and higher detection significance. We see no 
particular trend in the best-fit $\alpha$ values with 
$\Delta \chi^2$ in the two top panels. 
The red stars in Figure~\ref{fig:ezmock_bao_alphas} 
indicate the values obtained in real data. The error in $\aperp$ in 
the data is typical of what is found in mocks, although for $\apara$ 
it is found at the extreme of the mocks distribution. As discussed 
in Section~\ref{sec:mocks_description} and displayed in
Figure~\ref{fig:multipoles_data_versus_mocks}, the BAO peak 
amplitude in the data multipoles is slightly larger than the one 
seen in this {\sc EZmock} sample. A similar behaviour is observed 
in the eBOSS QSO sample \citep{hou_2020, neveux_2020} who also 
use {\sc EZmocks} from \citet{zhao_2020} and in the BOSS DR12 CMASS sample (see Figure 12 of
\citealt{ross_clustering_2017}).

Table~\ref{tab:bao_ezmock_stats_errors} presents a statistical summary 
of the fits performed on the {\sc EZmocks}. We tested several changes 
to our baseline analysis: include the hexadecapole, change the 
separation range $[r_{\rm min}, r_{\rm max}]$, allow BAO damping 
parameters $\Sigma_\perp$ and $\Sigma_\parallel$ to vary within a 
Gaussian prior ($5.5 \pm 2$\hmpc), and fit the pre-reconstruction 
multipoles. We remove realisations with fits that did not converge or
with extreme error values (more than 5$\sigma$ of their distribution, where $\sigma$ is defined as the half the range covered by 68 per cent of values). The total number of valid realisations is given by $N_{\rm good}$ in Table~\ref{tab:bao_ezmock_stats_errors}.
In most cases studied, the observed standard deviation of the 
best-fit parameters $\sigma(\alpha)$ is consistent with the average 
per-mock error estimates $\langle \sigma_\alpha \rangle$, indicating 
that our errors are correctly estimated. We also see that the dispersion 
of dilation parameters is not significantly reduced when adding the 
hexadecapole $\xi_4$ to the BAO fits, showing that most of the
BAO information is contained in the monopole and quadrupole at this level 
of precision. The mean and dispersion of the pull parameter, defined as $Z_\alpha = (\alpha - \langle \alpha \rangle)/\sigma_\alpha$, are consistent with an unit Gaussian for almost all cases, which further validates our error estimates.

\begin{table*}
    \centering
      \caption{Statistics on errors from BAO fits on 1000 {\sc EZmocks} realisations. 
      All results are based on post-reconstruction correlation functions. 
      $\sigma$ is the scatter of best-fit values $x_i$ amongst the $N_{\rm good}$
      realisations with confident detection or non-extreme values or errors (out of the 1000), $\langle \sigma_i \rangle$ is the mean estimated error per
      mock, $Z = (x_i - \langle x_i \rangle)/\sigma_i$ is the pull 
      quantity for which we show the mean $\langle Z_i \rangle$ and standard deviation $\sigma(Z)$. First row corresponds to our baseline analysis. }
    \begin{tabular}{lccccccccccccc}
    \hline
    \hline
    Analysis & 
 $N_{\rm good}$ & & \multicolumn{4}{c}{$\aperp$} & &
             \multicolumn{4}{c}{$\apara$}  \\
 & & &
$\sigma$ &  $\langle \sigma_i \rangle$  &  $\langle Z_i \rangle $ & $\sigma(Z_i)$ &  &
$\sigma$ &  $\langle \sigma_i \rangle$  &  $\langle Z_i \rangle $ & $\sigma(Z_i)$ \\
\hline
   baseline & 990 & & 0.022 & 0.023 & -0.02 & $0.99 $ &  & 0.035 & 0.036 & -0.03 & $0.96 $ \\
  $\ell_{\rm max}= 4$ & 995 & & 0.022 & 0.023 & -0.02 & $0.99 $ &  & 0.035 & 0.035 & -0.03 & $0.97 $ \\
  pre-recon & 968 & & 0.030 & 0.030 & -0.05 & $1.07 $ &  & 0.055 & 0.056 & -0.06 & $0.97 $ \\
  pre-recon $\ell_{\rm max}= 4$ & 968 & & 0.029 & 0.028 & -0.03 & $1.04 $ &  & 0.054 & 0.054 & -0.07 & $1.02 $ \\
  $r_{\rm min} = 20$\hmpc & 979 & & 0.023 & 0.026 & -0.01 & $0.93 $ &  & 0.035 & 0.040 & 0.04 & $1.26 $ \\
  $r_{\rm min} = 30$\hmpc & 987 & & 0.023 & 0.024 & -0.02 & $0.95 $ &  & 0.036 & 0.038 & -0.02 & $0.92 $ \\
  $r_{\rm min} = 40$\hmpc & 995 & & 0.022 & 0.023 & -0.02 & $0.98 $ &  & 0.035 & 0.036 & -0.02 & $0.94 $ \\
  $r_{\rm max} = 160$\hmpc & 989 & & 0.022 & 0.023 & -0.02 & $0.99 $ &  & 0.036 & 0.036 & -0.03 & $0.96 $ \\
  $r_{\rm max} = 170$\hmpc & 989 & & 0.022 & 0.023 & -0.02 & $0.99 $ &  & 0.036 & 0.036 & -0.03 & $0.96 $ \\
  $r_{\rm max} = 180$\hmpc & 990 & & 0.022 & 0.023 & -0.02 & $0.98 $ &  & 0.035 & 0.036 & -0.03 & $0.95 $ \\
  Prior $\Sigma_{\perp,\parallel}$ & 993 & & 0.022 & 0.023 & -0.02 & $1.00 $ &  & 0.035 & 0.035 & -0.03 & $0.96 $ \\
\hline

    \end{tabular}
    \label{tab:bao_ezmock_stats_errors}
\end{table*}

All the tests performed in this section show that our BAO analysis is unbiased and provides correct error estimates. We apply our baseline analysis to the real data and report results in Section~\ref{sec:results_bao}. 

\subsection{Systematics in the RSD analysis}
\label{sec:systematics_rsd}

We present in this section the systematic error budget of the 
full-shape RSD analysis. Particularly, we discuss the impact of the choice of scales used in the fit, the bias introduced by each model, the bias introduced by varying the fiducial cosmology, the bias associated to the choice of the LRG halo occupation distribution model, and the impact of observational effects. These are quantified through the analysis of the various sets of mocks with both TNS and CLPT-GS models, which are described in Section~\ref{sec:rsd_modelling}.


\subsubsection{Optimal fitting range of scales}

We first study the optimal range of scales in the fit for the two RSD models considered in this work (see Section~\ref{sec:method}). It is worth noting that the optimal range of scales is not necessarily the same for the two models. Generally, full-shape RSD analyses use scales going from tens of \hmpc\ to about $130-150\,$\hmpc. Including smaller scales potentially increases the precision of the constraints but at the expense of stronger biases on the recovered parameters. This is related to the limitations of current RSD models to fully describe the non-linear regime. On the other hand, including scales larger than $\sim 130$ \hmpc\ does not significantly improve the precision, since the variations of the model on those scales are small.

\begin{figure}
    \centering
    \includegraphics[width=\columnwidth]{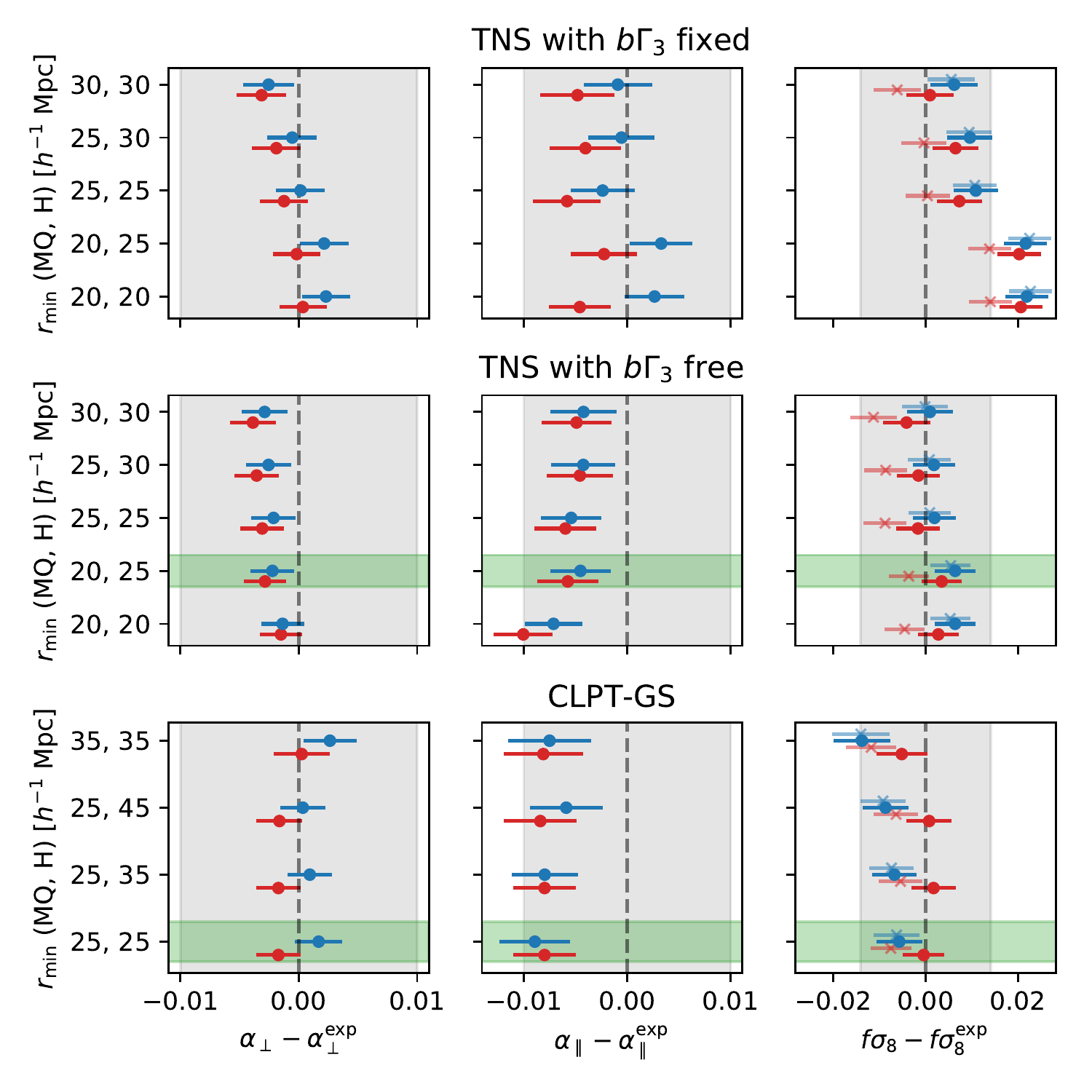}

    \caption{Biases in the measurement of $\fsig, \apara, \aperp$ obtained from full-shape fits to the average of 84 multipoles from the \textsc{Nseries} mocks as a function of the separation range used. The y-axis displays the value of the minimal separation $r_{\rm min}$ used in fits of the monopole, quadrupole (MQ) and hexadecapole (H). Top and mid rows display results for the TNS model when fixing or letting free the parameter $b\Gamma_3$ respectively. Bottom row presents results for the CLPT-GS model. The blue circles correspond to the analysis using $\Omega_m^{\rm fid} = 0.286$ (the true value of simulations) while the red squares correspond to $\Omega_m^{\rm fid} = 0.31$. The gray shaded areas correspond to 1 per cent errors in $\aperp, \apara$ and to 3 per cent in $\fsig$. The green shared area shows our choice for baseline analysis for TNS and CLPT-GS models.
}

    \label{fig:rmin_impact}
\end{figure}

In order to determine the optimal range of scales for our RSD models, we performed fits to the mean correlation function of the \textsc{Nseries} mocks, which are those that most accurately predict the expected RSD in the data. Figure~\ref{fig:rmin_impact} shows the best-fit values of $\fsig$, $\apara$, and $\aperp$ as a function of the minimum scale used in the fit, $r_{\rm min}$. In each panel, the grey bands show 1 per cent errors in $\aperp, \apara$ and 3 per cent errors in $\fsig$ for reference. 
Top panels present the measurements from 
the TNS model when the parameter $b_{\Gamma 3}$ fixed to the value given by Eq.~\ref{eq:nllbg3}, while in the mid panels this parameter is let free. 
Bottom panels show best-fit values for the CLPT-GS model as studied 
in \citet{icaza-lizaola_clustering_2020}. 
As noted in \citet{zarrouk_clustering_2018}, the hexadecapole is more
sensitive to the difference between the true and fiducial cosmologies 
and is generally less well modelled on small scales compared to the monopole and quadrupole. We therefore consider the possibility of having a different minimum fitting scale for the hexadecapole with respect to the monopole and quadrupole that share the same $r_{\rm min}$. For consistency with the other systematic tests, we performed this analysis using two choices of fiducial cosmologies, $\Omega^{\rm fid}_m = 0.286$ (blue) and $\Omega^{\rm fid}_m = 0.31$ (red). The maximum separation in all cases is $r_{\rm max} = 130$\hmpc, as we find that using larger $r_{\rm max}$ has a negligible impact on the recovered parameter values and associated errors.

In the case of the TNS model, we consider two different cases that correspond to when $b_{\Gamma 3}$ is fixed to its Lagrangian prediction and when $b_{\Gamma 3}$ is allowed to vary. In the case of $\Omega^{\rm fid}_m = 0.286$ and when $b_{\Gamma 3}$ is fixed, in the top panels of Figure~\ref{fig:rmin_impact}, we can see that $f\sigma_8$ is overestimated by 1.5 per cent when using scales above 25\hmpc\ 
and by 2 per cent below. Using $r_{\rm min}>25$\hmpc\ reduces the bias to about 1 per cent on $f\sigma_8$. For $\alpha_\parallel$ and $\alpha_\perp$ parameters, biases range from 0.3 to 0.5 per cent and are all statistically consistent with zero. When $b_{\Gamma 3}$ is let free, in the mid panels of Figure~\ref{fig:rmin_impact}, the model provide more robust measurements of $f\sigma_8$ at all tested ranges. The biases in $f\sigma_8$ over all ranges does not exceed 0.6$\sigma$, compared to approx 2.5$\sigma$ for the fixed $b_{\Gamma 3}$ case. We also remark that letting $b_{\Gamma_3}$ free also provides a better fit to the BAO amplitude and the hexadecapole on the scales of $20-25$\hmpc. We see a 1 per cent bias on $\apara$ when $r_{\rm min} = 20$\hmpc\ for all three multipoles. This bias is however reduced by increasing the hexadecapole minimum scale to $r_{\rm min}=25$\hmpc. The most optimal configuration for the TNS model is to let $b\Gamma_3$ free and fit the monopole and quadrupole in the range  $20 \leq r \leq 130$\hmpc\ and the hexadecapole in the range $25  \leq r \leq 130$\hmpc, as marked by the green band in Figure~\ref{fig:rmin_impact}. If we use $\Omega^{\rm fid}_m=0.31$, the trends and quantitative results are similar to the case with $\Omega^{\rm fid}_m=0.286$.

For the CLPT-GS model, an exploration of the optimal fitting range was done in \citet{icaza-lizaola_clustering_2020}. Two sets of tests have been performed. The first set consisted of fitting the mean of the mocks  when varying $r_{\rm min}$ and the second,  fitting  the 84 individual mocks and measuring the bias and variance of the best fits when varying $r_{\rm min}$. We revisit the first set of tests, but this time performing a full MCMC analysis to determine best fits and errors.
The bottom panels of Figure~\ref{fig:rmin_impact} summarise the results. 
In the case of $\Omega^{\rm fid}_m = 0.286$, we see that using $r_{\rm min} = 25$\hmpc\ for all multipoles yields to biases  of 0.1, 1.1 and 1.6 per cent in $\aperp$, $\apara$, and  $f\sigma_8$.
Increasing $r_{\rm min}$ for the hexadecapole while fixing $r_{\rm min} = 25$\hmpc\ for the monopole and quadrupole, does not change the results significantly, the biases are 0.1 per cent for all ranges in $\aperp$, and 1 per cent also for all ranges in $\apara$. For $f\sigma_8$ variations of 0.1-0.2 per cent arises when varying the range, but this variation in statistically consistent with zero. In the case of $\Omega^{\rm fid}_m = 0.31$, we find very similar trends. Using $r_{\rm min} = 25$\hmpc\ for all multipoles yields biases of 0.2, 0.9 and 1.6 in $\aperp$, $\apara$ and $f\sigma_8$ respectively. When we decrease the range of the fits, the biases on $(\aperp,\apara,\fsig)$ varies by (0.1-0.2, 0.2-0.3, 0.3-0.4) per cent. These variations are not significant and we decide to keep the lowest considered minimum scales on the hexadecapole in the fits. 


Compared with previous BOSS full-shape RSD analysis in configuration space, we used for CLPT-GS model the same minimum scale for the monopole and quadrupole \citep{ satpathy_clustering_2017,alam_clustering_2017}. The hexadecapole was not included in BOSS analyses. The exploration for the optimal minimum scale to be used for the hexadecapole was done in \cite{icaza-lizaola_clustering_2020} and revisited in this work. The systematic error associated to the adopted fitting range is also consistent with previous results for the case where only the monopole and quadrupole are used, as reported in \cite{icaza-lizaola_clustering_2020}. 
The TNS model was not used in configuration space for analysing previous SDSS samples. However, as we describe in section \ref{section:syserr}, the bias associated with both models when using their optimal fitting range is consistent between them, as well as consistent with previous BOSS results.

Overall, these tests performed on the {\sc Nseries} mocks allow us to define the optimal fitting ranges of scales for both RSD models. Minimizing the bias of the models while keeping $r_{\rm min}$ as small as possible, we eventually adopt the following optimal ranges:
\begin{itemize}
    \item TNS model: $20 < r < 130$\hmpc\ for $\xi_0$ and $\xi_2$, and $25 < r < 130$\hmpc\ for $\xi_4$ 
    \item CLPT-GS model: $25 < r < 130$\hmpc\ for  all multipoles,
\end{itemize}
which serve as baseline in the following. We compare the performance of the two models using these ranges in the following sections.

\subsubsection{Systematic errors from RSD modeling and adopted fiducial cosmology}\label{section:syserr}

\begin{figure}
    \centering
    \includegraphics[width=\columnwidth]{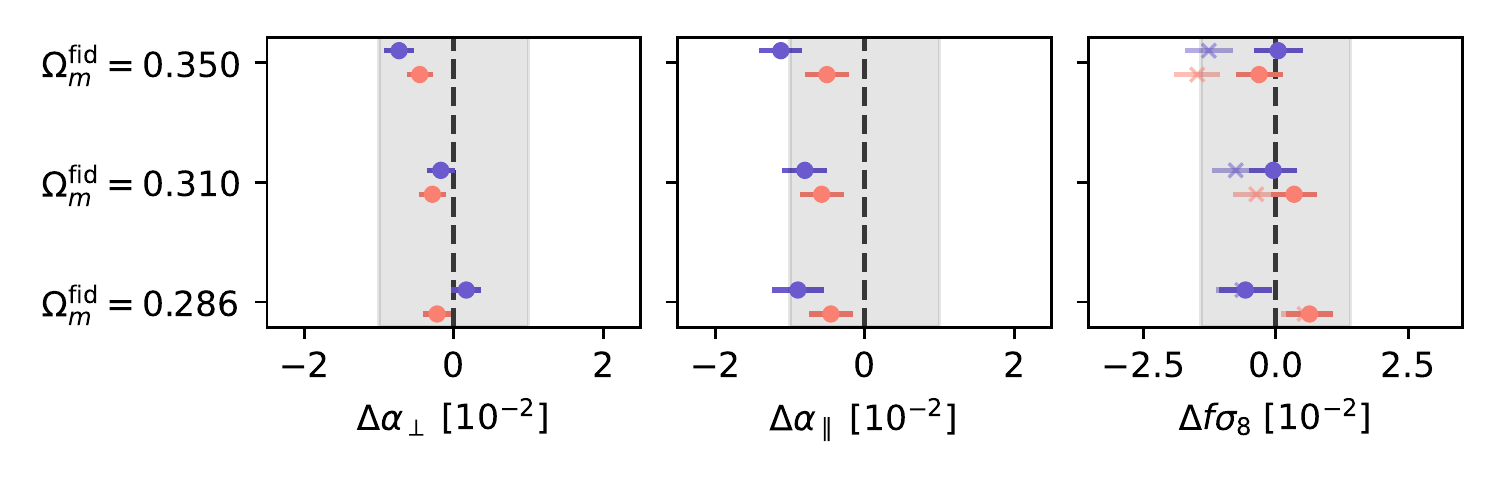}
    \caption{Biases in best-fit parameters for both CLPT-GS (blue) and TNS (red) 
    models from fits to the average multipoles of 84 {\sc Nseries} mocks.
    Shaded grey areas show the equivalent of 1 per cent error for $\aperp,\apara$
    and 3 per cent for $\fsig$. In the right panel, crosses indicate $\fsig$ values
    when $\sigma_8$ is not recomputed as described in Section~\ref{sec:fs8_scaling}. 
    The true cosmology of the mocks is
    $\Omega_m = 0.286$. 
    For reference, the errors on our data sample are $\sim$ 2, 3 and 10 per cent for $\aperp, \apara, \fsig$ respectively. 
    } 
    \label{fig:rsd_nseries_cosmology}
\end{figure}

We quantify in this section the systematic error introduced by the RSD
modelling and the choice of fiducial cosmology. For this, we used the 
{\sc Nseries} mocks\footnote{Given the mismatch between the clustering of 
the \textsc{MockChallenge} mocks and data, and its larger cosmic variance 
compared to {\sc Nseries} mocks, we decided to use {\sc MockChallenge} 
only for the quantification of systematic errors related to the halo occupation models.}. The measurements of $\aperp$, $\apara$ and 
$f\sigma_8$ from fits to the average multipoles are given in
Table~\ref{tab:rsd_ns_stats_bias} and shown in Figure~\ref{fig:rsd_nseries_cosmology}. The shaded area in the figure corresponds to 1 per cent deviation for $\aperp, \apara$ expected values and 3 per cent for $\fsig$ expected value. We used both TNS (red) and CLPT-GS (blue) models and consider three choices of fiducial cosmologies parameterised by their value of $\Omega_m^{\rm fid}$. Note that, as for the BAO analysis, we only test flat $\Lambda$CDM models close to the most probable one. We expect the full-shape analysis to be biased if the fiducial cosmology is too different from the truth (the parametrisation with $\aperp$ and $\apara$ would not fully account for the distortions and the template power spectrum would differ significantly). 

\begin{table}
  \centering
  \caption{Performance of the two full-shape models on the 
  {\sc Nseries} mocks. Fits were performed on the
  the average of 84 multipoles. We report the shifts
  of best-fit parameters relative to their expected values.
  For $\Omega_m^{\rm fid} = 0.286$
  we expect that both the $\alpha$ parameters are equal to 1. 
  For $\Omega_m^{\rm fid} = 0.31$, $\aperp^{\rm exp} = 0.9788$, 
  $\apara^{\rm exp}= 0.9878$ while for $\Omega_m^{\rm fid} = 0.35$ 
  we expect $\aperp^{\rm exp} = 0.9623$, $\apara^{\rm exp}= 0.9851$. 
  Since the growth rate of structures does not depend on the assumed
  cosmology, we expect to recover $f\sigma_8^{\rm exp} = 0.469$ for
  all cases. 
  }
  \begin{tabular}{lccccc}
    \hline
    \hline
Model & $\Omega_m^{\rm fid}$ &  $\Delta \aperp \ [10^{-2}]$ & $\Delta \apara \ [10^{-2}]$ &  $\Delta f\sigma_8 \ [10^{-2}]$ \\ 
\hline
{\sc clpt-gs} & 0.286& $0.2 \pm 0.2$  & $-0.9 \pm 0.3$  & $-0.6 \pm 0.5$   \\
{\sc clpt-gs} & 0.31& $-0.2 \pm 0.2$  & $-0.8 \pm 0.3$  & $-0.0 \pm 0.5$   \\
{\sc clpt-gs} & 0.35& $-0.7 \pm 0.2$  & $-1.1 \pm 0.3$  & $0.0 \pm 0.5$   \\
{\sc tns} & 0.286& $-0.2 \pm 0.2$  & $-0.5 \pm 0.3$  & $0.6 \pm 0.4$   \\
{\sc tns} & 0.31& $-0.3 \pm 0.2$  & $-0.6 \pm 0.3$  & $0.3 \pm 0.4$   \\
{\sc tns} & 0.35& $-0.5 \pm 0.2$  & $-0.5 \pm 0.3$  & $-0.3 \pm 0.4$   \\
\hline
\hline
  \end{tabular}
  \label{tab:rsd_ns_stats_bias}
\end{table}

We find that both RSD models are able to recover the true parameter values within these bounds. We estimate the systematic errors related to RSD modelling using Eq.~\ref{eq:syst_error1} and \ref{eq:syst_error2} by considering the shifts for the case where $\Omega_m^{\rm fid} = 0.286$ which is the true cosmology of the {\sc Nseries} mocks. 
We obtain, for $\aperp$, $\apara$ and $\fsig$, respectively: 
\begin{eqnarray}
    {\rm {\sc CLPT-GS}}: \sigma_{\rm syst, model} = (0.4, \ 0.9, \ 1.0) \times 10^{-2} \\
    {\rm TNS}: \sigma_{\rm syst, model} = (0.4, \ 0.6,  \ 0.9) \times 10^{-2}.
\end{eqnarray}

The biases on the recovered parameters shown in
Figure~\ref{fig:rsd_nseries_cosmology} induced by the choice 
of fiducial cosmology remain within 1, 1, and 3 per cent
for $\aperp, \apara$, and $\fsig$ respectively. 
For $\aperp$, both CLPT-GS and TNS models produces biases 
lower than 2$\sigma$ for all cosmologies except 
$\Omega_m^{\rm fid}=0.35$, which is the most distant value 
from the true cosmology of the simulation $\Omega_m = 0.286$.
For $\apara$, all biases are consistent with zero at 2$\sigma$
level for the TNS model, while CLPT-GS shows biases slightly 
larger than 2$\sigma$ for all $\Omega_m^{\rm fid}$.

The right panel of Figure~\ref{fig:rsd_nseries_cosmology} shows
the measured $\fsig$ when using the original value of $\sigma_8$ from the 
template (crosses) and when recomputing it with the scaling of 
$R=8$\hmpc\ by the isotropic dilation factor $\alpha = \aperp^{(2/3)}\apara^{(1/3)}$ (filled circles) as described in 
Section~\ref{sec:fs8_scaling}.
Both TNS and CLPT-GS models show a consistent dependency with
$\Omega_m^{\rm fid}$ when $\sigma_8$ is not re-evaluated: larger $\Omega_m^{\rm fid}$ yields smaller $\fsig$.
This is also found in the Fourier-space analysis of
\citet{gil-marin_2020} and in Figure 14 of \citet{smith_2020}.
As we recompute $\sigma_8$, this dependency is considerably 
reduced, which in turn reduces the contribution of the choice of fiducial 
cosmology to the systematic error budget.
Using Eq.~\ref{eq:syst_error1} and \ref{eq:syst_error2}, with
the entries of Table~\ref{tab:rsd_ns_stats_bias} (with $\sigma_8$ re-computed) where
$\Omega_m^{\rm fid} \neq 0.286$ compared to the entries where $\Omega_m^{\rm fid} = 0.286$, we obtain the following systematic errors associated with the choice of fiducial 
cosmology for $\aperp$, $\apara$ and $\fsig$, respectively: 
\begin{eqnarray}
    {\rm CLPT-GS}: \sigma_{\rm syst, fid} = (0.9, \ 1.0, \ 1.4 ) \times 10^{-2} \\
    {\rm TNS}: \sigma_{\rm syst, fid} = (0.5, \ 0.8,  \ 1.2) \times 10^{-2}
\end{eqnarray}
These systematic errors would be twice as large if $\sigma_8$ was
not recomputed as described in Section~\ref{sec:fs8_scaling}.

\subsubsection{Systematic errors from HOD}

We quantify in this section the potential systematic errors
introduced by the models with respect to how LRGs occupy dark
matter halos. This is done by analysing mock catalogs produced with different halo occupation distribution (HOD) models that mimic different underlying galaxy clustering properties. The same input dark matter field is used when varying the HOD model. We use the \textsc{OuterRim} mocks described in 
Section~\ref{sec:mocks_description} and in \citet{rossi_2020}.
Specifically, we analysed the mocks constructed using the 
``Threshold 2'' for the HOD models from
\citet{leauthaud_theoretical_2011, tinker_evolution_2013} and 
\citet{hearin_beyond_2015} and performed fits to the average
multipoles over the 27 realisations available for each HOD model.

\begin{figure}
    \centering
     \includegraphics[width=\columnwidth]{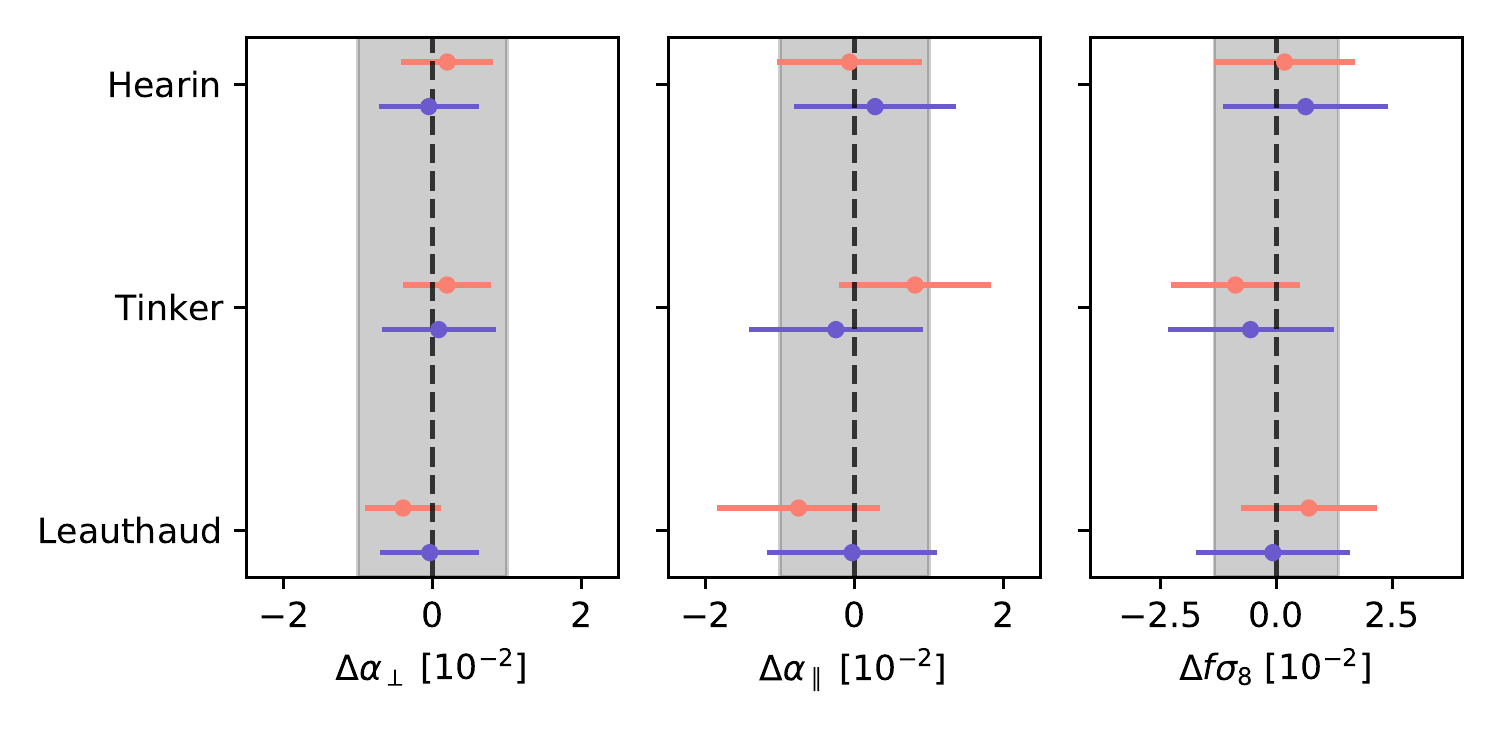}
    \caption{Best-fit values of $\aperp$, $\apara$ and $\fsig$ from 
    fitting the average multipoles of the \textsc{OuterRim} mocks 
    compared to their average over all HOD models. 
    Blue points show results for the CLPT-GS model and red points show results for the TNS model. The shaded area shows 1\% error for 
    $\aperp, \apara$ and 3\% for $\fsig$. 
    }
    \label{fig:OR_stats}
\end{figure}

Figure~\ref{fig:OR_stats} and Table~\ref{tab:rsd_or_stats_bias}
shows the results. In this figure, each best-fit parameter 
is compared to the average best-fit over all HOD models in order to quantify the relative impact of each HOD (instead of comparing with their true value). The biases with 
respect to the true values were quantified in the previous 
section. The shaded regions represent 1 per cent error for $\aperp$
and $\apara$, and 3 per cent error for $f\sigma_8$. 

We find that the biases for both RSD models are all within 1$\sigma$ 
from the mean, although statistical errors are quite large (around
one per cent for $\aperp, \apara$)
compared to {\sc Nseries} mocks for instance.  Also, the 
observed shifts are all smaller than the systematic errors 
estimated in the previous section. If we were to use the same
definition for the systematic error introduced in
Section~\ref{sec:robustness}, the relatively 
large errors from these measurements would produce a significant
contribution to the error budget. Therefore we consider that HOD 
has a negligible  contribution to the total systematic error budget.

\begin{table}
  \centering
  \caption{Performance of the full-shape analyses on the 
  {\sc Outerim} mocks produced using different HOD recipes. 
  For each HOD \citep{leauthaud_theoretical_2011,
  tinker_evolution_2013, hearin_beyond_2015},
  we display results obtained from our two RSD models (CLPT-GS and TNS). 
  All results are from fits to the average multipoles of 27 realisations.
  Each row displays the shift of best-fit parameters 
  with respect to the average parameters over the three HOD models: $\Delta x = x - \langle x \rangle_{\rm HOD}$.
  We found that these shifts are not significant and therefore do not 
  contribute to systematic errors. 
  }
  \begin{tabular}{llccc}
\hline
\hline
HOD  &  Model &   $\Delta \aperp$ [$10^{-2}$] & 
   $\Delta \apara$   [$10^{-2}$]& 
   $\Delta \fsig$  [$10^{-2}$] \\
  
\hline

L11 & {\sc clpt-gs} & $0.0 \pm 0.7$  & $0.0 \pm 1.1$  & $-0.1 \pm 1.7$  \\
T13 & {\sc clpt-gs} & $0.1 \pm 0.8$  & $-0.2 \pm 1.2$  & $-0.6 \pm 1.8$   \\
H15 & {\sc clpt-gs} & $0.0 \pm 0.7$  & $0.3 \pm 1.1$  & $0.6 \pm 1.8$  \\
L11 & {\sc tns}  &  $-0.4 \pm 0.5$  & $-0.7 \pm 1.1$   & $0.7 \pm 1.5$   \\
T13 & {\sc tns}  &  $0.2 \pm 0.6$  & $0.8 \pm 1.0$ &  $-0.9 \pm 1.4$   \\
H15 & {\sc tns}  &  $0.2 \pm 0.6$  & $-0.1 \pm 1.0$   & $0.2 \pm 1.5$   \\
\hline
  \end{tabular}
  \label{tab:rsd_or_stats_bias}
\end{table}

\subsubsection{Systematic errors from observational effects}

We investigate in this section the observational systematics. 
We used a set of 100 {\sc EZmocks} to quantify their impact on our measurements. 
From the same set, we added different observational effects. 
For simplicity, those samples were made from mocks reproducing only the 
eBOSS component of the survey, neglecting the CMASS component. 
We consider that the systematic errors estimated this way can be 
extrapolated to the full eBOSS+CMASS sample by assuming that their 
contribution is the same over the CMASS volume. We thus produced the 
following samples:
\begin{itemize}
    \item[1.] no observational effects included, which we use as reference,
    \item[2.] including the effect of the radial integral constraint 
    \citep[RIC,][]{de_mattia_integral_2019}, where the redshifts of the random catalog are randomly chosen from the redshifts of the data catalog,
    \item[3.] including RIC and all angular features: fiber collisions, redshift failures, and photometric corrections.
\end{itemize}

For each set, we computed the average multipoles and 
fitted them using our two RSD models. The covariance matrix is held fixed between cases. Table~\ref{tab:rsd_ez_stats_bias} 
summarises the biases in $\aperp, \apara, \fsig$ caused by the different observational effects. The shifts are relative to results of mocks without observational effects. We find that the radial integral constraint produces the greatest effect, particularly for the CLPT-GS model for which the deviation on $\fsig$ is slightly larger than $2\sigma$.  Indeed, the quadrupole for mocks with RIC 
has smaller absolute amplitude, which translates into small $\fsig$ 
values. However, when adding 
angular observational effects the shifts are all broadly consistent with 
zero, which indicates that the two effects partially cancel each other.

Using values from the Table~\ref{tab:rsd_ez_stats_bias} and Eqs.~\ref{eq:syst_error1} 
and \ref{eq:syst_error2}, we derive the following systematic errors 
from observational effects for $\aperp$, $\apara$ and $\fsig$, respectively: 
\begin{eqnarray}
    {\rm CLPT-GS}: \sigma_{\rm syst, obs} &=& (0.9, \ 1.2, \ 1.7 ) \times 10^{-2} \\
    {\rm TNS}: \sigma_{\rm syst, obs} &=& (1.0, \ 1.3, \ 1.8 )   \times 10^{-2}
\end{eqnarray}
These systematic errors are about 50 per cent of the statistical 
errors for each parameter, which corresponds to the  
most significant contribution to the systematic error budget.    



\begin{table}
  \centering
  \caption{Impact of observational effects on the full-shape analysis using 
  {\sc Ezmocks}. 
  Each row displays the shifts of best-fit parameters with respect to the 
  case without observational effects (``no syst''): 
  $\Delta x = x - x^{\rm no \ syst}$. Fits are performed on 
  the average multipoles of 100 realisations. We test the cases of mocks
  with radial integral constraint (RIC) and mocks with the combination of RIC and all angular observational effects (fiber collisions, redshift failures and photometric fluctuations). The angular effects introduced in mocks are corrected using the same procedure used in data. For simplicity, the mocks used here are only for the eBOSS part of the survey.
  }
  \begin{tabular}{ccccc}
\hline
\hline
Type &  
Model & 
$\Delta \aperp$ [$10^{-2}$] & 
$\Delta \apara$ [$10^{-2}$]&  
$\Delta f\sigma_8$ [$10^{-2}$] \\
\hline
RIC & {\sc clpt-gs}          & $-0.3 \pm 0.5$  & $1.1 \pm 0.6$  & $-1.7 \pm 0.8$   \\
$+$Ang. Sys. & {\sc clpt-gs}  & $0.0 \pm 0.4$  & $0.3 \pm 0.6$   & $0.0 \pm 0.9$   \\
RIC & {\sc tns}           & $0.6 \pm 0.5$  & $-0.1 \pm 0.7$  & $-0.8 \pm 0.9$   \\
$+$Ang. Sys. & {\sc tns}   & $0.8 \pm 0.5$  & $-0.2 \pm 0.7$  & $0.1 \pm 0.9$   \\
\hline
\hline

  \end{tabular}
  \label{tab:rsd_ez_stats_bias}
\end{table}

\subsubsection{Total systematic error of the full-shape RSD analysis}

Table~\ref{tab:rsd_total_systematic} summarises all systematic error contributions to the full-shape measurements discussed in the previous sections. We show the results for our two configuration-space RSD models TNS and CLPT-GS and for the Fourier space analysis of \citet{gil-marin_2020}. We compute the total systematic error $\sigma_{\rm syst}$ by summing up all the contributions in quadrature, assuming that they are all independent. By comparing the systematic errors with the statistical error from the baseline fits to the data (see Section~\ref{sec:results_rsd}), we find that the systematic errors are far from being negligible: 
more than 50 per cent of the statistical errors for all parameters. The systematic errors are in quadrature to the 
diagonal of the covariance of each measurement. 
We do not attempt to compute the covariance between systematic errors and this approach is more conservative (it does not underestimate errors).
\begin{table}
  \centering
  \caption{Summary of systematic errors obtained from tests with mock
  catalogs. The total systematic error $\sigma_{\rm syst}$ is the 
  quadratic sum of each contribution. We compare the systematic errors 
  to the statistical errors from our baseline fits on real data. The last rows
  display the final error which is a quadratic sum of statistical and
  systematic errors. 
  }
  \begin{tabular}{llccc}
    \hline
    \hline
Type & Model &  $\sigma_{\aperp}$ & $\sigma_{\apara}$ & $\sigma_{\fsig}$ \\
\hline
 \multirow{2}{*}{Modelling}  & {\sc clpt-gs} & 0.004 & 0.009 & 0.010 \\
 & {\sc tns} & 0.004 & 0.006 & 0.009 \\
\multirow{2}{*}{Fid. cosmology}  & {\sc clpt-gs} & 0.009 & 0.010 & 0.014 \\
  & {\sc tns} & 0.005 & 0.008 & 0.012 \\
\multirow{2}{*}{Obs. effects}  & {\sc clpt-gs} & 0.009 & 0.012 & 0.017 \\
  & {\sc tns} & 0.010 & 0.014 & 0.018 \\
\hline
                    & {\sc clpt-gs} & 0.013 & 0.018 & 0.024 \\
$\sigma_{\rm syst}$ & {\sc tns} & 0.012 & 0.017 & 0.023 \\
                    & $P_\ell$ & 0.012 & 0.013 & 0.024 \\
                    \hline
                    & {\sc clpt-gs}  & 0.020  & 0.028  & 0.045  \\
$\sigma_{\rm stat}$ & {\sc tns}  & 0.018  & 0.031  & 0.040  \\
                    & $P_\ell$  & 0.027  & 0.036  & 0.042  \\
                    \hline
                                      & {\sc clpt-gs}  & 0.66  & 0.63  & 0.54  \\
$\sigma_{\rm syst}/\sigma_{\rm stat}$ & {\sc tns}  & 0.65  & 0.55  & 0.58  \\
                                      & $P_\ell$  & 0.43  & 0.37  & 0.58  \\
                                      \hline
                 & {\sc clpt-gs}  & 0.024  & 0.033  & 0.051  \\
$\sigma_{\rm tot} = \sqrt{\sigma_{\rm syst}^2+\sigma_{\rm stat}^2}$ 
                 & {\sc tns}  & 0.021  & 0.035  & 0.046  \\
                 & $P_\ell$  & 0.029  & 0.038  & 0.048  \\
\hline
\hline
  \end{tabular}
  \label{tab:rsd_total_systematic}
\end{table}

\subsection{Statistical properties of the LRG sample}
\label{sec:statistical_properties}

We can also use the {\sc EZmocks} for evaluating the statistical properties of the LRG sample, in particular to quantify how typical is our data compared with {\sc EZmocks}, but also for measuring the correlations among the different methods and globally validating our error estimation.

\begin{figure*}
    \centering
    \includegraphics[width=0.49\textwidth]{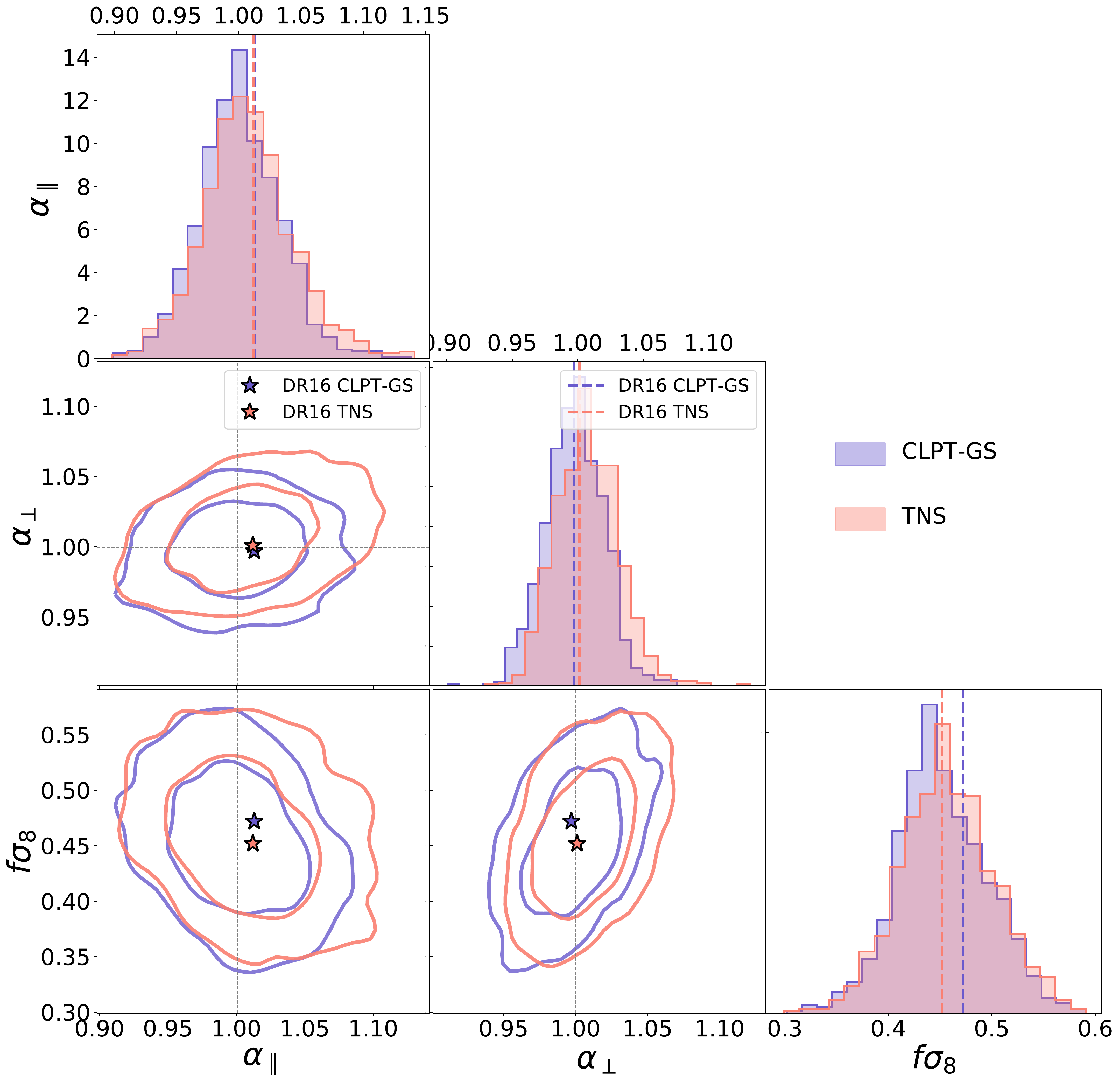}
    \includegraphics[width=0.49\textwidth]{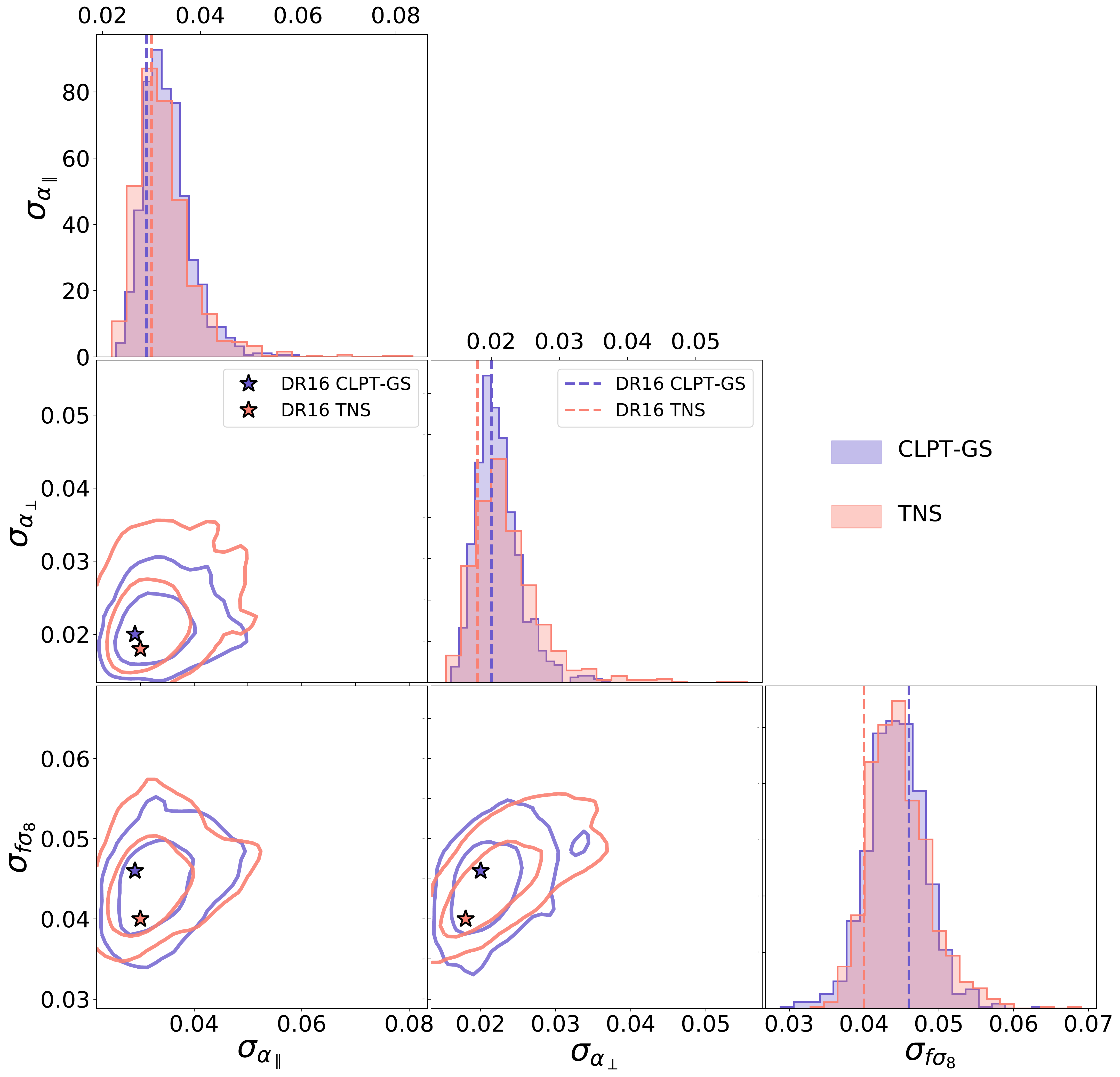}
    \caption{Comparison between best-fit values (left panels) and estimated errors
    (right panel) for $(\aperp, \apara, \fsig)$ using 1000 realisations of 
    {\sc EZmocks} fitted with the TNS and CLPT-GS models. The values obtained with 
    real data are indicated by stars in each panel or as coloured vertical lines 
    in the histograms. The thin black dashed line on the 2D plots refer to the true
    values each parameter in the {\sc EZmocks}.
    }
    \label{fig:Triangle_params_errors_EZ}
\end{figure*}

The left panel of the Figure~\ref{fig:Triangle_params_errors_EZ} presents a comparison between the best-fit $(\aperp, \apara, \fsig)$ and their estimated errors from fits of the TNS and the CLPT-GS models. The confidence contours contain approximately 68 per cent and 95  per cent of the results around the mean. The contours and histograms reveal a good agreement for the two models. Stars indicate the corresponding best fit values obtained from the data. 
The correlations between best-fit parameters of both models are 86, 83 and 93 per cent for 
$\aperp$, $\apara$ and $\fsig$ respectively. 
A similar comparison for the errors is presented in the right panel of the Figure~\ref{fig:Triangle_params_errors_EZ}.
The errors inferred from the data analysis, shown as stars, are in good agreement with the 2D distributions from the mocks, lying within the 68 per cent contours. The histograms comparing the distributions of errors for both methods also show a good agreement, in particular for $\apara$ and $f\sigma_8$. For $\aperp$, we observe that the distribution from CLPT-GS is slightly peaked towards smaller errors, while for TNS the error distribution has a larger dispersion for this parameter. The correlation coefficients between estimated errors from the two models are: 56, 38, and 39 per cent 
for $\aperp, \apara, \fsig$, respectively.

Table \ref{tab:ezmock_stats_errors} summarizes the statistical properties 
of errors for ${\aperp, \apara, \fsig}$ for both BAO and full shape RSD 
analysis in configuration space (noted $\xi_\ell$). We also include for 
reference the results from Fourier space analysis of \citet{gil-marin_2020}, 
noted $P_\ell$. For each parameter we show the standard deviation of the best 
fits values, $\sigma$, the mean estimated error $\langle \sigma \rangle$, the 
mean of the pull, $Z_i = (x_i - \langle x \rangle)/\sigma_x$ where 
$x={\aperp, \apara, \fsig}$, and its standard deviation $\sigma(Z)$. 
 If errors are correctly estimated and follow a Gaussian distribution, 
we expect that $\sigma = \langle \sigma_i \rangle$, $\langle Z_i \rangle = 0$ 
and $\sigma(Z) = 1$. 
For method, we remove results from non-converged chains and 5$\sigma$ outliers 
in both best-fit values and errors 
(with $\sigma$ defined as half of the range covered by the central 68 per cent values).
Table~\ref{tab:ezmock_stats_errors} also shows the results from combining 
different methods employing the procedure described in 
Section~\ref{sec:combining_analysis}.
For each combination, we create the covariance matrix $C$ 
(Eq.~\ref{eq:full_covariance}) from the correlation 
coefficients obtained from 1000 {\sc EZmocks} fits, 
with small adjustments to account for the
observed errors of a given realisation. 
The correlation coefficients (before this 
adjustement) is shown in Figure~\ref{fig:correlation_matrix} for all 
five methods. The BAO measurements from configuration and Fourier spaces 
are 87 and 88 per cent correlated for $\aperp$ and $\apara$, respectively.
In RSD analyses these correlations reduce to slightly less than 80 per cent
between $\aperp, \apara$ of both spaces, while $\fsig$ correlations reach 
84 per cent. The fact that these correlations are not exactly 100 per cent 
indicates that there is potential gain combining them.  

For the BAO results (top three rows of Table~\ref{tab:ezmock_stats_errors}), 
we see good agreement between $\sigma_x$ and $\langle \sigma \rangle$ 
for all the parameters in both the spaces. 
The mean of the pull  
$\langle Z_i \rangle$ is consistent with zero 
(their errors are roughly 0.02) and the standard deviation 
$\sigma(Z_i)$ is slightly smaller than unity for all variables, 
indicating that errors might be slightly overestimated. 
The combined BAO results of
$(\xi_\ell+P_\ell)$ have errors slightly reduced to 2.2\% for 
$\aperp$ and 3.4\% in $\apara$ (based on the scatter $\sigma$ of the 
best-fit values). The $\sigma(Z_i)$ are both closer to 1.0,
indicating better estimate of errors for the combined case. 
As a conservative approach, the BAO errors on data
(Section~\ref{sec:results_bao}) are therefore not corrected
by this overestimation. 

\begin{figure}
    \includegraphics[width=\columnwidth]{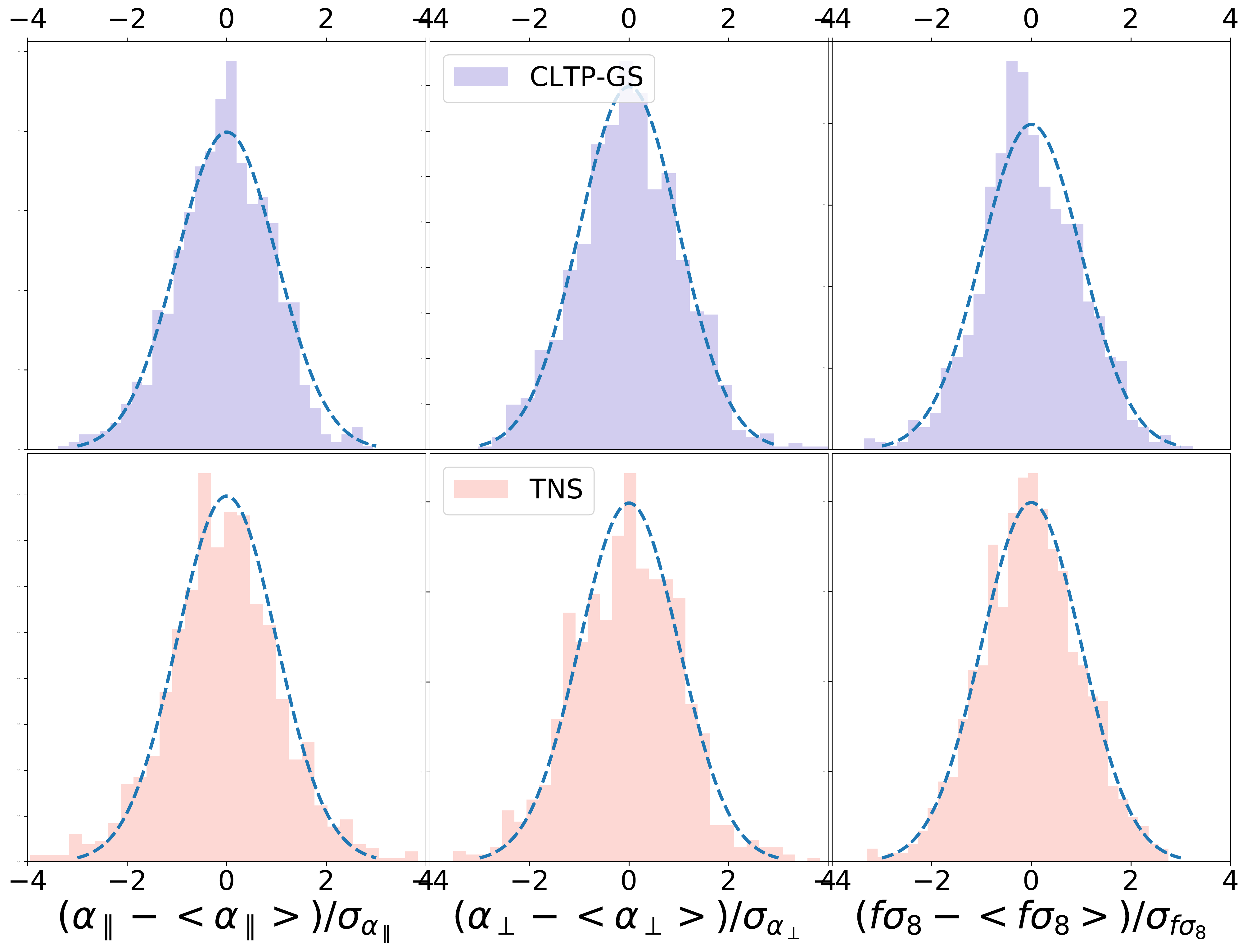}
    \caption{Normalized distributions of the pull for the $\apara$, $\aperp$ and $\fsig$ from fits of TNS and CLPT-GS models on {\sc EZmocks}. 
    The blue dashed lines represent the centered normalized Gaussian distribution.
    }
    \label{fig:Unitary_errors_EZmocks}
\end{figure}

\begin{figure}
    \centering
    \includegraphics[width=\columnwidth]{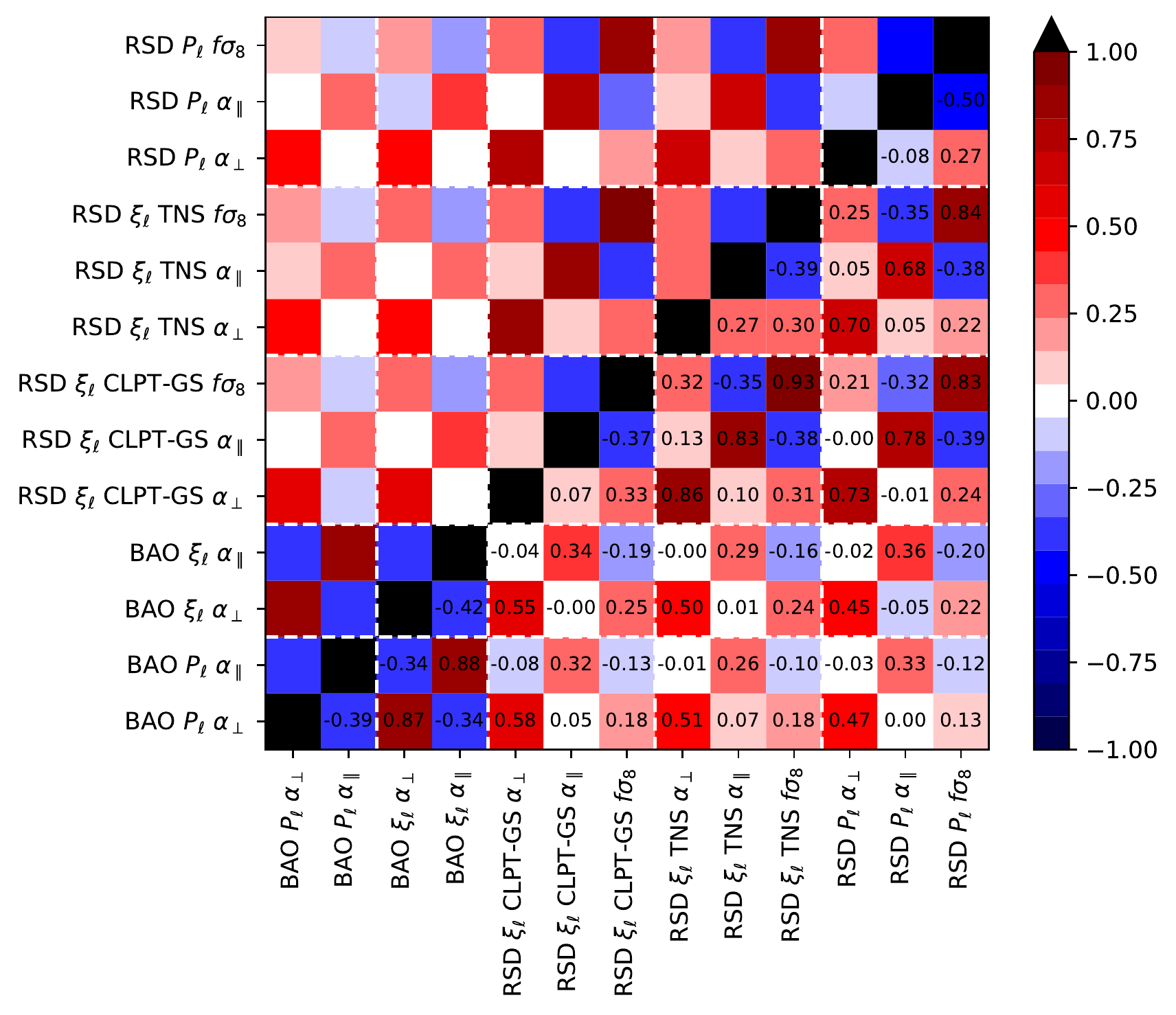}
    \caption{Correlation coefficients between $\aperp, \apara, \fsig$ for all methods and models obtained from fits to 1000 {\sc EZmock} realisation of the eBOSS LRG+CMASS
    sample.
    The values of $\fsig$ have been corrected with the procedure described in Section~\ref{sec:fs8_scaling}.}
    \label{fig:correlation_matrix}
\end{figure}

 Full shape RSD results (4th to 8th rows in 
Table~\ref{tab:ezmock_stats_errors}) 
also show good agreement between $\sigma_x$ and $\langle \sigma \rangle$
for all the parameters for both models and both spaces. 
Figure~\ref{fig:Unitary_errors_EZmocks} shows the pull distributions for 
both CLPT-GS and TNS models.  
The mean of the pull for $\aperp$ and $\fsig$ are 
consistent with zero in all cases though the mean pull for $\apara$ is 
negative, indicating a slightly skewed 
distribution. The $\sigma(Z_i)$
values for CLPT-GS and TNS models are consistent with one for $\aperp$ 
and slightly different than one for $\apara$ and $\fsig$. Their 
combination (6th row) with inverse variance weighing slightly compensates
for these differences, yielding better
estimated errors, with $\sigma(Z_i)$ closer to one for all three parameters.
The full-shape measurements in Fourier space (7th row) show similar 
behaviour than the ones in configuration space, with errors larger than 
measurements in configuration space. This is due to the larger number of 
nuisance parameters in the Fourier space analysis and to the choice
of scales used in the Fourier space fits 
($0.02 \leq k \leq 0.15 \ h {\rm Mpc}^{-1}$), 
which do not exactly translate to the range in 
separation used in our fits $(25 < r< 130$~\hmpc), 
and may contain less information in average. 
The combined $\xi_\ell+P_\ell$ full-shape results in the 8th row
present smaller dispersion on all parameters relative to each individual method.
The pull values indicating slightly overestimated errors, 
which we do not attempt to correct.

\begin{table*}
\centering
\caption{Statistics on errors from consensus results on 1000 {\sc EZmocks}
realisations. For each parameter, we show the standard deviation 
of best-fit values, $\sigma(x_i)$, the mean estimated error 
$\langle \sigma_i \rangle$, the mean of the pull, $Z_i = (x_i - \langle x_i 
\rangle)/\sigma_i$ and its standard deviation $\sigma(Z_i)$.
$N_{\rm good}$ shows the number of valid realisations for each case 
after removing extreme values and errors at 5$\sigma$ level.
}
\begin{tabular}{lccccccccccccccc}
\hline
\hline
Observable & $N_{\rm good}$ & \multicolumn{4}{c}{$\aperp$} & &
             \multicolumn{4}{c}{$\apara$} & & 
             \multicolumn{4}{c}{$\fsig$}  \\
             & &
$\sigma$ &  $\langle \sigma_i \rangle$  &  $\langle Z_i \rangle $ & $\sigma(Z_i)$ &  &
$\sigma$ &  $\langle \sigma_i \rangle$  &  $\langle Z_i \rangle $ & $\sigma(Z_i)$ &  &
$\sigma$ &  $\langle \sigma_i \rangle$  &  $\langle Z_i \rangle $ & $\sigma(Z_i)$ \\
\hline
{\sc bao} $\xi_\ell$
 & 
987
 & 0.023  & 0.023  & -0.02  & 0.98   &  & 0.036  & 0.035  & -0.02  & 0.96   &  & - & - & - & - \\
{\sc bao} $P_\ell$
 & 
978
 & 0.024  & 0.024  & -0.02  & 0.95   &  & 0.039  & 0.040  & 0.00  & 0.90   &  & - & - & - & - \\
{\sc bao} $\xi_\ell+P_\ell$
 & 
970
 & 0.022  & 0.022  & -0.02  & 1.01   &  & 0.034  & 0.034  & -0.02  & 0.97   &  & - & - & - & - \\
 \hline
{\sc rsd} $\xi_\ell$ {\sc clpt}
 & 
819
 & 0.023  & 0.021  & 0.01  & 1.03   &  & 0.033  & 0.033  & -0.04  & 0.95   &  & 0.046  & 0.045  & -0.01  & 0.97  \\
{\sc rsd} $\xi_\ell$ {\sc tns}
 & 
951
 & 0.024  & 0.023  & -0.05  & 1.03   &  & 0.037  & 0.033  & -0.05  & 1.07   &  & 0.046  & 0.045  & -0.01  & 0.95  \\
{\sc rsd} $\xi_\ell$
 & 
781
 & 0.021  & 0.021  & -0.01  & 0.99   &  & 0.031  & 0.032  & -0.03  & 0.96   &  & 0.042  & 0.045  & -0.01  & 0.95  \\
{\sc rsd} $P_\ell$
 & 
977
 & 0.025  & 0.026  & 0.02  & 0.94   &  & 0.037  & 0.036  & -0.04  & 1.00   &  & 0.046  & 0.046  & 0.01  & 0.96  \\
{\sc rsd} $\xi_\ell+P_\ell$
 & 
767
 & 0.019  & 0.020  & 0.00  & 0.98   &  & 0.030  & 0.031  & -0.03  & 0.97   &  & 0.041  & 0.043  & -0.00  & 0.97  \\
 \hline
{\sc bao}$+${\sc rsd} $\xi_\ell$
 & 
772
 & 0.018  & 0.019  & -0.01  & 1.00   &  & 0.024  & 0.025  & -0.03  & 0.97   &  & 0.043  & 0.040  & -0.02  & 1.06  \\
{\sc bao}$+${\sc rsd} $P_\ell$
 & 
955
 & 0.019  & 0.020  & 0.00  & 0.96   &  & 0.028  & 0.029  & -0.03  & 0.96   &  & 0.044  & 0.042  & -0.01  & 1.05  \\
 {\sc bao}$\times${\sc rsd} $P_\ell$
 & 
 986
 & 0.019  & 0.019  & 0.03  & 0.99   &  & 0.029  & 0.028  & -0.06  & 1.02   &  & 0.041  & 0.045  & -0.01  & 0.92  \\
 \hline
{\sc bao} ($\xi_\ell+P_\ell$) + 
 & 
747
 & 0.017  & 0.018  & -0.01  & 1.00   &  & 0.024  & 0.025  & -0.03  & 0.97   &  & 0.042  & 0.039  & -0.02  & 1.09  \\
 {\sc rsd}  ($\xi_\ell+P_\ell$) \\
({\sc bao}$+${\sc rsd}) $\xi_\ell$ +
 & 
747
 & 0.017  & 0.018  & -0.01  & 1.01   &  & 0.024  & 0.025  & -0.03  & 0.99   &  & 0.042  & 0.039  & -0.02  & 1.09  \\
 ({\sc bao}$+${\sc rsd}) $P_\ell$ \\
  \hline
 \hline
\end{tabular}
\label{tab:ezmock_stats_errors}
\end{table*}

The 9th and 10th row of Table \ref{tab:ezmock_stats_errors} 
show results of combining BAO and full-shape RSD results for a given 
space, $\xi_\ell$ or $P_\ell$, while fully accounting for their 
large covariance as described in Section~\ref{sec:combining_analysis}. 
We see that the scatter of $\aperp$ 
and $\apara$ is reduced by $\sim$ 20 and 30 per cent, respectively,
relative to their BAO-only analyses. For $\fsig$ the scatter of best-fit
values is the same as the full-shape-only analyses, as expected 
(BAO only do not provide extra information on 
$\fsig$). The values of $\sigma(Z_i)$ for the combined results 
are consistent with one for $\aperp, \apara$, though for $\fsig$ they 
are more than 5 per cent larger than unity for both configuration and Fourier
space. This would indicate that our 
combination procedure from Section~\ref{sec:combining_analysis} produces slightly underestimated errors for $\fsig$.
In \citet{gil-marin_2020}, an alternative method was suggested to extract the consensus results from BAO and RSD analysis: a simultaneous fit.  Both BAO and RSD models are fitted simultaneously to the
concatenation of the pre- and post-reconstruction data vectors. 
This  fit requires the full covariance matrix between pre- and post-reconstruction multipoles and is estimated from 1000 {\sc EZmocks}. 
Results of simultaneous fits on mocks are shown in the 11th row of
Table~\ref{tab:ezmock_stats_errors} and are noted ``BAO$\times$RSD $P_\ell$''. These are to be compared with our usual method of combining posteriors, noted ``BAO$+$RSD'' and shown in the 10th row. 
First, we see good agreement between the scatter of best-fit values
of all three parameters between BAO$\times$RSD and BAO$+$RSD. 
However, the simultaneous fit overestimates the errors in
$\fsig$ by 8 per cent, based on its $\sigma(Z_i)$ value. 
While in theory the simultaneous fit is a better procedure, accounting
for all correlations, in practice we only use 1000 mocks to estimate
a larger covariance matrix with large off-diagonal terms. 
Therefore we cannot conclude from this test, which method leads to 
better estimated errors. We use BAO$+$RSD entries for the consensus results.

 The last two rows of Table \ref{tab:ezmock_stats_errors} show
statistics on the final consensus results from the LRG sample when combining 
BAO and full shape from both Fourier and configuration spaces. 
These results reflect the full statistical power of the LRG sample.
The excellent agreement between the statistics of these two rows 
shows that the order of combination does not impact results.
The dispersion $\sigma$ on $\aperp$ and $\apara$ are reduced to 
1.8 and 2.6 per cent respectively while we had 2.2 and 3.4 per cent 
for BAO only, and 2.0 and 3.2 per cent for full-shape only. 
The pull distributions for $\aperp$ and $\apara$ are consistent with 
a Gaussian distribution. 
The scatter in $\fsig$ is not reduced compared to individual methods,
which is expected since BAO does not add information on this parameter,
so the consensus error should be equal to the one obtained from the full-shape fits.
However, the $\sigma(Z_i)$ for $\fsig$ indicates that our consensus 
errors on this parameter might be underestimated by 10 per cent.
While this seems to be significant, this result can be a consequence 
of the Gaussian assumption of all individual likelihoods
not holding for all realisations, or the combination procedure itself 
might lead to underestimated errors 
(as seen with $\fsig$ in the 9th and 10th rows),
though we would need more mocks to test these hypotheses carefully. 

For this work, we consider the underestimation on $\fsig$ consensus errors (last two rows of Table~\ref{tab:ezmock_stats_errors})
as another source of systematic error. The simplest correction to this underestimation is to scale the estimated errors of $\fsig$ in each realisation by $\sigma(Z_i) = 1.09$. We proceed to apply this correction factor to the consensus $\fsig$ errors with our data sample. This factor is to be applied only to statistical errors. In Section~\ref{sec:consensus_results} we describe how we apply with this scaling in the presence of systematic errors. 

\section{Results} \label{sec:results}

We provide in this section the results of the BAO analysis, 
the full-shape RSD analysis and the combination of the two 
for the eBOSS LRG sample. The analysis assumes an effective 
redshift for the sample of $z_{\rm eff}=0.698$. 

\subsection{Result from the BAO analysis}
\label{sec:results_bao}

\begin{figure}
    \centering
    \includegraphics[width=0.8\columnwidth]{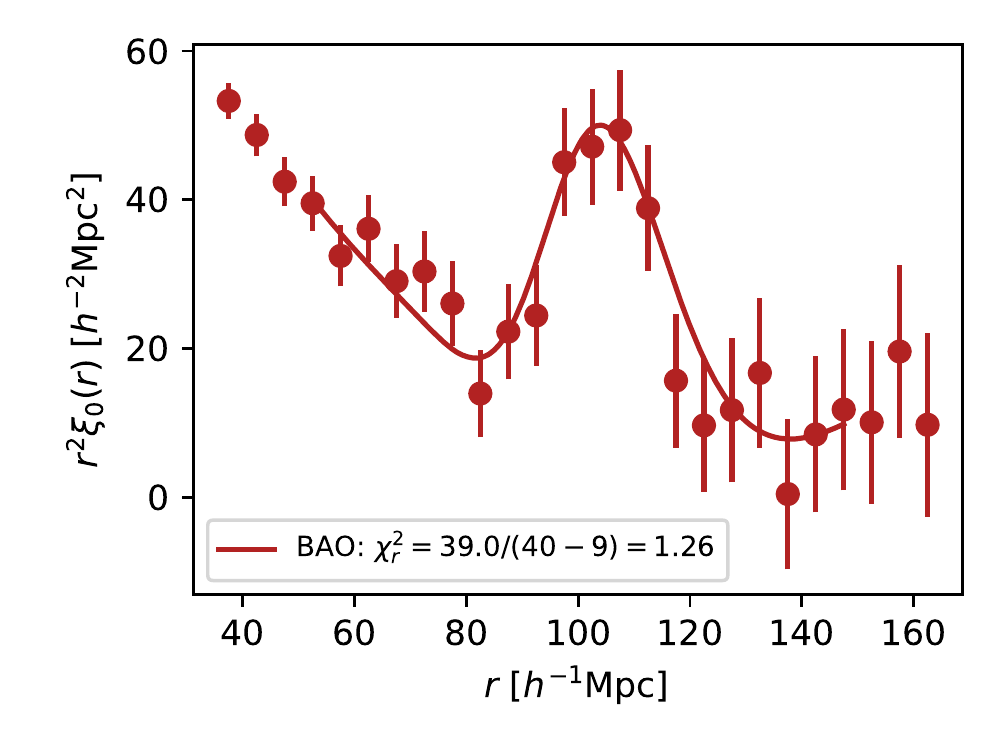}
    \includegraphics[width=0.8\columnwidth]{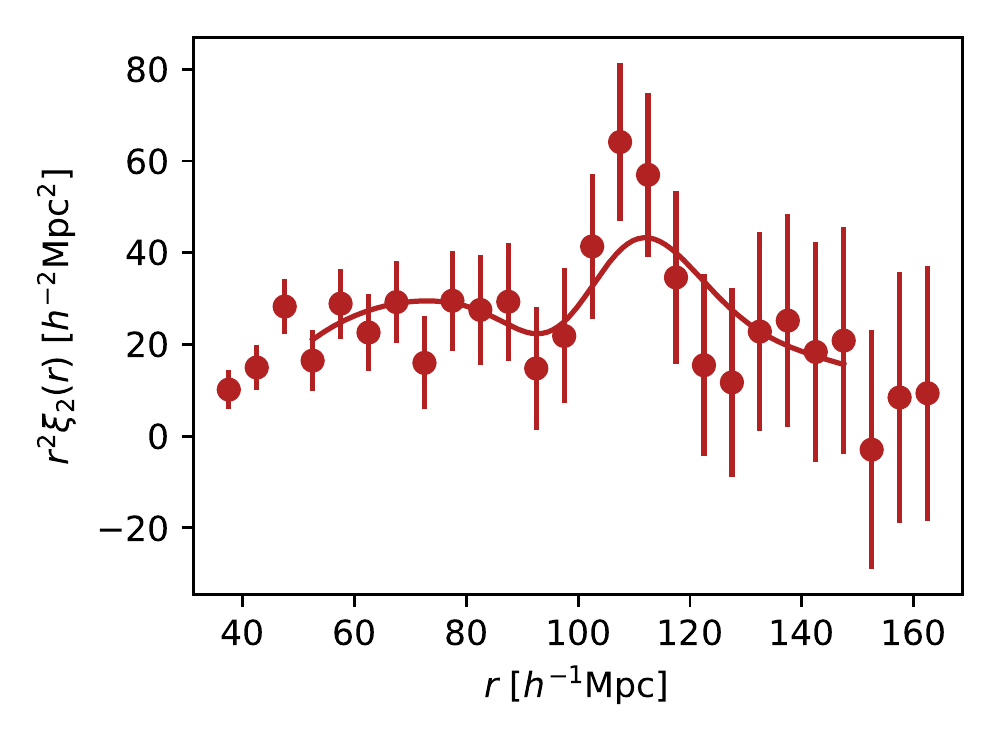}
    \caption{Best-fit BAO model to the monopole (top) and quadrupole (bottom) of the post-reconstruction correlation function of the eBOSS + CMASS LRG sample. The
    legend displays the $\chi^2$ value of the fit.}
    \label{fig:bao_bestfit}
\end{figure}

We present in Figure~\ref{fig:bao_bestfit} our best-fit BAO model 
to the post-reconstruction eBOSS LRG multipoles. The associated 
reduced chi-squared is $\chi^2/{\rm dof} = 39/(40-9) = 1.26$. By scaling the 
resulting $\aperp$ and $\apara$ by $(D_M/r_d)^{\rm fid}$ and
$(D_H/r_d)^{\rm fid}$, respectively (Eqs.~\ref{eq:aperp} 
and \ref{eq:apara}), we obtain:
 \begin{equation}
 \mathbf{D}_{{\rm BAO},{\xi_\ell}} =
 \begin{pmatrix}
D_M/r_d \\
D_H/r_d 
 \end{pmatrix}=
 \begin{pmatrix}
  17.86 \pm 0.33 \\
  19.34 \pm 0.54 
  \end{pmatrix}
 \end{equation}
and the covariance matrix is
 \begin{equation}
\mathbf{C}_{{\rm BAO},{\xi_\ell}} = \begin{blockarray}{cc}
D_M/r_d & D_H/r_d  \vspace{1mm} \\
\begin{block}{(cc)}
1.11 \times 10^{-1} & -5.86 \times 10^{-2} &  \\
 -  & 2.92 \times 10^{-1} &  \\
\end{block}
\end{blockarray}
\end{equation}
The errors correspond to a BAO measurement at 1.9 per cent in 
the transverse direction and 2.8 per cent in the radial direction, the best constraints ever obtained from $z>0.6$ galaxies. The correlation coefficient between both parameters is -0.33. 

\begin{figure}
    \centering
    \includegraphics[width=0.8\columnwidth]{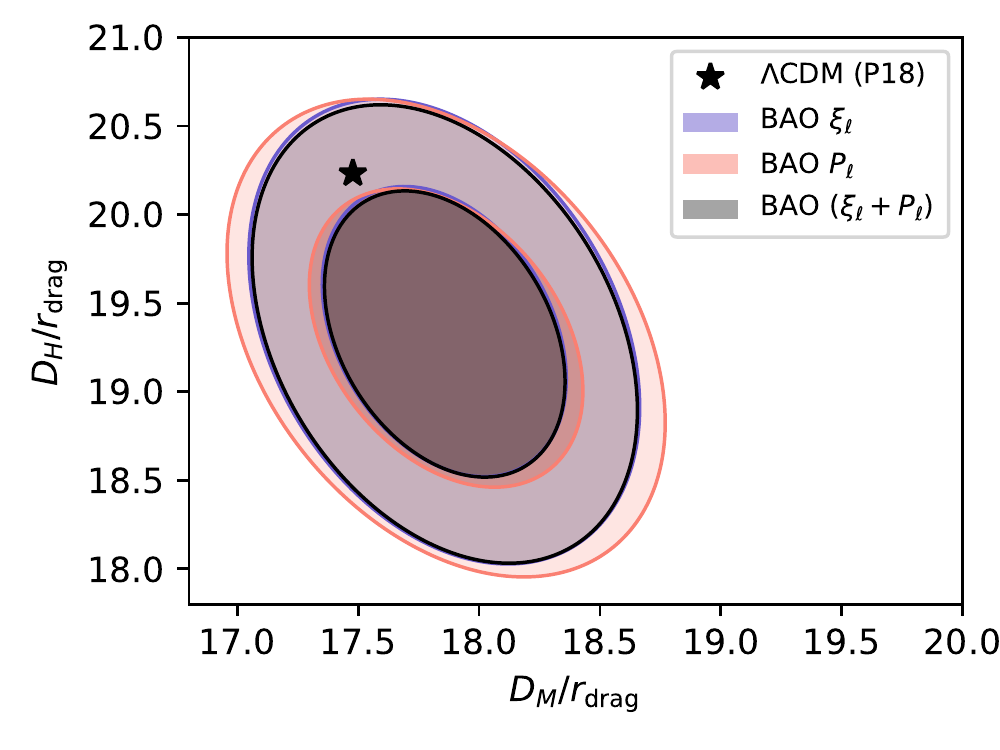}
    \caption{Constraints on $D_M/r_d$ and $D_H/r_d$ at 
    $z_{\rm eff}=0.698$ from the BAO analysis of 
    the eBOSS LRG sample post-reconstruction. Contours show 68 and 95 per cent 
    confidence regions for the configuration space analysis in blue (this work), the Fourier space analysis from \citet{gil-marin_2020} in salmon, and the consensus BAO result in grey. 
    The expected values in a flat $\Lambda$CDM 
    model with Planck 2018 best-fit parameters, shown by the black star,
    lies at 1.26$\sigma$ from our best-fit parameters of the configuration space analysis.  }
    \label{fig:consensus_bao}
\end{figure}

Figure~\ref{fig:consensus_bao} shows in blue the 68 and 95\% confidence 
contours in the $(D_M/r_d, D_H/r_d)$ space for the BAO measurement in 
configuration space. Our best-fit values are consistent within 
1.26$\sigma$ to the prediction of a flat $\Lambda$CDM model
given by Planck 2018 best-fit parameters 
\citep{planck_collaboration_planck_2018} assuming a $\chi^2$ distribution with two degrees of freedom. This measurement is also in 
excellent agreement with the BAO analysis performed in Fourier space 
\citep{gil-marin_2020}, shown as red contours in Figure~\ref{fig:consensus_bao}. 
Since Fourier and configuration space analyses use the same data, 
final measurements are highly correlated. 
Based on measurements of the same 1000 realisations of {\sc EZmocks}, 
we obtain correlation coefficients of 0.86 for both $D_M/r_d$ and $D_H/r_d$. 
As these correlations are not unity, 
there is some gain, in combining both measurements. Using
the methods presented in Section~\ref{sec:combining_analysis}, we compute
the combined BAO measurements between Fourier and configuration space. The 
result is displayed as grey contours in Figure~\ref{fig:consensus_bao} 
and in Table~\ref{tab:finalresults} as ``BAO $\xi_\ell+P_\ell$''. 
The error of the combined result is only 2\% smaller than the error 
of the configuration space analysis alone. 

Table~\ref{tab:bao_results_tests} shows the impact on 
the BAO results in configuration space of different
modifications in the methodology around the baseline configuration. 
The middle part of the table shows that our result is reasonably 
insensitive to some of these changes. Setting all systematic 
weights to unity causes only mild shifts to best-fit parameters
while estimated errors are unchanged. Removing the corrections
by weights  significantly distorts the broad shape of the 
correlation function. The fact that our BAO results are 
insensitive to these corrections proves that practically all information
comes uniquely from the BAO peak and 
not from the full-shape of the correlation function. 
This is a strong robustness validation of our BAO measurement. 
When leaving BAO damping parameters $(\Sigma_\perp,\Sigma_\parallel)$ free or 
constrained within a Gaussian prior, the best-fit values barely change while 
their errors are smaller than our baseline analysis. As observed on mocks, 
some realisations present sharper peaks due to noise and a sharper model 
could be considered as a better fit. However, we prefer to be conservative 
and not allow for this artificial increase in precision in our BAO analysis. 
Including the hexadecapole or changing the fiducial cosmology shifts alphas 
by less than one error bar, which is consistent to what is observed in mocks. 
We performed the BAO fits using the methods used in the BOSS DR12 analysis
\citep{alam_clustering_2017} which gives results in excellent agreement with 
our baseline method, with a slight better $\chi^2$. 
In the third part of Table~\ref{tab:bao_results_tests} we present the 
pre-reconstruction result with similar best-fit $\aperp$ and $\apara$ 
but with errors larger by factors of 1.3 and 1.5, which is typical as seen 
in mocks (Figure~\ref{fig:ezmock_bao_alphas}). 
Pre-reconstruction BAO-only fits using our methodology show biases of about 
1 per cent in the mocks, therefore we do not recommend using 
pre-reconstruction results without accounting for these biases.
The NGC and SGC results are two independent samples and their 
best-fit $\aperp$ and $\apara$ are 0.25 and 0.53$\sigma$ from 
each other respectively, therefore not representing a significant 
difference among hemispheres. 

\begin{table}
\caption{The BAO measurement with the DR16 eBOSS+CMASS LRG dataset using the 
standard pipeline described in Section~\ref{sec:bao_modelling} and other analysis 
choices. Note that for cases with different $\Omega_m^{\rm fid}$, we scale the obtained $\aperp, \apara$ by the distance ratios in order to
make them comparable
with the case where $\Omega_m^{\rm fid} = 0.31$.}
\begin{center}
\begin{tabular}{cccc}
case & $\alpha_\perp$ & $\alpha_\parallel$ & $\chi^2/{\rm d.o.f.}$ \\
\hline
\hline
Baseline & $1.024 \pm 0.019$ & $0.956 \pm 0.023$ & $39.0/(40-9)$ \\
\hline
$w_{\rm sys}w_{\rm cp}w_{\rm noz} = 1$ & $1.022 \pm 0.018$ & $0.954 \pm 0.023$ & $30.1/(40-9)$ \\
$\Sigma_\perp, \Sigma_\parallel$ free & $1.027 \pm 0.016$ & $0.947 \pm 0.019$ & $31.9/(40-11)$ \\
$\Sigma_\perp, \Sigma_\parallel$ prior & $1.025 \pm 0.017$ & $0.952 \pm 0.021$ & $36.2/(40-11)$ \\
 $+\xi_4$ & $1.031 \pm 0.019$ & $0.949 \pm 0.024$ & $53.5/(60-12)$ \\
$\Omega_m^{\rm fid} = 0.27$ & $1.026 \pm 0.020$ & $0.950 \pm 0.023$ & $33.3/(40-9)$ \\
$\Omega_m^{\rm fid} = 0.35$ & $1.026 \pm 0.019$ & $0.951 \pm 0.022$ & $39.6/(40-9)$ \\
DR12 method & $1.023 \pm 0.019$ &  $0.955 \pm 0.024$ &  $34.5/(40-10)$  \\
\hline
Pre-recon & $1.035 \pm 0.025$ & $0.957 \pm 0.035$ & $42.0/(40-9)$ \\
NGC only & $1.038 \pm 0.024$ & $0.943 \pm 0.024$ & $42.3/(40-9)$ \\
SGC only & $0.993 \pm 0.032$ & $0.982 \pm 0.070$ & $44.5/(40-9)$ \\
\hline
\hline
\end{tabular}
\end{center}
\label{tab:bao_results_tests}
\end{table}

\subsection{Results from the full-shape RSD analysis}
\label{sec:results_rsd}

\begin{table*}
\caption{The full-shape measurements with the DR16 eBOSS+CMASS LRG dataset 
from our baseline analysis described in Section~\ref{sec:rsd_modelling} 
followed by results from other analysis 
choices. The presented errors are purely statistical and do not include systematic errors. }
\centering
\begin{tabular}{lccccc}
\hline
\hline
Model & Analysis & $\alpha_\perp$ & $\alpha_\parallel$ & $f\sigma_8$ &$\chi^2/{\rm d.o.f.}$ \\
\hline
CLPT-GS &  baseline & 
$0.997 \pm 0.020$ & 
$1.013 \pm 0.028$ & 
$0.471 \pm 0.045$ &   $83.7/(63-6)=1.47$ \\
CLPT-GS & $r_{\rm min} = 35$\hmpc\ for $\xi_4$ & 
$1.017 \pm 0.022$ & 
$0.971 \pm 0.031$ & 
$0.499 \pm 0.046$ & 
  $79.3/(61-6)=1.44$ \\
CLPT-GS &  NGC only & 
$1.015 \pm 0.025$ & 
$1.009 \pm 0.031$ & 
$0.464 \pm 0.055$ &  $81.1/(63-6)=1.40$ \\
CLPT-GS &  SGC only & 
$0.985 \pm 0.036$ & 
$1.041 \pm 0.062$ & 
$0.439 \pm 0.078$ & $71.3/(63-6)=1.25$ \\
TNS & baseline & 
$ 1.001 \pm 0.018 $ & 
$ 1.013 \pm 0.031 $ & 
$ 0.451 \pm 0.040 $ & $85.2/(65-7)  = 1.47 $ \\
TNS & $r_{\rm min} = 35$\hmpc\ for $\xi_4$ & 
$ 1.013 \pm 0.016 $ & 
$ 0.976 \pm 0.027 $ & 
$ 0.458 \pm 0.036 $ & $73.7/(63-7)= 1.32$ \\
TNS & Without $\xi_4$ & 
$1.019 \pm 0.019$ & 
$0.963 \pm 0.035$ & 
$0.472 \pm 0.044$& $50.1/(44-7)= 1.35$ \\
TNS & NGC only & 
$ 1.024 \pm 0.029 $ & 
$ 1.013 \pm 0.036 $ & 
$ 0.436 \pm 0.053 $ & $80.6/(65-7) = 1.39$ \\
TNS & SGC only & 
$ 0.993 \pm 0.034 $ & 
$ 1.076 \pm 0.070 $ & 
$ 0.423 \pm 0.076 $ & $69.1/(65-7)= 1.19$ \\
\hline
\hline
\end{tabular}
\label{tab:rsd_results_variations}
\end{table*}%

\begin{figure*}
    \centering
    \includegraphics[width=0.32\textwidth]{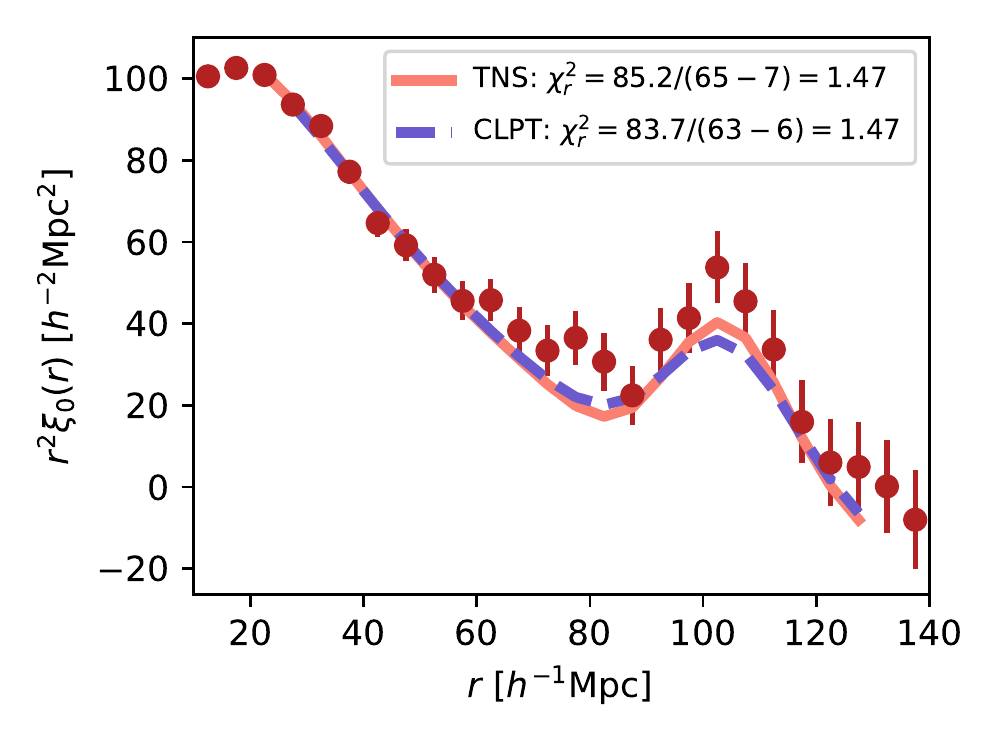}
    \includegraphics[width=0.32\textwidth]{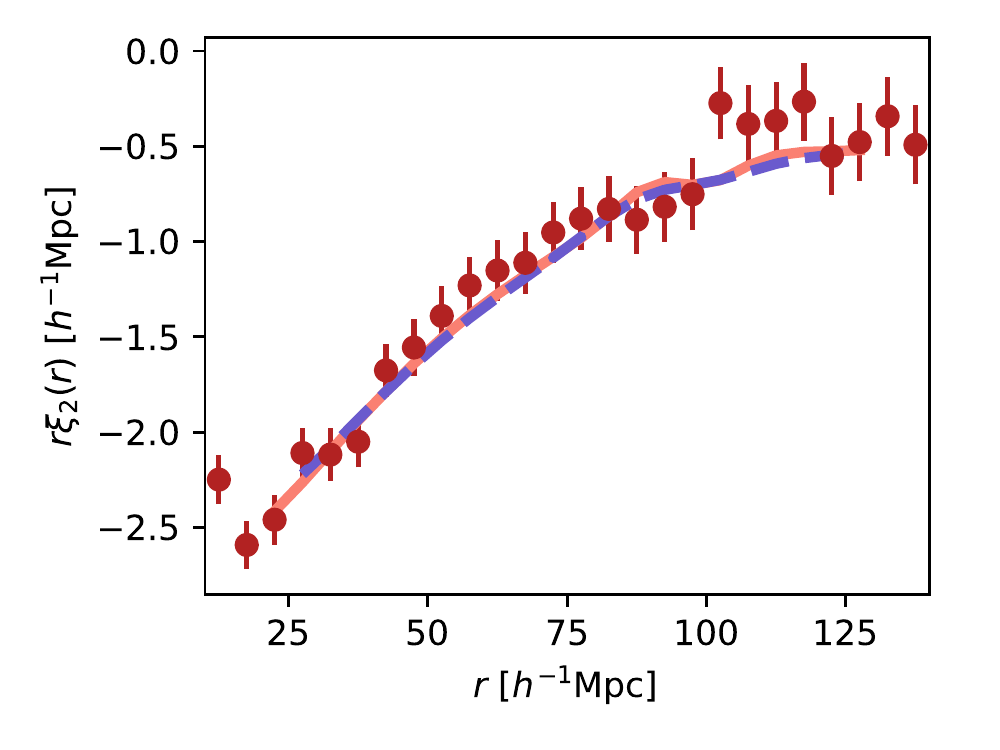}
    \includegraphics[width=0.32\textwidth]{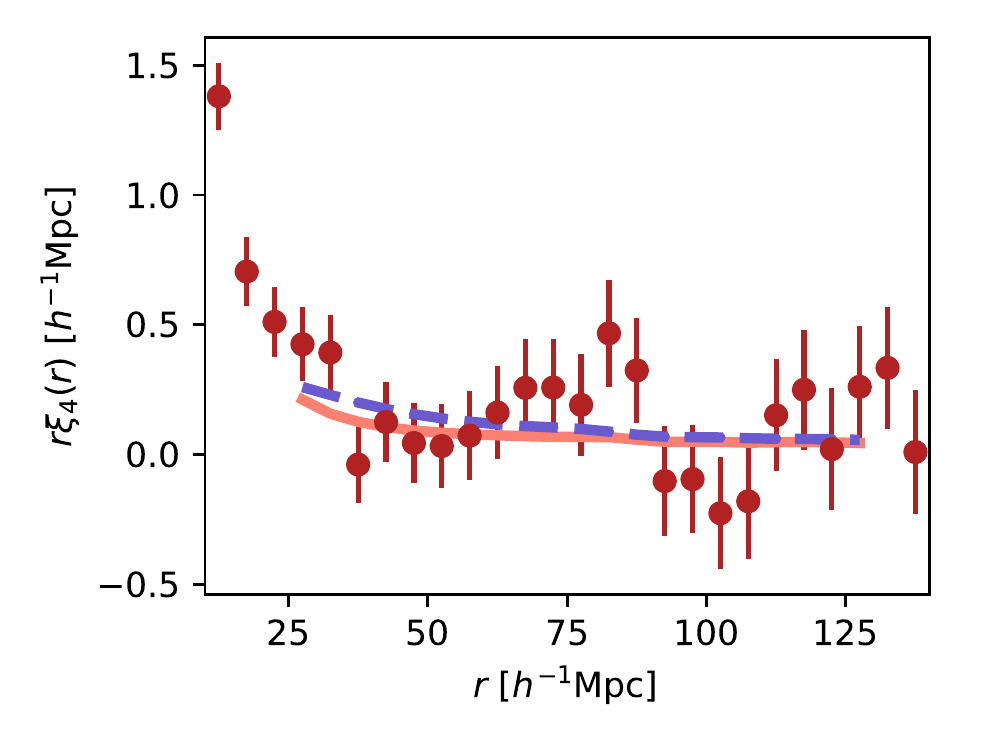}

    \caption{Best-fits full-shape models to the eBOSS + CMASS 
    multipoles. Left, mid and right panel display mono, quad and
    hexadecapole, respectively. The monopole is scaled by $r^2$ while the other 
    two are scaled by $r$.
    The CLPT-GS model is shown by the blue dashed line while the TNS
    model is shown by the red solid line. Note the baseline ranges
    used for each model are slightly different (see Figure~\ref{fig:rmin_impact}).
    }
    \label{fig:rsd_bestfit}
\end{figure*}

We present in Figure~\ref{fig:rsd_bestfit} the best-fit 
TNS (red) and CLPT-GS (blue) RSD models to the pre-reconstruction 
eBOSS LRG multipoles. The associated reduced chi-squared values 
are 
$\chi^2/{\rm dof} = 85.2/(65-7) = 1.47$ for TNS and 
$\chi^2/{\rm dof} = 83.7/(63-6) = 1.47$ for CLPT-GS. 
While these values are unlikely explained by statistical fluctuations, 
we verified that the values reported for the $\chi^2$ 
for both models are within {\sc EZmock} $\chi^2$ distributions.
Both models perform similarly on data, but some differences are 
visible in Figure~\ref{fig:rsd_bestfit}.
The TNS model produces a slightly sharper BAO peak than the 
CLPT-GS model, clearly visible in the monopole. This is due 
the fact that, intrinsically, the CLPT-GS model tends to predict 
a slighter higher BAO damping compared to Eulerian perturbation 
theory, as implemented here in the TNS model with {\sc RESPRESSO} 
prescription. The CLPT-GS model have a slightly higher hexadecapole
amplitude than the TNS model but both models seems to underestimate 
the hexadecapole amplitude below 35 \hmpc\ by 1$\sigma$ of the 
statistical uncertainties of the data. This underestimation in 
the amplitude of the hexadecapole is also present in the mocks 
for both the \textsc{Nseries} and \textsc{EZmocks} and was already 
reported in \citet{icaza-lizaola_clustering_2020} explaining 
the relative high $\chi^{2}$ of the data.

Table~\ref{tab:rsd_results_variations} shows the impact of different 
modifications in the methodology around the baseline configuration. 
First, if we change the range of scales used in the hexadecapole by 
changing $r_{\rm min}$ from  25\hmpc\ to 35\hmpc. We see a decrease
of the reduced chi-squared as we remove these scales from the 
hexedecapole, which are underestimated by the models. Removing those 
scales impact the measured cosmological parameters, particularly
$\apara$, which is shifted by about 1 $\sigma$. We performed
the same cuts on the analysis of {\sc EZmocks}, finding that 
such a shift lies at about 2.3$\sigma$ of the shifts observed in 1000 mocks
(see details in Appendix~\ref{app:hexadec}). The NGC and SGC fields are 
two independent samples and we find that their individual best-fit 
$\aperp$ and $\apara$, and $f\sigma_8$ are 0.7$\sigma$, $0.5\sigma$
and 0.3$\sigma$ from each other respectively for CLPT-GS and 
0.7$\sigma$, $0.8\sigma$ and 0.1$\sigma$ for TNS, which is not a 
significant difference. 

\begin{figure}
    \centering
    \includegraphics[width=\columnwidth]{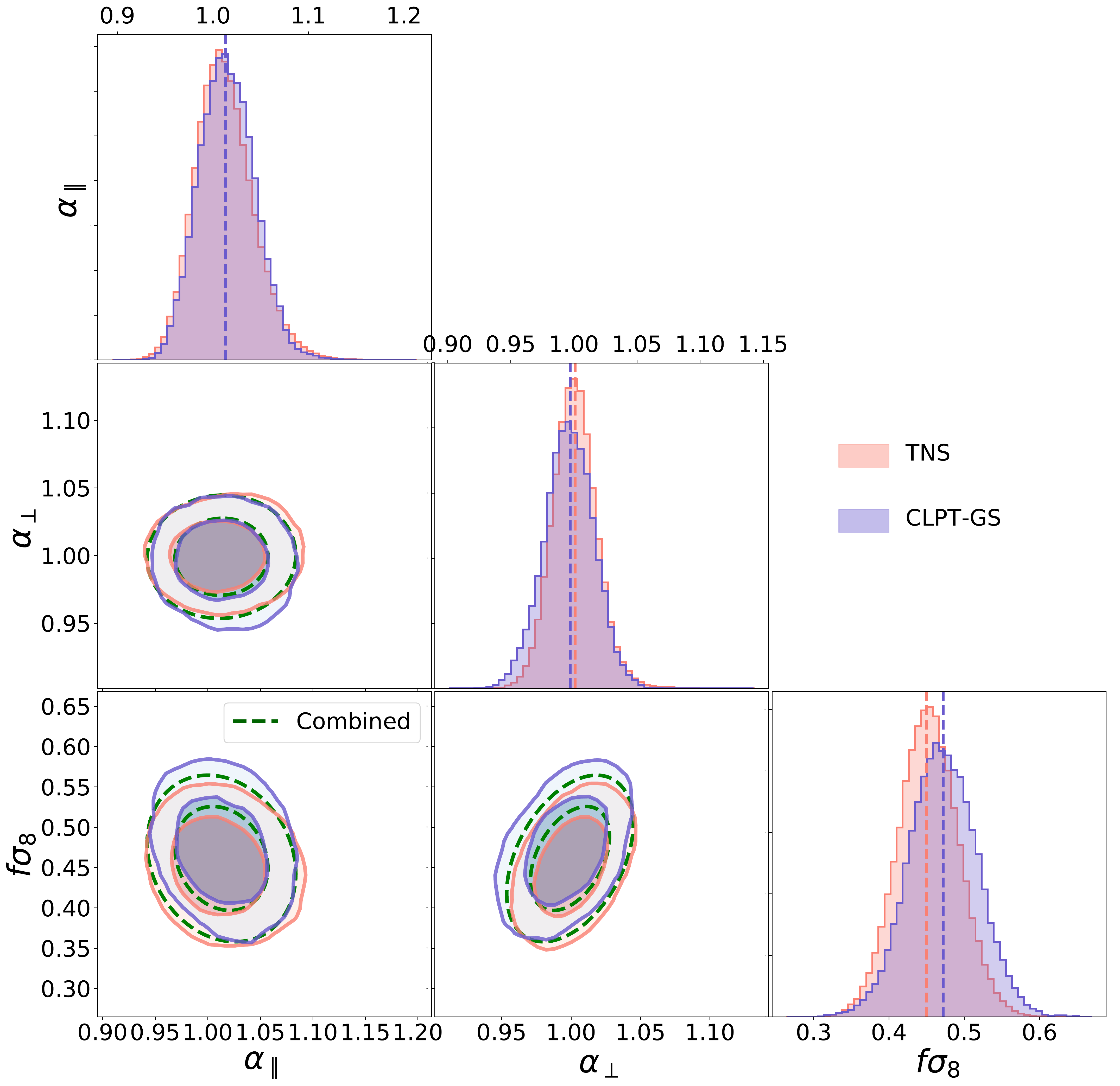}
    \caption{Comparison between the TNS and CLPT-GS final posterior
    distributions over the three main parameters using the DR16 data. The 
    distributions are in good agreement for the two models. The vertical dashed 
    lines on the 1D distributions refer to the mean. Dashed line contours
    show the combined result from the two models, assumming Gaussian 
    errors. The full posteriors including nuisance parameters can be found in Appendix~\ref{app:posteriors}.
    }
    \label{fig:rsd_likelihood}
\end{figure}

The marginal posteriors on $\aperp, \apara, \fsig$ and associated 
68\% and 95\% confidence contours are shown in 
Figure~\ref{fig:rsd_likelihood}. The posteriors obtained from 
both models are in good agreement. Entries denoted as 
``RSD $\xi_\ell $ CLPT-GS'' and ``RSD $\xi_\ell$ TNS'' in 
Table~\ref{tab:finalresults} gives the best-fit parameters 
and 1$\sigma$ error (including systematic errors), translated 
into $D_M/r_d, D_H/r_d, \fsig$. We find an excellent agreement 
in the best-fit parameters and errors between the two RSD models, 
as expected from the posteriors. The full posteriors including 
all nuisance parameters can be found in Appendix~\ref{app:posteriors}.

We combine the results from our two RSD models using a weighted average 
based on the individual covariance matrices 
(see Section~\ref{sec:combining_analysis}). The combined 
measurement is indicated by ``RSD $\xi_\ell$'' in 
Table~\ref{tab:finalresults} and shown with dashed 
contours in Figure~\ref{fig:rsd_likelihood}. Central 
values and errors of the combined result fall approximately 
in between the values of each individual measurement. 

The combined best-fit parameters and covariance matrix of 
the full-shape RSD analysis in configuration space, 
including systematic errors, are 
 \begin{equation}
 \mathbf{D}_{{\rm RSD},{\xi_\ell}} =  
 \begin{pmatrix}
D_M/r_d \\
D_H/r_d \\
f\sigma_8 
 \end{pmatrix}=
 \begin{pmatrix}
 17.42 \pm 0.40 \\
 20.46 \pm 0.70 \\
 0.460 \pm 0.050
 \end{pmatrix}
 \end{equation}
  \begin{equation}
\mathbf{C}_{{\rm RSD},{\xi_\ell}} = \begin{blockarray}{ccc}
D_M/r_d & D_H/r_d & \fsig \vspace{1mm} \\
\begin{block}{(ccc)}
1.59 \times 10^{-1} & 6.28 \times 10^{-3} & 6.13 \times 10^{-3} &  \\
 -  & 4.88 \times 10^{-1} & -4.83 \times 10^{-3} &  \\
 -  &  -  & 2.46 \times 10^{-3} &  \\
\end{block}
\end{blockarray}
\end{equation}
This corresponds to a 2.3 and 3.4 per cent measurements of the 
transverse and radial dilation parameters and a 11 per cent 
measurement of the growth rate of structure times $\sigma_8$.
The errors on $D_M/r_d$ and $D_H/r_d$ are slightly larger than 
the ones from the BAO-only analysis, as expected, but the 
correlation coefficient between them is reduced from $-0.33$ 
to $0.02$. This happens because information on dilation parameters
also come from the full-shape of the correlation function, 
rather than just the BAO peak. For instance, the correlation 
coefficient between $\fsig$ and $D_M/r_d$ is 0.31 and between 
$\fsig$ and $D_H/r_d$ is -0.14. 

\begin{figure}
    \centering
    \includegraphics[width=\columnwidth]{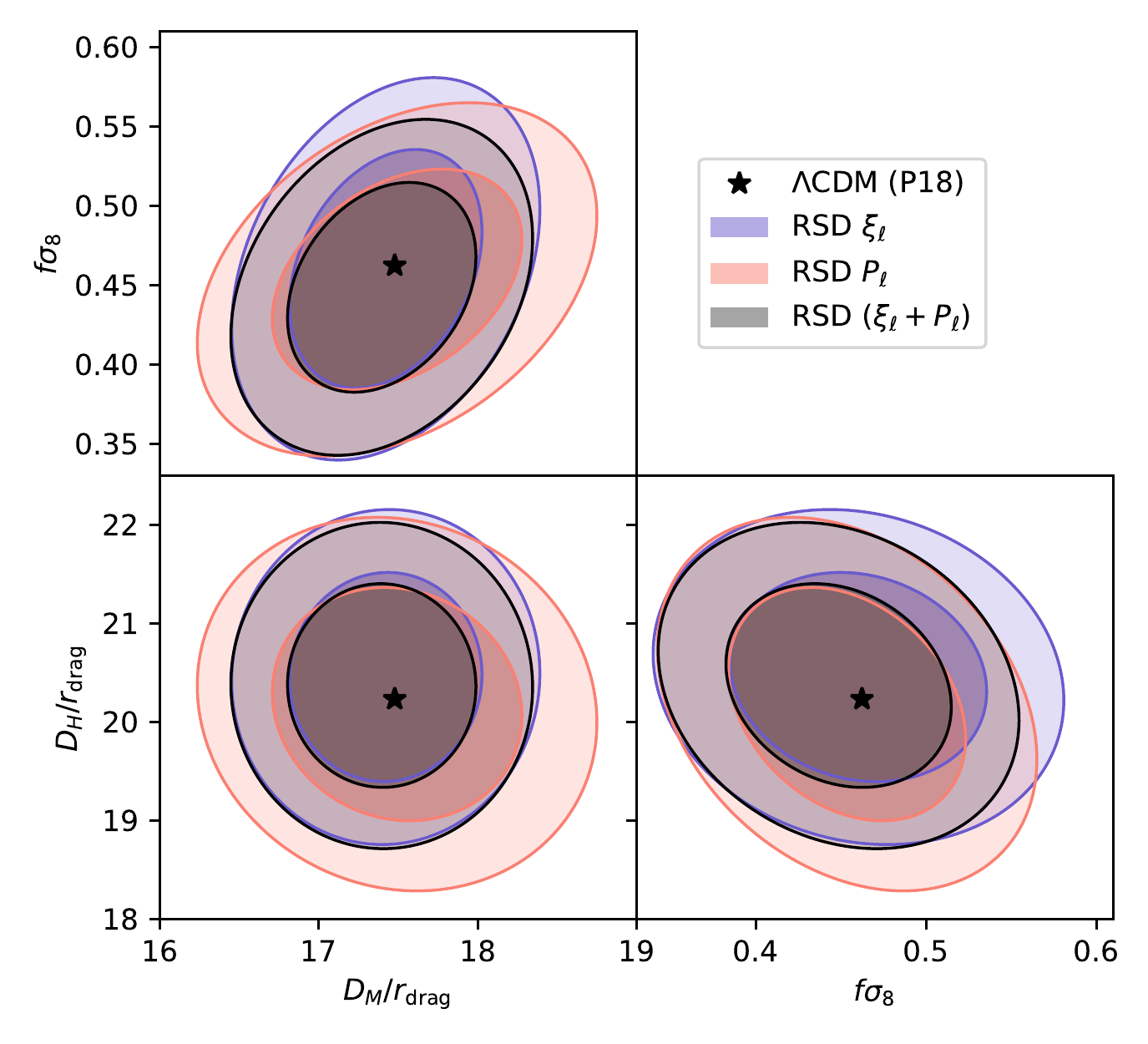}
    \caption{Constraints on $D_M/r_d, D_H/r_d$ and $\fsig$ $z_{\rm eff} = 0.698$ 
    from the full-shape 
    RSD analysis of the completed eBOSS LRG sample pre-reconstruction.
    Contours show 68 and 95 per cent confidence regions for the analyses in 
    configuration space (blue), Fourier space (red) 
    and the combined (grey). 
    The expected values in a flat $\Lambda$CDM 
    model with best-fit parameters from Planck 2018 results 
    is indicated as a black star.}
    \label{fig:consensus_rsd}
\end{figure}

\subsection{Consensus Results}
\label{sec:consensus_results}

\begin{figure}
    \centering
    \includegraphics[width=\columnwidth]{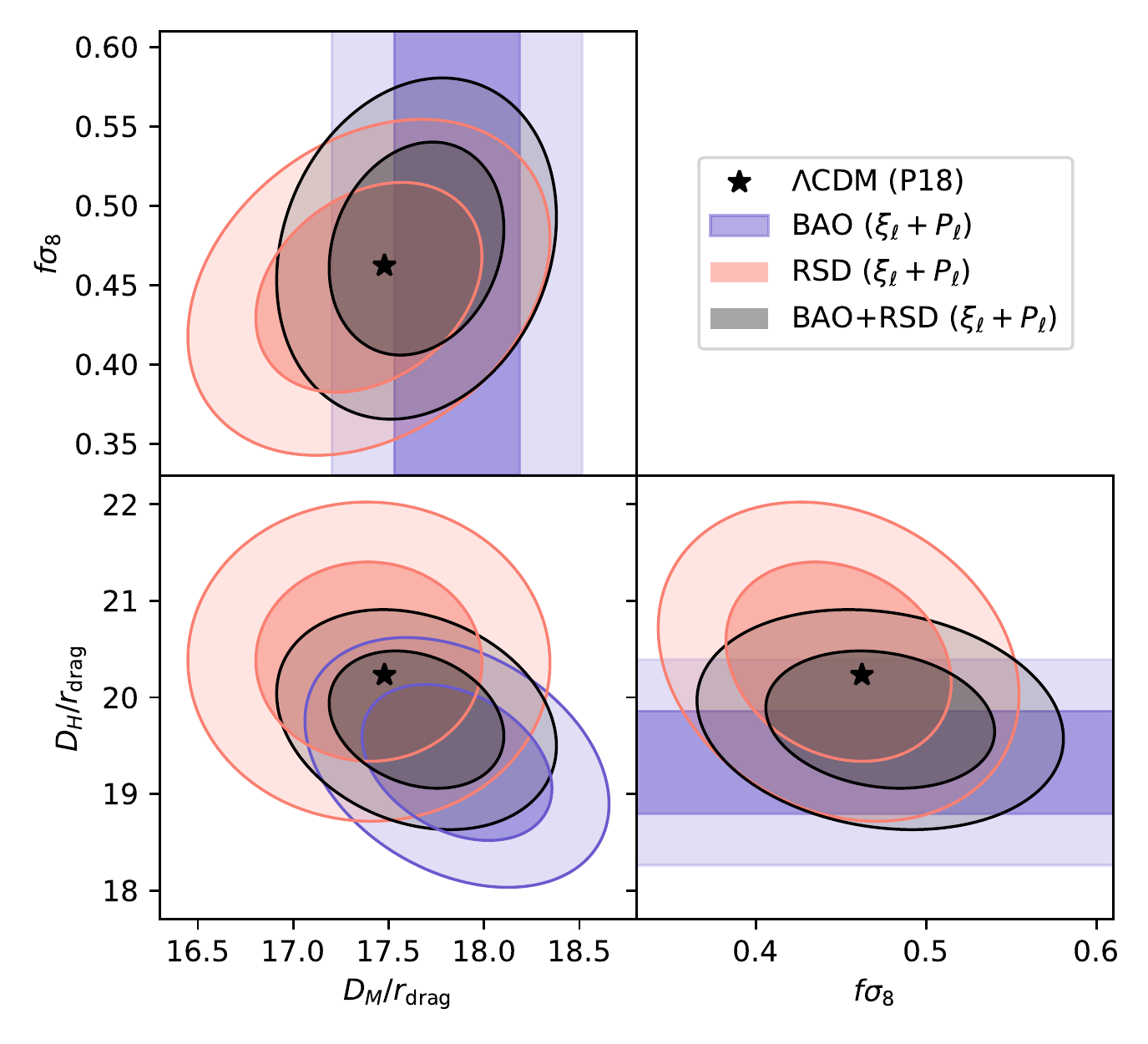}
    \includegraphics[width=\columnwidth]{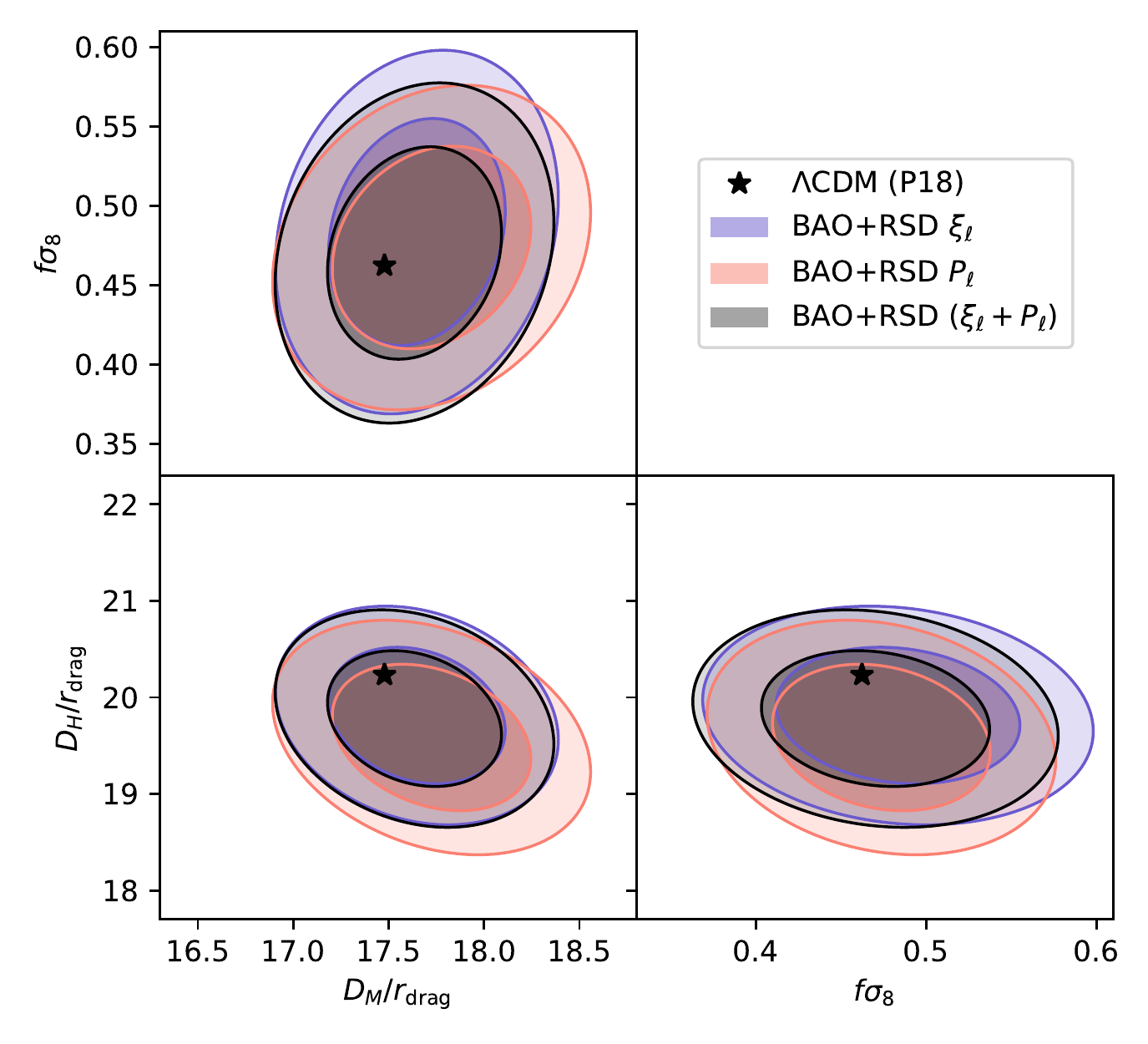}
    \caption{Final measurements of $D_M/r_d, D_H/r_d, \fsig$ from the 
    completed eBOSS LRG sample at $z_{\rm eff}=0.698$. 
    Top and bottom panels show two possible procedures for obtaining
    the final result. 
    The grey contours show the final results, which virtually the same 
    in both panels (two bottom lines in Table~\ref{tab:finalresults}).
    The black star indicates the prediction in a flat $\Lambda$CDM 
    model with parameters from Planck 2018 results. } 
    \label{fig:consensus_bao_rsd}
\end{figure}

We present in Figure~\ref{fig:consensus_bao_rsd} the final results of this work obtained by the combination of BAO and full-shape RSD analyses in both configuration and Fourier spaces. Accounting for all sources of systematic error discussed in Section~\ref{sec:systematics_rsd} and \ref{sec:statistical_properties}, the best-fit parameters and associated covariance matrix are:
 \begin{equation}
 \mathbf{D}_{{\rm LRG}} = 
 \begin{pmatrix}
     D_M/r_d  \\ 
     D_H/r_d  \\
     \fsig
 \end{pmatrix} = 
 \begin{pmatrix}
 17.65 \pm 0.30 \\
 19.77 \pm 0.47 \\
 0.473 \pm 0.044
 \end{pmatrix}
 \label{eq:consensus_vector}
 \end{equation}
  \begin{equation}
\mathbf{C}_{{\rm LRG}} = \begin{blockarray}{ccc}
D_M/r_d & D_H/r_d & \fsig \vspace{1mm} \\
\begin{block}{(ccc)}
9.11 \times 10^{-2} & -3.38 \times 10^{-2} & 2.47 \times 10^{-3} &  \\
 -  & 2.20 \times 10^{-1} & -3.61 \times 10^{-3} &  \\
 -  &  -  & 1.96 \times 10^{-3} &  \\
\end{block}
\end{blockarray}
\label{eq:consensus_covmatrix}
\end{equation}
which translate into a 1.7 and 2.4 per cent measurement of $D_M/r_d$ 
and $D_H/r_d$ respectively. The correlation between these two is 
$-24$ per cent. The error on $\fsig$ is 9.4 per cent, which is 
the most precise measurement to date in this redshift range. 
We note that this final measurement is not sensitive to the order 
of combinations, as seen in the second panel of 
Figure~\ref{fig:consensus_bao_rsd} and in the last row of 
Table~\ref{tab:finalresults}. Those measurements agree well 
with the predictions from \citet{planck_collaboration_planck_2018-1}, 
which predict at this redshift: 
17.48, 20.23 and 0.462, respectively, for a flat $\Lambda$CDM model 
assuming gravity is described by General Relativity.
These values are shown as 
stars in Figure~\ref{fig:consensus_bao_rsd}.

Systematic errors originating from observational effects, modelling and 
combination methods were carefully included in our measurements and 
are responsible for inflating final errors by 6, 13 and 20 per cent,
respectively, on $D_M/r_d, D_H/r_d$ and $\fsig$.
In Section~\ref{sec:statistical_properties}, we found
that our statistical errors on the consensus $\fsig$ were slightly 
underestimated. To apply this correction on the data consensus,
we proceed as follows. First, we compute consensus with and without
accounting for systematic errors from Table~\ref{tab:rsd_total_systematic}.
The difference between their error matrices gives us the additive systematic
matrix. Then, we scale the statistical errors on $\fsig$ by 1.09 and we 
add back the additive systematic matrix. 
This procedure yields the results reported in Eq.~\ref{eq:consensus_vector} 
and \ref{eq:consensus_covmatrix}.

\begin{table}
\caption{
Summary table with results from this work, from \citet{gil-marin_2020}, 
and their combination. All reported errors include the systematic component.
The effective redshift of all measurements is $z_{\rm eff} = 0.698$.
 }
\begin{center}
\begin{tabular}{lccc}
\hline
\hline
Method & $D_M/r_d$  & $D_H/r_d$ & $f\sigma_8$ \\ 
\hline
 {\sc bao} $\xi_\ell$
 & $17.86 \pm 0.33$ & $19.34 \pm 0.54$ & -   \\
{\sc bao} $P_\ell$
 & $17.86 \pm 0.37$ & $19.30 \pm 0.56$ & -   \\
{\sc bao} $\xi_\ell+P_\ell$
 & $17.86 \pm 0.33$ & $19.33 \pm 0.53$ & -   \\
 \hline
{\sc rsd} $\xi_\ell$ {\sc clpt}
 & $17.39 \pm 0.43$ & $20.46 \pm 0.68$ & $0.471 \pm 0.052$  \\
{\sc rsd} $\xi_\ell$ {\sc tns}
 & $17.45 \pm 0.38$ & $20.45 \pm 0.72$ & $0.451 \pm 0.047$  \\
{\sc rsd} $\xi_\ell$
 & $17.42 \pm 0.40$ & $20.46 \pm 0.70$ & $0.460 \pm 0.050$  \\
{\sc rsd} $P_\ell$
 & $17.49 \pm 0.52$ & $20.18 \pm 0.78$ & $0.454 \pm 0.046$  \\
{\sc rsd} $\xi_\ell+P_\ell$
 & $17.40 \pm 0.39$ & $20.37 \pm 0.68$ & $0.449 \pm 0.044$  \\
 \hline
{\sc bao}$+${\sc rsd} $\xi_\ell$
 & $17.65 \pm 0.31$ & $19.81 \pm 0.47$ & $0.483 \pm 0.047$  \\
{\sc bao}$+${\sc rsd} $P_\ell$
 & $17.72 \pm 0.34$ & $19.58 \pm 0.50$ & $0.474 \pm 0.042$  \\
 \hline
{\sc bao} ($\xi_\ell+P_\ell$) $+$ 
 & $17.65 \pm 0.30$ & $19.77 \pm 0.47$ & $0.473 \pm 0.044$  \\
 {\sc rsd}  ($\xi_\ell+P_\ell$) \\
({\sc bao}$+${\sc rsd}) $\xi_\ell$ $+$ 
 & $17.64 \pm 0.30$ & $19.78 \pm 0.46$ & $0.470 \pm 0.044$  \\
 ({\sc bao}$+${\sc rsd}) $P_\ell$ \\
 \hline
 \hline
\end{tabular}
\end{center}
\label{tab:finalresults}
\end{table}

\subsection{Comparison with previous results}
\label{sec:comparison_previous}

Our final consensus result for the DR16 LRG sample is shown in
Eqs.~\ref{eq:consensus_vector} and \ref{eq:consensus_covmatrix}, 
and used a total of 402,052 (weighted) galaxies over 9,463~deg$^2$ (with 4,242~deg$^2$ observed by eBOSS). 
\citet{bautista_sdss-iv_2018} and \citet{icaza-lizaola_clustering_2020}
describe, respectively, the BAO and full-shape RSD measurements using the 
DR14 LRG sample, that contains 126,557 galaxies over 1,844~deg$^2$. 
In the DR14 sample, CMASS galaxies outside of the eBOSS footprint were 
not used. Because of that, the effective redshift of the DR14 
measurements is slightly higher, at $z_{\rm eff} = 0.72$. 

\citet{bautista_sdss-iv_2018} reported a 2.5 per cent measurement of the ratio of the spherically averaged distance to the sound horizon scale, $D_V(z=0.72)/r_d = 16.08^{+0.41}_{-0.40}$. This result was obtained with isotropic fits to the monopole of the post-reconstruction 
correlation function. The statistical power of the DR14 sample is relatively low for anisotropic BAO constraints and has large non-Gaussian errors. Converting our DR16 anisotropic measurement of 
Eq.~\ref{eq:consensus_vector} into spherically averaged distances we 
obtain: $D_V(z=0.698)/r_d = 16.26 \pm 0.20$, which is well within 
1$\sigma$ from the DR14 value. The error on $D_V$ has reduced by a factor 
of two, slightly more than the square-root of the increase in 
effective volume, which gives a factor of 
$\sqrt{ V_{\rm eff, DR16}/V_{\rm eff, DR14}} = \sqrt{2.73/0.9} \sim 1.74$.
Note that in DR16 we combine BAO and full-shape analysis in Fourier and 
configuration spaces, which maximizes the amount of extracted cosmological information. 

\citet{icaza-lizaola_clustering_2020} presented the full-shape RSD analysis in the DR14 LRG sample in configuration space, yielding 
$f \sigma_8 = 0.454 \pm 0.134$,
$D_M/r_d = 17.07 \pm 1.55$, and 
$D_H/r_d = 19.17 \pm 2.84$. All values are consistent within 1$\sigma$ of DR16 results, even though errors for DR14 are quite large given the even lower significance of the BAO peak in the pre-reconstruction multipoles. The error on the growth rate of structure $\fsig$ reduces by a factor of 3 in DR16 compared to DR14, clearly benefiting from the larger sample and the combination with post-reconstruction BAO results that help breaking model degeneracies.

Our DR16 LRG results at $0.6 < z < 1.0$ supersede the highest redshift results of the DR12 BOSS sample at $0.5 < z < 0.75$, which has an effective redshift of $z_{\rm eff} = 0.61$. \citet{alam_clustering_2017} report a 1.4, 2.2 and 7.8 per cent measurements of $D_M/r_d, D_H/r_d$ and $\fsig$ respectively. While the errors in the high-redshift bin are slightly smaller than our DR16 result, it has a large correlation with the intermediate-redshift bin at $0.4 < z <0.6$. Our DR16 measurement is thus virtually independent of the first two DR12 BOSS redshift bins, and has effectively more weight in the final joint cosmological constraints. The cosmological implications of our DR16 LRG measurements are fully described in \citet{mueller_2020}.

\section{Conclusion} 
\label{sec:conclusion}

This work presented the cosmological analysis of the configuration-space anisotropic clustering in the DR16 eBOSS LRG sample, which is used for the final cosmological analysis of the completed eBOSS survey. We extracted and model the BAO and RSD features from the galaxy two-point correlation function monopole, quadrupole, and hexadecapole moments. We used the reconstruction technique to sharpen the BAO peak and mitigate associated non-linearies. The pre- and post-reconstruction multipole moments were used to perform a full-shape RSD analysis and a BAO-only analysis, respectively. In the RSD analysis, we considered two different RSD models, which results were later combined to increase the robustness and accuracy of the measurements. The combination of the BAO-only and full-shape RSD analyses allowed us to derive joint constraints on the three cosmological parameter combinations: $D_H(z)/r_d$, $D_M(z)/r_d$ and $f\sigma_8(z)$. This analysis is complementary to that performed in Fourier space and presented in Gil-Marin et al. 2020. We found an excellent agreement between the inferred parameters in both spaces, both for BAO-only and full-shape RSD analyses. After combining the results with those from that in Fourier space, we obtain the following final constraints: 
     $D_M/r_d =  17.65 \pm 0.30$,
     $D_H/r_d = 19.77 \pm 0.47$, 
     $\fsig = 0.473 \pm 0.044$,
which are currently the most accurate at $z_{\rm eff}=0.698$.

The adopted methodology has been extensively tested on a set of realistic simulations and shown to be very robust against systematics. In particular, we investigated different potential sources of systematic errors: inaccuracy in the modelling of both BAO/RSD and intrinsic galaxy clustering, arbitrary choice of reference cosmology, and systematic errors from observational effects such as redshift failures, fiber collision, incompleteness, or the radial integral constraint. We quantified the associated systematic error contributions and included them on the final cosmological parameter constraints. Overall, we found that the total systematic error inflate
errors by 6, 13 and 20 per cent for $\aperp$, $\apara$ and $\fsig$.

The cosmological parameters inferred from the DR16 eBOSS LRG sample are in good agreement with the predictions from General Relativity in a flat $\Lambda$CDM cosmological model with parameters set to Planck 2018 results. These measurements complement those obtained from the other eBOSS tracers \citep{raichoor_2020, de_mattia_2020, hou_2020, neveux_2020, du_mas_des_bourboux_2020}. The full cosmological interpretation of all eBOSS tracer results and combined with previous BOSS results is presented in \citet{mueller_2020}.

Future large spectroscopic surveys such as DESI or Euclid will probe much larger volumes of the Universe. This will allow reducing the statistical errors on the cosmological parameters considerably, at the percent level or below. For those it will be crucial to control the level of systematics at a extremely low level. This is today a challenge and the work presented here has shown the current state-of-the-art methodology, which will have to be further developed and improved in view of the optimal exploitation of next-generation surveys. 

\section*{Data Availability}
The correlation functions, covariance matrices, and resulting likelihoods for cosmological parameters are (will be made) available (after acceptance) via the SDSS Science Archive Server (\href{https://sas.sdss.org/}{https://sas.sdss.org/}), with the exact address tbd.

\section*{Acknowledgements}
RP, SdlT, and SE acknowledge the support from the French National Research Agency (ANR) under contract ANR-16-CE31-0021, eBOSS. SdlT and SE acknowledge the support of the OCEVU Labex (ANR-11-LABX-0060) and the A*MIDEX project (ANR-11-IDEX-0001-02) funded by the "Investissements d’Avenir" French government program managed by the ANR. MVM and SF are partially supported by Programa de Apoyo a Proyectos de Investigaci\'on e Inovaci\'on ca Teconol\'ogica (PAPITT)  no. IA101518, no. IA101619, Proyecto LANCAD-UNAM-DGTIC-319 and LANCAD-UNAM-DGTIC-136. HGM acknowledges the support from la Caixa Foundation (ID 100010434) which code LCF/BQ/PI18/11630024. SA is supported by the European Research Council through the COSFORM Research Grant (\#670193). GR, PDC and JM acknowledge support from the National Research Foundation of Korea (NRF) through Grants No. 2017R1E1A1A01077508 and No. 2020R1A2C1005655 funded by the Korean Ministry of Education, Science and Technology (MoEST), and from the faculty research fund of Sejong University. 
 
Numerical computations were done on the Sciama High Performance Compute (HPC) cluster which is supported by
the ICG, SEPNet and the University of Portsmouth.
This research used resources of the National Energy Research Scientific Computing Center, a DOE Office of Science User Facility supported by the Office of Science of the U.S. Department of Energy under Contract No. DE-AC02-05CH11231.
This research also uses resources of the HPC cluster ATOCATL-IA-UNAM M\'exico. This project has received funding from the European Research Council (ERC) under the European Union’s Horizon 2020 research and innovation program (grant agreement No 693024).

Funding for the Sloan Digital Sky Survey IV has been provided by the Alfred P. Sloan Foundation, the U.S. Department of Energy Office of Science, and the Participating Institutions. SDSS-IV acknowledges
support and resources from the Center for High-Performance Computing at
the University of Utah. The SDSS web site is www.sdss.org.

SDSS-IV is managed by the Astrophysical Research Consortium for the 
Participating Institutions of the SDSS Collaboration including the 
Brazilian Participation Group, the Carnegie Institution for Science, 
Carnegie Mellon University, the Chilean Participation Group, the French Participation Group, Harvard-Smithsonian Center for Astrophysics, 
Instituto de Astrof\'isica de Canarias, The Johns Hopkins University, Kavli Institute for the Physics and Mathematics of the Universe (IPMU) / 
University of Tokyo, the Korean Participation Group, Lawrence Berkeley National Laboratory, 
Leibniz Institut f\"ur Astrophysik Potsdam (AIP),  
Max-Planck-Institut f\"ur Astronomie (MPIA Heidelberg), 
Max-Planck-Institut f\"ur Astrophysik (MPA Garching), 
Max-Planck-Institut f\"ur Extraterrestrische Physik (MPE), 
National Astronomical Observatories of China, New Mexico State University, 
New York University, University of Notre Dame, 
Observat\'ario Nacional / MCTI, The Ohio State University, 
Pennsylvania State University, Shanghai Astronomical Observatory, 
United Kingdom Participation Group,
Universidad Nacional Aut\'onoma de M\'exico, University of Arizona, 
University of Colorado Boulder, University of Oxford, University of Portsmouth, 
University of Utah, University of Virginia, University of Washington, University of Wisconsin, 
Vanderbilt University, and Yale University.




\bibliographystyle{mnras}
\bibliography{MyLibrary.bib} 

\appendix

\section{Impact of scales used from the hexacadecapole}

\label{app:hexadec}

\begin{figure}
    \centering
    \includegraphics[width=\columnwidth]{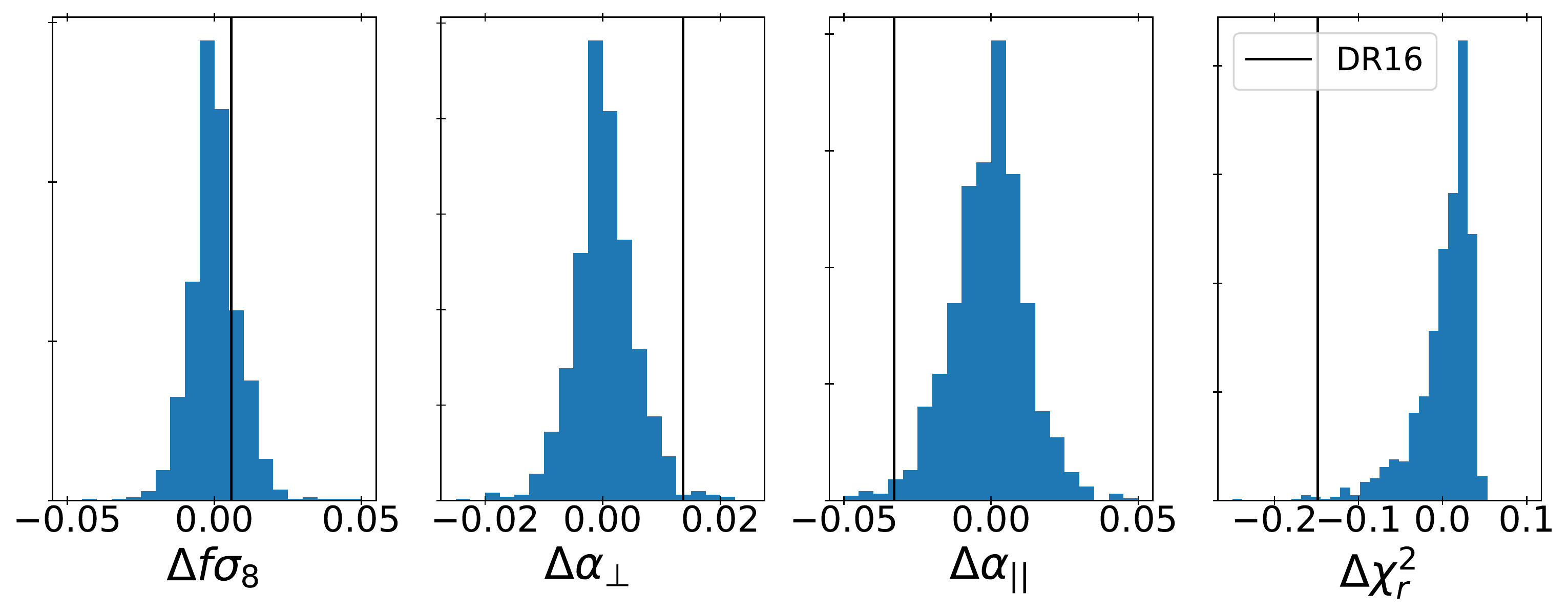}
    \caption{Variation of the cosmological parameters and the reduced chi-squared as a function of the truncation scale of the hexadecapole for the TNS model. The normalisazed distributions corresponds to 1000 \textsc{EZmocks} while the vertical lines corresponds to the shift for the data.}
    \label{fig:25_vs_35}

\end{figure}

\begin{figure}
    \centering
    \includegraphics[width=\columnwidth]{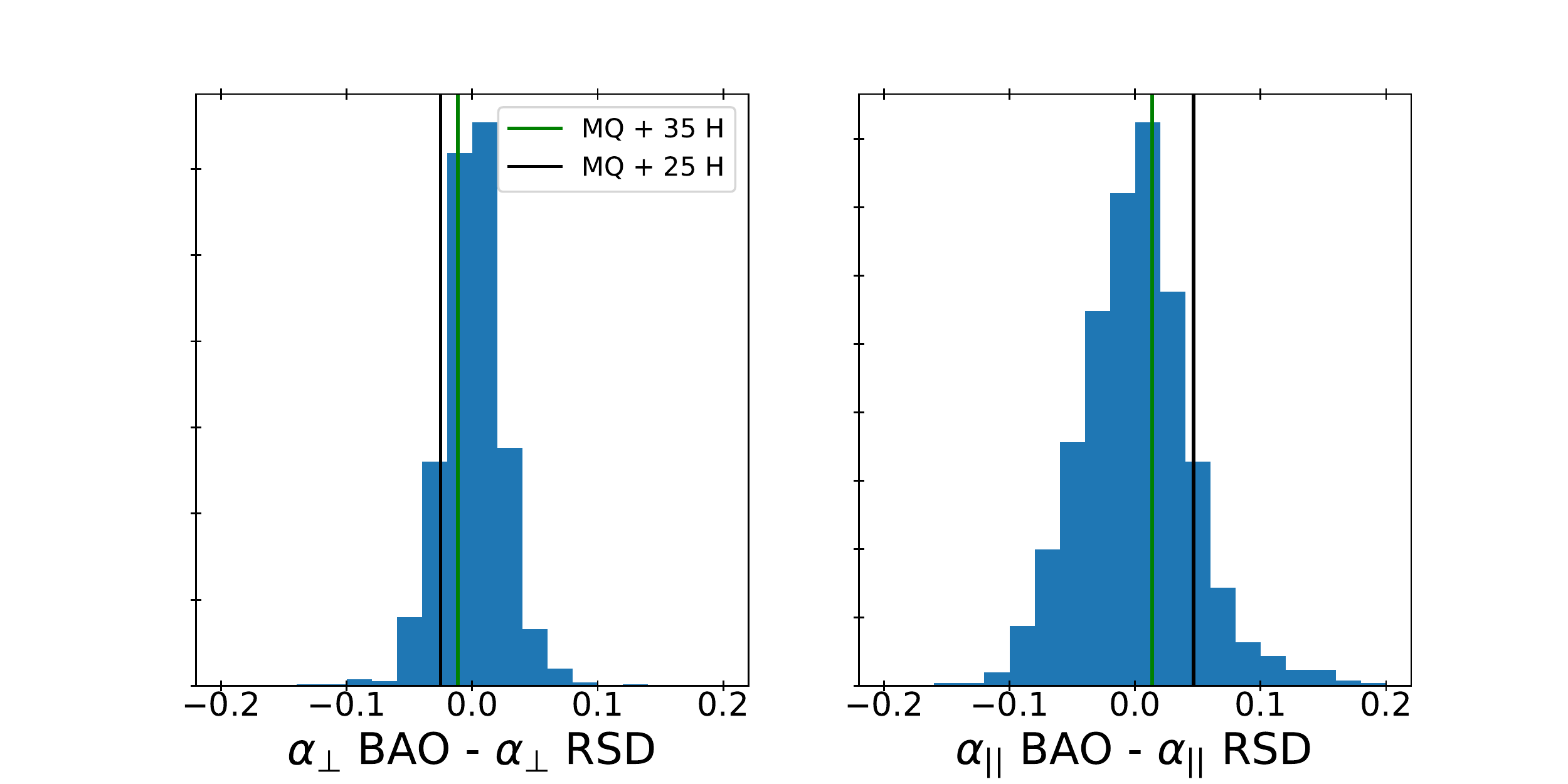}
    \caption{Absolute difference between TNS and BAO post recon constraint on the alphas. The normalisazed distributions corresponds to 1000 \textsc{EZmocks}, while to the two vertical lines corresponds to the two differents truncation scales for the hexadecapole.}
    \label{fig:bao_vs_rsd}

\end{figure}

We present in Figure \ref{fig:25_vs_35} the distribution in the 
\textsc{EZmocks} of the difference on parameter constraints and 
reduced $\chi^2$ induced by including or not the smallest scales 
of the hexadecapole in the fit. We find that the distribution 
for each of the parameters is centered on zero with a standard 
deviation of 0.006, 0.015 and 0.01 for $\alpha_\perp$, 
$\alpha_\parallel$ and $f\sigma_{8}$ respectively, which 
correspond to 0.3, 0.4 and 0.2 $\%$ of the uncertainties on the 
RSD TNS measurements in the data. This demonstrates that 
cosmological constraints are stable to the choice of the 
truncation scale for the hexadecapole. The vertical line shows 
the corresponding shift in the data. This shift remains within 
1 $\sigma$ of the \textsc{EZmocks} distribution for $f\sigma_8$ 
and reaches up to 2.3 $\sigma$ for $\alpha_\parallel$.  For the 
reduced $\chi^2$, the observed difference is on the edge of the
\textsc{EZmocks} distribution. Even if few \textsc{EZmocks} 
realisations exhibit the same variation as in the data, the 
observed shifts are still statistically consistent.

Figure \ref{fig:bao_vs_rsd} displays the difference on the geometrical distortion parameters between the BAO post-reconstruction and RSD TNS measurements in the \textsc{EZmocks}. Similarly as in the previous figure, the vertical line shows the difference found in the data. While the differences for $\alpha_\perp$ and $\alpha_\parallel$ are smaller when the smallest scales of the hexadecapole are removed from the RSD fits, both measurements seem to be consistent with BAO post-reconstruction measurements similarly as in the mocks. Both distributions are centered on zero with a standard deviation between BAO and RSD measurements of 0.04 and 0.06 for $\alpha_\perp$ and $\alpha_\parallel$ respectively. This shows that as expected RSD measurements provide a more uncertain determination of the geometrical distortion parameters than BAO measurements.

\section{Impact of fiber collision correction scheme}
\label{app:pip_weights}

\begin{table}
    \centering
    \caption{Impact of the choice of fiber collision correction scheme on the recovered $\aperp$, $\apara$, and $\fsig$ parameters in the eBOSS LRG sample without CMASS galaxies.}
    \label{tab:pip_weights}
    \begin{tabular}{cccc}
    \hline
    \hline
        Model & Par & Base & PIP   \\
        \hline
            & $\aperp$ &   $1.189 \pm 0.062$ &  $1.199 \pm 0.070$  \\
        BAO & $\apara$ &  $0.850 \pm 0.071$ & $0.843 \pm 0.074$ \\
            & $\chi^2/{\rm dof}$ & $47.7/48 = 0.99$ & $51.7/48 = 1.08$ \\
        \hline
                & $\aperp$ & $1.009 \pm 0.046$ &  $0.980 \pm 0.044$ \\
        CLPT-GS & $\apara$ & $1.027 \pm 0.056$ &  $1.035 \pm 0.055$ \\
                & $\fsig$ &  $0.473 \pm 0.066$ & $0.446 \pm 0.066$ \\
                & $\chi^2/{\rm dof}$ & $67.8/54 = 1.26$ & $71.5/54 = 1.32$ \\
        \hline
                & $\aperp$ & $1.024 \pm 0.044$ & $1.001 \pm 0.041$ \\
        TNS     & $\apara$ & $1.038 \pm 0.050$ & $1.032 \pm 0.046$  \\
                & $\fsig$ &  $0.451 \pm 0.068$ & $0.420 \pm 0.065$ \\
                & $\chi^2/{\rm dof}$ &  $71.1/58 = 1.23$ & $74.8/58 = 1.29$  \\
    \hline
    \hline
    \end{tabular}
\end{table}

\citet{mohammad_2020} present an improved correction scheme for fiber collisions for the eBOSS LRG sample but without CMASS galaxies. It is based on the method of \citet{bianchi_unbiased_2017}, commonly referred to as the pair inverse probability (PIP) weighting. We performed fits of our BAO and full-shape RSD models to the multipoles for this restricted sample. Note that this sample is about two thirds of the full sample used in our
work. Table~\ref{tab:pip_weights} compares the results on $\aperp$, $\apara$, and $\fsig$ obtained with our baseline fiber collision correction to those using PIP weights. We see that the changes are small 
compared to the statistical errors for all methods and models. This is expected since PIP weights mostly impact the clustering on the smallest scales not used in our analysis ($r<20$\hmpc), while the baseline correction already well accounts for large-scale effects.

\section{Full parameter space posterior distributions}
\label{app:posteriors}

\begin{figure*}

    \centering
    \includegraphics[width=0.8\textwidth]{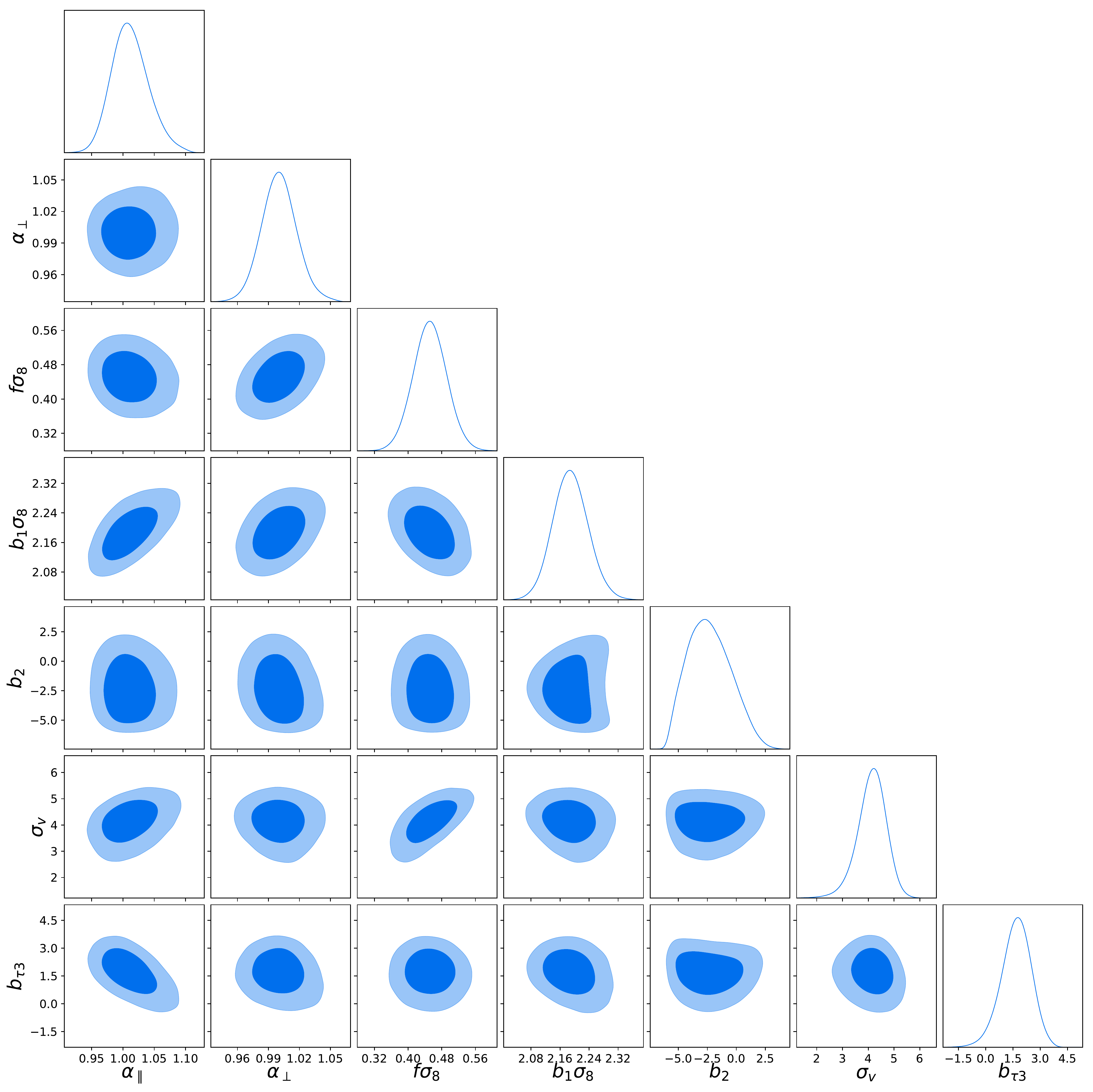}
    \caption{Full posterior distribution of the MCMC chain for the TNS model.}
    \label{fig:chain_TNS}

\end{figure*}

\begin{figure*}

    \centering
    \includegraphics[width=0.8\textwidth]{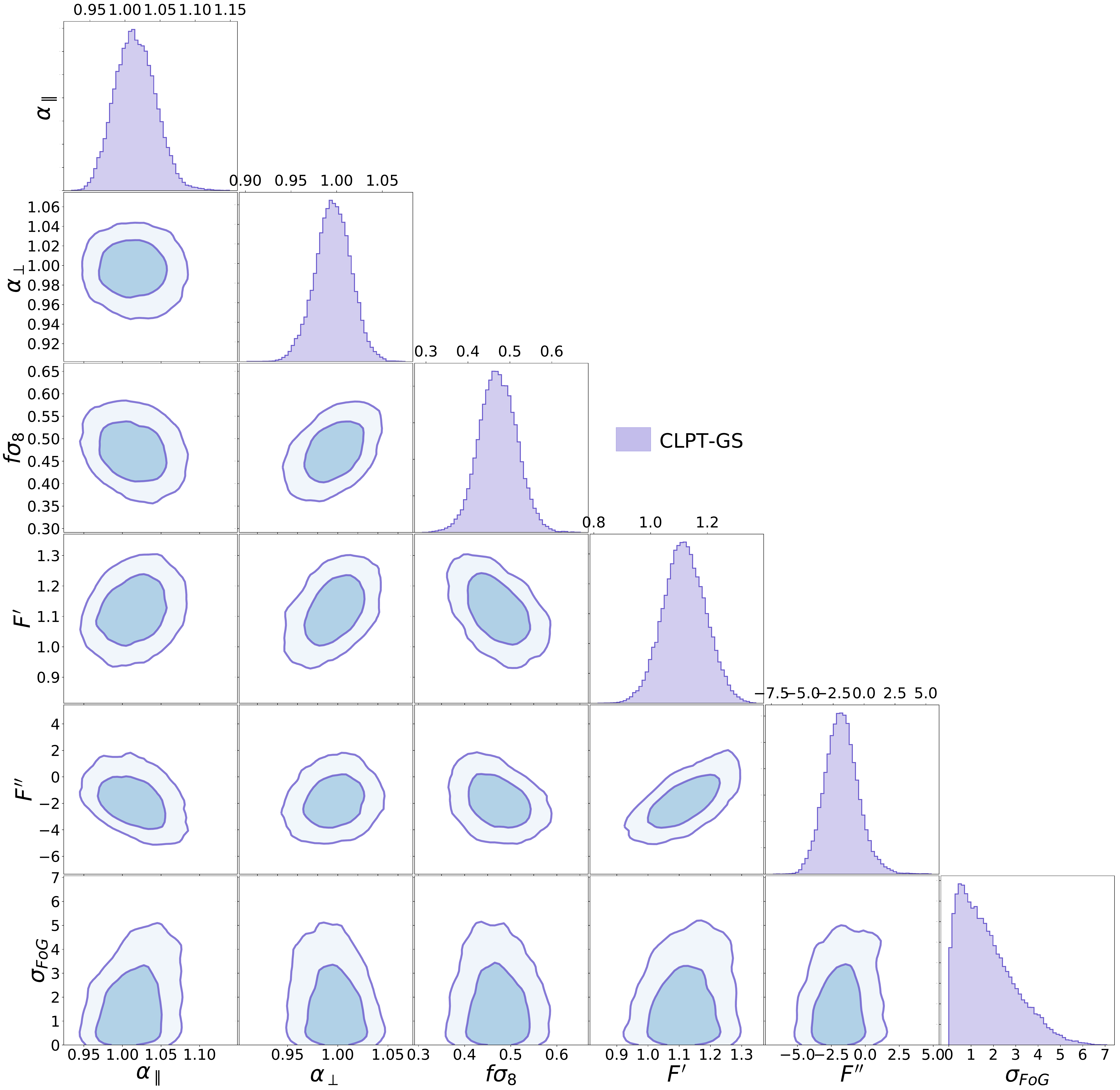}
    \caption{Full posterior distribution of the MCMC chain for the CLPT-GS model.}
    \label{fig:chain_CLPT}

\end{figure*}

For the sake of readability we do not present the full parameter space posterior distribution of our chains in the main text. However, it could be of interest to see the full information. Due to some differences between parameters in the two models we use, we separately present the full posterior distributions for TNS and CLPT-GS in figures \ref{fig:chain_TNS} and \ref{fig:chain_CLPT} respectively.

\label{app:hexadec}




\bsp	
\label{lastpage}
\end{document}